\documentclass[11pt]{article}
\pdfoutput=1

\usepackage[utf8]{inputenc}
\usepackage{lipsum}

\usepackage{amsmath, amssymb,amsfonts,amsthm,mathtools}
\numberwithin{equation}{section}
\usepackage{physics}  
\usepackage{xspace,graphicx,relsize,bm}
\usepackage[dvipsnames]{xcolor}
\usepackage[breakable,skins]{tcolorbox}
\usepackage{soul} 
\usepackage{titletoc}  
\usepackage{suffix}  

\usepackage[skip=0.3\baselineskip]{parskip}  
\usepackage{titlesec}  
\titlespacing{\paragraph}{0pt}{2ex}{.5em}  

\usepackage{libertinus}  
\usepackage[T1]{fontenc}
\usepackage{dsfont}  
\usepackage[zerostyle=a,scaled=.97]{newtxtt}  
\usepackage[libertine]{newtxmath}  
\usepackage[cal=stix, calscaled=.97]{mathalpha}  
\usepackage{anyfontsize}

\usepackage{xurl} 
\usepackage[
    backend=biber,
    style=alphabetic,
    sorting=anyt,
    minalphanames=3,
    maxalphanames=3,
    maxnames=99,
    backref=true
    ]{biblatex}
\DefineBibliographyStrings{english}{%
  backrefpage = {page},
  backrefpages = {pages},
}

\usepackage[singlespacing]{setspace}
\usepackage{hyperref}
\definecolor{linkcol}{rgb}{0.0,0.55,0.7}
\definecolor{citecol}{rgb}{0.0, 0.6, 0.45}
\definecolor{urlcol}{rgb}{0.7, 0.0, 0.55}
\hypersetup{
	colorlinks,
	linkcolor={linkcol},
	citecolor={citecol},
	urlcolor={urlcol}
}
\usepackage[margin=1in]{geometry}
\usepackage[hang,flushmargin]{footmisc}  
\usepackage{fancyhdr} 
\usepackage{subcaption}
\usepackage{mleftright}
\usepackage{multirow}  
\usepackage{array}  
\usepackage{changepage}  

\usepackage{xfrac}  

\usepackage{authblk}  

\usepackage{enumitem}  

\usepackage{cleveref} 

\def\01{\{0,1\}}
\RenewDocumentCommand{\ev}{}{\!\expectationvalue}
\RenewDocumentCommand{\mel}{}{\!\matrixelement}
\newcommand{\mc}[1]{\mathcal{#1}}
\newcommand{\mb}[1]{\mathbb{#1}}
\newcommand{\mrm}[1]{\mathrm{#1}}
\newcommand{\mds}[1]{\mathds{#1}}
\newcommand{\mtt}[1]{\mathtt{#1}}
\newcommand{\mbf}[1]{\mathbf{#1}}

\newcommand{\dualnorm}[1]{\norm{#1}_\star}
\newcommand{\midvert}{\;\middle\vert\;}
\newcommand{\cdotspace}{\,\cdot\,}
\newcommand{\doublecolon}{::}

\DeclareMathOperator*{\Ex}{\mathbf{E}}
\let\Pr\relax
\DeclareMathOperator*{\Pr}{\mathbf{Pr}}
\DeclareMathOperator*{\Var}{\mathbf{Var}}
\DeclareMathOperator*{\Cov}{\mathbf{Cov}}
\DeclareMathOperator*{\argmax}{arg\,max}
\DeclareMathOperator*{\argmin}{arg\,min}

\newcommand{\addsectionfooter}[1]{
    \rfoot{\footnotesize{\Cref{#1}\textcolor{linkcol}{:} \nameref{#1}}}
    }   
\newcommand{
    \clearsectionfooter}{\rfoot{}
    }   
\newcommand{\addsectionheader}[1]{
    \rhead{\footnotesize{\Cref{#1}\textcolor{linkcol}{:} \nameref{#1}}}
    }   

\newcommand{\eps}{\epsilon}

\newcommand{\A}{\ensuremath{\mathcal{A}}}
\newcommand{\D}{\ensuremath{\mathcal{D}}}

\renewcommand{\O}{\ensuremath{\mathcal{O}}}

\newcommand{\U}{\ensuremath{\mathrm{U}}}

\newcommand{\Cl}{\ensuremath{\mathrm{Cl}}}

\newcommand{\QCA}{\ensuremath{\mathrm{QCA}}}

\newcommand{\mubrqc}[1][]{\mu_{\,\mathrm{brqc}#1}}
\newcommand{\Cbrqc}[1][]{\mathcal{C}_{\,\mathrm{bw}#1}}
\newcommand{\Qbrqc}[1][]{\mathcal{Q}_{\,\mathrm{bw}#1}}
\newcommand{\Dbrqc}[1][]{\mathcal{D}_{\,\mathrm{bw}#1}}

\newcommand{\mucl}[1][]{\mu_{\,\Cl#1}}

\newcommand{\met}{\ensuremath{\mathrm{d}}}

\newcommand{\learn}{\mathtt{Learn}}
\newcommand{\dec}{\mathtt{Dec}}

\renewcommand{\eval}{\mathtt{Eval}}
\newcommand{\stat}{\mathtt{Stat}}
\newcommand{\qstat}{\mathtt{QStat}}
\newcommand{\tqstat}{\mathtt{2QStat}}
\newcommand{\mqstat}[1]{\mathtt{MQStat}^{#1}}
\newcommand{\kqstat}{\mqstat{k}}
\newcommand{\csq}{\mathtt{CSQ}}
\newcommand{\mcsq}{\mathtt{MCSQ}}
\newcommand{\qcsq}{\mathtt{QCSQ}}
\newcommand{\qusq}{\mathtt{QUSQ}}
\newcommand{\rd}{\mathtt{rd}}
\newcommand{\rdl}{\mathtt{rdl}}
\newcommand{\dl}{\mathtt{dl}}
\newcommand{\triv}[1][\mc Z,\mu]{\mathtt{triv}\qty(#1)}
\newcommand{\qnt}[1][\mu,\tau]{q_\mathtt{nt}\qty(#1)}
\WithSuffix\newcommand\qnt*[1][\mu,\tau]{q_\mathtt{nt}(#1)}

\newcommand{\poly}{\mathrm{poly}}
\newcommand{\overpoly}[1][n]{\sfrac{1}{\poly(#1)}}  

\newcommand{\kfrac}{\mathrm{frac}}

\newcommand{\tv}{\mathrm{d}_\mathrm{TV}}
\newcommand{\id}{\ensuremath{\mathds{1}}}

\newcommand{\stacksim}[1]{\stackrel{#1}{\sim}}

\newcommand{\set}[1]{\qty{#1}}

\DeclarePairedDelimiter\ceil{\lceil}{\rceil}
\DeclarePairedDelimiter\floor{\lfloor}{\rfloor}

\DeclarePairedDelimiter\lrangle{\langle}{\rangle}
\makeatletter
\let\oldlrangle\lrangle
\def\lrangle{\@ifstar{\oldlrangle}{\oldlrangle*}}
\makeatother

\newlist{lossproperties}{enumerate}{1} 
\setlist[lossproperties]{ref=L\arabic*,
                 label=\bfseries L\arabic*:,
                 left= 1cm .. 2.5cm}

\crefname{losspropertiesi}{property}{properties}  
\Crefname{losspropertiesi}{Property}{Properties}  

\newlist{narrowgorge}{enumerate}{1} 
\setlist[narrowgorge]{ref=NG\arabic*,
                 label=\bfseries NG\arabic*:,
                 left= 1cm .. 2.5cm}

\crefname{narrowgorgei}{property}{properties}  
\Crefname{narrowgorgei}{Property}{Properties}  

\newlist{barrenplateau}{enumerate}{1} 
\setlist[barrenplateau]{ref=BP,
                 label=\bfseries BP:,
                 left= 1cm .. 2.5cm}

\crefname{barrenplateaui}{property}{properties}  
\Crefname{barrenplateaui}{Property}{Properties}  

\newlist{barrenplateauprime}{enumerate}{1} 
\setlist[barrenplateauprime]{ref=GBP,
                 label=\bfseries GBP:,
                 left= 1cm .. 2.5cm}

\crefname{barrenplateauprimei}{property}{properties}  
\Crefname{barrenplateauprimei}{Property}{Properties}  

\Crefname{myexample}{Example}{Examples}  

\def\myBoxleftRightSpace{1.8mm}
\def\myBoxTopBottomSpace{2mm}

\newtcolorbox{envbox}[2][]{
    colback=#2,
    colframe=#2,
    arc=5pt,
    boxsep=0mm,
    grow to left by=\myBoxleftRightSpace,
    grow to right by=\myBoxleftRightSpace,
    left=\myBoxleftRightSpace,
    right=\myBoxleftRightSpace,
    top=\myBoxTopBottomSpace,
    bottom=\myBoxTopBottomSpace,    
    boxrule=0mm,
    opacityfill=1.,
    breakable=true,
    skin=enhanced,
    #1,
}

\newtheoremstyle{mydefinitionsty}
{10pt}
{10pt}
{}
{}
{}
{}
{.5em}
{\textbf{\thmname{#1}~\thmnumber{#2}:  }\thmnote{(#3)}}
\theoremstyle{mydefinitionsty}
\newtheorem{mydefinition}{Definition}[section]
\newtheorem{myexample}{Example}[section]
\newtheorem{remark}{Remark}[section]

\newtheorem{observation}{Observation}[section]

\newtheorem{assumption}{Assumption}

\crefname{mydefinition}{definition}{definitions}  
\Crefname{mydefinition}{Definition}{Definitions}  

\newenvironment{definitionbox}{
    \begin{envbox}[opacityfill=.08]{TealBlue}
        \begin{mydefinition}
}{
        \end{mydefinition}
    \end{envbox}
}

\newenvironment{examplebox}{
    \begin{envbox}[opacityfill=.12]{MidnightBlue}
        \begin{myexample}
}{
        \end{myexample}
    \end{envbox}
}

\newtheoremstyle{myproblemsty}
{10pt}
{10pt}
{}
{}
{}
{}
{.5em}
{\textbf{\thmname{#1}~\thmnumber{#2}:  }\thmnote{(#3)}}
\theoremstyle{myproblemsty}
\newtheorem{myproblem}{Problem}

\crefname{myproblem}{problem}{problems}  
\Crefname{myproblem}{Problem}{Problems}  

\newenvironment{problembox}{
    \begin{envbox}[opacityfill=.15]{Maroon}
        \begin{myproblem}
}{
        \end{myproblem}
    \end{envbox}
}

\newtheoremstyle{mypers}
{10pt}
{10pt}
{\itshape}
{}
{}
{}
{.5em}
{\textbf{\thmname{#1}\thmnumber{#2}:}\thmnote{(#3)}}
\theoremstyle{mypers}
\newtheorem*{perspective}{Perspective}

\newtheoremstyle{mythmsty}
{10pt}
{10pt}
{\itshape}
{}
{}
{}
{.5em}
{\textbf{\thmname{#1}~\thmnumber{#2}:  }\thmnote{(#3)}}
\theoremstyle{mythmsty}

\newtheorem{mytheorem}{Theorem}[section]
\newtheorem{mylemma}[mytheorem]{Lemma}
\newtheorem{mycorollary}[mytheorem]{Corollary}

\crefname{mytheorem}{theorem}{theorems}  
\Crefname{mytheorem}{Theorem}{Theorems}  
\crefname{mylemma}{lemma}{lemmas}  
\Crefname{mylemma}{Lemma}{Lemmas}  
\crefname{mycorollary}{corollary}{corollaries}  
\Crefname{mycorollary}{Corollary}{Corollaries}  

\def\mythmcolor{Goldenrod!30}

\newenvironment{theorembox}{
    \begin{envbox}{\mythmcolor}
        \begin{mytheorem}
}{
        \end{mytheorem}
    \end{envbox}
}

\newenvironment{lemmabox}{
    \begin{envbox}{\mythmcolor}
        \begin{mylemma}
}{
        \end{mylemma}
    \end{envbox}
}

\newenvironment{corollarybox}{
    \begin{envbox}{\mythmcolor}
        \begin{mycorollary}
}{
        \end{mycorollary}
    \end{envbox}
}

\newenvironment{definition}{
    \begin{definitionbox}
}{
    \end{definitionbox}
}

\newenvironment{example}{
    \begin{examplebox}
}{
    \end{examplebox}
}

\newenvironment{problem}{
    \begin{problembox}
}{
    \end{problembox}
}

\newenvironment{theorem}{
    \begin{theorembox}
}{
    \end{theorembox}
}

\newenvironment{lemma}{
    \begin{lemmabox}
}{
    \end{lemmabox}
}

\newenvironment{corollary}{
    \begin{corollarybox}
}{
    \end{corollarybox}
}

\newenvironment{proofof}[1][\unskip]{%
\par
\noindent
\textbf{\textit{Proof of} #1:}
\noindent}
{\qed}

\definecolor{alexcolor}{rgb}{0.0, 0.47, 0.75}   
\definecolor{todocol}{rgb}{0.8,0.2,0.0}

\definecolor{fixmecol}{rgb}{1.0,0.3,0.3}

\definecolor{ideacol}{rgb}{1.0,0.75,0.0}

\definecolor{probcol}{rgb}{1.0,0.1,0.0}

\definecolor{questioncolor}{rgb}{0.36, 0.54, 0.66}

\definecolor{aurometalsaurus}{rgb}{0.43, 0.5, 0.5}

\definecolor{green-munsell}{rgb}{0.0, 0.66, 0.47}

\title{Unifying (Quantum) Statistical and Parametrized (Quantum) Algorithms}  %

\author[1]{Alexander Nietner
\thanks{Email:
\href{mailto:a.nietner@fu-berlin.de}{a.nietner@fu-berlin.de}}}

\affil[1]{\small Dahlem Center for Complex Quantum Systems, Freie Universit\"at Berlin, Germany}
 
\date{}

\begin{document}

%
%

\maketitle

\begin{abstract}
    \noindent
    Kearns' \emph{statistical query} (SQ) oracle (STOC'93) lends a unifying perspective for most classical machine learning algorithms.
    This ceases to be true in quantum learning,
    where many settings do not admit, neither an SQ analog nor a \emph{quantum statistical query} (QSQ) analog.
    In this work, we take inspiration from Kearns' SQ oracle and Valiant's \emph{weak evaluation} oracle (TOCT'14) and establish a unified perspective bridging the statistical and parametrized learning paradigms in a novel way.
    We explore the problem of learning from an \emph{evaluation} oracle, which provides an estimate of function values,
    and introduce
    an extensive yet intuitive framework that yields unconditional lower bounds for learning from evaluation queries and characterizes the query complexity for learning linear function classes.
    The framework is directly applicable to the QSQ setting and virtually all algorithms based on loss function optimization.

    Our first application is
    to extend prior results on the learnability of output distributions of quantum circuits and Clifford unitaries from the SQ to the (multi-copy) QSQ setting, implying exponential separations between learning stabilizer states from (multi-copy) QSQs versus from quantum samples. 
    Our second application is to analyze 
    some popular quantum machine learning (QML) settings. 
    We gain an intuitive picture of the hardness of many QML tasks which goes beyond existing methods such as barren plateaus and the statistical dimension, and contains crucial setting-dependent implications. 
    Our framework not only unifies the perspective of cost concentration with that of the statistical dimension in a unified language but exposes their connectedness and similarity.
\end{abstract}

\clearpage

\pagestyle{plain}  
\addtocontents{toc}{\protect\hypertarget{toc}{}}
\tableofcontents

\clearpage
\pagestyle{fancy}  

%
%

\section{Introduction}\label{sec:introduction}
\addsectionfooter{sec:introduction}
\addsectionheader{sec:introduction}

Modern machine learning algorithms are predominantly understood within two pa\-ra\-digms: statistical and para\-metrized algorithms. 
Statistical algorithms interact with the data through empirical averages and their formal study dates back to Kearns' seminal work on statistical query (SQ) learning \cite{kearns_efficient_1998}. 
Since then, the SQ toolbox has become a crucial technique in computational and statistical learning theory~\cite{feldman2017statistical,reyzinSQReview2020,aamari2021statistical,agarwal21b}.
In the paradigm of parametrized learning the algorithm is separated into a parametrized model and an optimization routine that fits the model regarding a loss function, where the loss function models the access to the data.
Examples of parametrized algorithms range from classics like linear regression and support vector machines to state-of-the-art methods like neural networks, deep learning and backpropagation.

Two of the main driving forces behind the success of the SQ toolbox undeniably are its ability to prove \emph{unconditional} lower bounds on the complexity of learning problems together with its near universal ability to capture most practical and theoretical learning techniques.
Indeed, with the famous exception of Gaussian elimination there exist SQ analogs for most learning algorithms ranging from differential privacy, evolvable algorithms, gradient descent, convex optimization, Markov-chain Monte Carlo, neural networks, backpropagation and ``\emph{pretty much everything else}''~\cite{reyzinSQReview2020}.
Thus, the framework of statistical queries deals as a unified framework for the statistical and the parametrized paradigm in classical machine learning.

Arunachalam et al. have generalized the SQ to the quantum statistical query oracle (QSQ), which models empirical expectation values of observables~\cite{arunachalamQSQ2020}.
However, not only does the SQ framework no longer contain many parametrized algorithms in the quantum sphere, 
but even the QSQ oracle fails to revive this property:
A typical classical parametrized loss function is 
$l(\vartheta)=\frac12\Ex_{x\sim P}[(f(x)-g(\vartheta,x))^2]=1-\Ex_{(x,y)\sim P_f}[yg(\vartheta,x)]$ which identifies the parametrized problem of learning an unknown $f$ with SQ learning an unknown distribution $P_f$.
In contrast, a common quantum loss function is 
$l(\vartheta)=\tr[\rho(\vartheta)O]$, where the unknown observable $O$ takes the role of $f$ and which does not correspond to a state learning problem of an unknown $\rho_O$.
As such there is no unified framework for (Q)SQ and parametrized learning algorithms in the general context of quantum learning~\cite{anshuSurveyComplexityLearning2023} and in particular for quantum machine learning (QML). 

In this work, we introduce the \emph{evaluation oracle} to establish a novel unified perspective bridging para\-metrized and statistical learning algorithms. 
The 
evaluation oracle for a function $f:\mc X\to M$ returns, when queried with some $x\in\mc X$ an approximation 
to $f(x)$ within some tolerance
and is inspired by Kearns' SQ oracle~\cite{kearns_efficient_1998} and closely related to Valiant's \emph{weak evaluation oracle}~\cite{valiantEvolvabilityRealFunctions2014}.
Within this new perspective, we introduce a comprehensive framework for unconditional lower bounds that brings the strengths of the SQ framework to quantum learning:
\begin{enumerate}
    \item When applied to classical statistical problems our results reduce to Feldman's well-established characterization of learning from statistical queries~\cite{feldman_general_2017}.
    \item It applies directly to parametrized learning problems without the need to rewrite the problem in a specific form, while the bounds are in general identical to statistical query bounds if there exists a statistical query analogue to the problem.
    \item The framework immediately applies to quantum statistical queries and generalizations thereof and generalizes Feldman's characterization to the quantum statistical query setting, and many \emph{linear learning problems}.
    \item Within our framework we can analyze, extend and unite so far unconnected approaches in QML such as barren plateaus~\cite{mccleanBarren2018}, narrow gorges~\cite{Arrasmith_2022} and cost concentration~\cite{nappQuantifyingBarrenPlateau2022a} with quantum correlational (QCSQ) and quantum unitary statistical queries (QUSQ)~\cite{anschuetz2022}. 
\end{enumerate}

Drawing from George Box's widely acknowledged quote, ``All models are wrong, but some are useful'', we believe that the evaluation oracle serves as a valuable solution, filling an essential gap in QML and enriching the current toolbox for statistical learning analysis.

\subsection{Overview of Formalism}\label{sec:formalism-overview}
\addsectionheader{sec:formalism-overview}
This section gives an overview of our formalism. See \Cref{sec:formalism} for a more detailed presentation.
We start by introducing general learning problems in \Cref{sec:intro-general-learning-problems}
and then introduce the evaluation oracle as a natural generalization of the SQ oracle in \Cref{sec:intro-eval-oracle}.
The actual formalism is then presented in \Cref{sec:intro-summary,sec:intro-decision-problems,sec:intro-verifiable,sec:intro-general-bounds,sec:intro-linear}.

\subsubsection{General Learning Problems}
\label{sec:intro-general-learning-problems}
\addsectionheader{sec:intro-general-learning-problems}

We define general learning problems, which are at the core of our formalism, similar to distributional search problems as defined by \cite{feldman2017statistical}. 

\vspace{-10pt}
\begin{mydefinition}[General learning Problems]
    Let $\mc S$ be a set of objects called \emph{source objects} 
    and let $\mc T$ be a set of objects called \emph{target objects}.
    Let $\mc Z:\mc S\to\mc P(\mc T)$ be a map from source objects $s\in\mc S$ to a subset of targets $\mc Z(s)\subseteq\mc T$, defined to be the valid solutions to $s$.
    The problem of learning $\mc Z$ over $\mc S$ and $\mc T$ from $q$ queries to an oracle $\mc O$ is to find a valid solution $t\in\mc Z(s)$ from $q$ queries to an oracle $\mc O(s)$ representing the access to the unknown $s\in\mc S$. 
\end{mydefinition}
\vspace{-7pt}

To gain some intuition let us consider the following learning problem called heavy Fourier search (HFS). 
For $\theta>0$ the task of HFS is to, given access to an unknown function $f:\mb F_2^n\to[-1,1]$, return a vector $z\in\mb F_2^n$ with Fourier coefficient $\hat{f}(z)$ of weight at least $\theta$, i.e. $\abs{\Ex_{x}[\chi_z(x)f(x)]}\geq\theta$, where $\chi_z(x)=(-1)^{x\cdot z}$ denotes the parity function corresponding to $z$\footnotemark.
\footnotetext{Throughout $\mb F_2=\{0,1\}$ denotes the finite field with two elements with modulo two addition and multiplication, respectively.}
The corresponding general learning problem is given by
\begin{align}
    \mc Z_{\mrm{HFS}}^\theta:[-1,1]^{\mb F_2^n}\to\mc P\qty(\mb F_2^n)\doublecolon f\mapsto \set{z\in\mb F_2^n\midvert \abs{\Ex_{x\sim\mb F_2^n}\qty[\chi_z(x)f(x)]}\geq\theta}\,.
\end{align}

Another important learning problem that we investigate closely following \cite{feldman_general_2017} is the generic many-to-one decision problem (see \Cref{ex:decissionproblem}).
For sets $\mc S$ and $\set{s^*}$ the decision problem $\dec(\mc S, s^*)$ asks to decide whether an unknown object $s\in\mc S\cup\set{s^*}$ is in $\mc S$. In particular, the problem is defined by the map 
\begin{align}
    \text{$\dec$}(\mc S, s^*):\mc S\cup\set{s^*}\to\set{\set{\mc S}, \set{\set{s^*}}}\subseteq\mc P(\set{\mc S, \set{s^*}})
    \doublecolon s\mapsto
    \begin{cases}
        \set{\mc S} &,\quad s\in\mc S\\
        \set{\set{s^*}} &,\quad s\notin\mc S\,.
    \end{cases}
\end{align}
Thus, $\dec(\mc S,s^*)$ is a learning problem over sources $\mc S\cup\set{s^*}$ and targets $\set{\mc S,\set{s^*}}$.

\subsubsection{Evaluation Oracle}
\label{sec:intro-eval-oracle}
\addsectionheader{sec:intro-eval-oracle}

General learning problems capture virtually any learning problem. 
Here, we are interested in settings, where the data access is of a specific structure, as formalized by the oracle $\mc O$.
Recall, that the SQ oracle $\stat_\tau(P)$ of a distribution $P$ over $\mc Y$ and tolerance $\tau>0$ returns, when queried with a function $\phi:\mc Y\to[-1,1]$
a value $v$ such that 
$\abs{v-\Ex_{x\sim P}\qty[\phi(x)]} \leq \tau$.
In other words, 
statistical algorithms as in \cite{feldman2017statistical} correspond to algorithms that access the data distribution only via coarse expectation values. 
This leads us to the following perspective.
\begin{envbox}[opacityfill=.08]{TealBlue}\begin{perspective}
    Statistical algorithms treat the distribution $P$ as a function 
    \begin{align*}
        P:[-1,1]^{\mc Y}\to[-1,1]\doublecolon \phi\mapsto P[\phi]=\Ex_{x\sim P}[\phi(x)]\,,
    \end{align*}
    and the SQ oracle implements a point-wise $\tau$-accurate approximation of this function.
\end{perspective}\end{envbox}
This reading leads to a natural generalization from statistical to evaluation queries.
\vspace{-10pt}
\begin{mydefinition}[Evaluation Oracle]
    Let $\tau>0$, $\mc X$ be a set and $(M,\met)$ be a metric space.
    For any $s\in M^\mc X$ we denote by $\eval_\tau(s)$ the \emph{Evaluation Oracle} of $s$ with tolerance $\tau$. 
    When queried with some $x\in\mc X$ it returns some $v\in M$ such that $\met(v,s(x))\leq\tau$.
\end{mydefinition}
\vspace{-7pt}

Let us list a few examples to highlight the different use cases.
We first consider the quantum statistical query (QSQ) oracle.
The QSQ oracle $\qstat_\tau(\rho)$ for an $N$-dimensional quantum state $\rho$ returns, when queried with a bounded Hermitian operator $O\in\mb H^{N\times N}$, $\norm{O}_\mrm{op}\leq1$ a value $v$ such that $\abs{v-\tr[\rho O]}\leq\tau$.

Secondly, let $l:\Theta\to[0,1]$ be a loss function of some parametrized learning problem with parameters $\vartheta\in\Theta$ and where we understand $[0,1]$ as equipped with the metric $\met(a,b)=\abs{a-b}$. 
Then, the evaluation oracle $\eval_\tau(l)$ returns, when queried with $\vartheta\in\Theta$ a value $v$ such that $\abs{v-l(\vartheta)}\leq\tau$.

Finally, let us mention the $\mathtt{1Bit}_\tau(P)$ oracle. When queried with a Boolean function $h:\mb F_2^n\to\01$ it returns a probability distribution on a single bit $\tilde{p}$ such that $\tv(\tilde{p},p)\leq\tau$.
Here, $p$ denotes the distribution $p(z)=\Pr_{x\sim P}[x\in h^{-1}(z)]$ corresponding to first sampling $x\sim P$ and then outputting $h(x)$.
As such, it corresponds to the evaluation oracle for a specific set of functions $\01^{\mb F_2^n}\to\Delta(2)$ with $\Delta(N)$ the $N$-dimensional simplex of discrete probability distributions endowed with the total variation distance $\tv$.
The $\mathtt{1Bit}$ oracle is tightly related with the $\mathtt{1\!\!-\!\!Stat}$  oracle as introduced in \cite{feldman2017statistical}. 

\subsubsection{Trivial and Non-Trivial Learning}\label{sec:intro-summary}
\addsectionheader{sec:intro-summary}

The notion of trivial and non-trivial learning will be a recurrent theme in our formalism.

\vspace{-10pt}
\begin{mydefinition}[Trivial and Non-Trivial Learning]
    For a learning problem $\mc Z$ over $\mc S\subseteq M^\mc X$ and $\mc T$ and $\mu$ a probability measure over $\mc S$ we denote by 
    \begin{align}
        \triv[\mc Z, \mu] = \sup_{t\in\mc T}\Pr_{s\sim\mu}\qty[s\in\mc Z_t]
    \end{align} 
    the \emph{Probability of Trivial Learning}. 
    Similarly, for any $s^*\in M^\mc X$ we refer to 
    $\kfrac(\mu,s^*,\tau)$
    as the \emph{Maximally Distinguishable Fraction} from $s^*$ with respect to $\mu$, where
    \begin{align}
        &\kfrac\qty(\mu,s^*,\tau) = \max_{x\in\mc X}\Pr_{s\sim\mu}\qty[\met(s(x), s^*(x))>\tau]\,.
    \end{align}
    Further $\qnt[\mu,\tau] = \max_{f\in M^\mc X}(\kfrac(\mu, f, \tau)^{-1})$ is referred to as the \emph{Query Complexity of Non-Trivial Learning} from evaluation queries with respect to $\mu$ and $\tau$.
\end{mydefinition}
\vspace{-7pt}

It is easy to see that $\triv[\mc Z,\mu]$ is the probability of success over $\mu$ of solving $\mc Z$ by the trivial algorithm, which without queries always returns a fixed $t$.
Similarly, $\kfrac(\mu,s^*,\tau)$ is the largest fraction with respect to $\mu$ which can be distinguished from $s^*$ from a single query $x$ to $\eval_\tau$.
As we will later see, the complexity of non-trivial learning $q_\mtt{nt}$ then asymptotically lower bounds the complexity for solving non-trivial learning problems, which are learning problems for which $\triv[\mc Z,\mu]$ is sufficiently small.
It is easy to see how to bound $\kfrac$ and $q_\mtt{nt}$ using Markov's and Chebyshev's inequality respectively.

\vspace{-10pt}
\begin{mycorollary}[Bounding $\kfrac$ and $q_\mtt{nt}$]
    The maximally distinguishable fraction is bounded by the average discrimination as 
    \begin{align}
        \kfrac(\mu,s^*,\tau) \leq \tau^{-1} \max_{x\in\mc X}\Ex_{s\sim\mu}[\met(s(x),s^*(x))]\,.
    \end{align}
    Further, if the variance of $s(x)$ with $s\sim\mu$ is well-defined for all $x\in\mc X$, it holds 
    \begin{align}
        \qnt[\mu,\tau] \geq \tau^2 \qty(\max_{x\in\mc X}\Var_{s\sim\mu}[s(x)])^{-1}\,.
    \end{align}
\end{mycorollary}
\vspace{-7pt}

\subsubsection{Characterizing Decision Problems}\label{sec:intro-decision-problems}
\addsectionheader{sec:intro-decision-problems}

We lower bound the query complexity of the many-to-one decision problem $\dec(\mc S, s^*)$ for random algorithms by (see \Cref{sec:decissionproblembounds}) 
\begin{align}
    q\geq  2\cdot(\alpha-\sfrac12)\cdot\kfrac\qty(\mu,s^*,\tau)^{-1}\,,
\end{align}
where $\alpha$ is the success probability with respect to the internal randomness of the algorithm.
Here, we assume that $\mc S\subset M^{\mc X}$, $s^*\in M^\mc X\setminus\mc S$\footnotemark  and $\mu\in\Delta(\mc S)$ is any probability measure over $\mc S$.
\footnotetext{Note that $\{\mc S\}$ is always a valid solution in case $s^*\in\mc S$.}

Similarly to the results from \cite{feldman_general_2017} this bound is tight in the sense that we accommodate it with the corresponding upper bound.
The many-to-one decision problem $\dec(\mc S, s^*)$ can be solved with probability at least $\alpha=1-\delta$ by a random algorithm using at most $q$ queries to $\eval_\tau$, where 
\begin{align}
    q\leq \ln\qty(\sfrac1\delta)\cdot \max_{\mu}\kfrac\qty(\mu,s^*,\tau)^{-1}\,.
\end{align}
The intuition of the lower bound follows the minimax philosophy or, in fact, the maximin advantage of solving $\dec(\mc S, s^*)$ due to a distinguishing query.
The prefactor $p=2(\alpha-\sfrac12)$ is the probability, that an algorithm that solves $\dec(\mc S, s^*)$  will correctly decide not out of luck, but due to a distinguishing query $x$ such that $\met(s(x),s^*(x))>\tau$. 
Thus, a successful algorithm for solving $\dec(\mc S,s^*)$ must have at least an advantage $p$ for the task of finding a distinguishing query between any $s\in\mc S$ and $s^*$.
We bound this advantage for any $\mu\in\Delta(\mc S)$ by
\begin{align*}
    p \leq \mrm{Adv} = \max_{\mc A}\min_{s\in\mc S} \Ex_{\mc A}\qty[L(\mc A(s), \mc A(s^*))] 
    \leq \max_{\mc A}\Ex_{\substack{s\sim\mu\\ \mc A}} \qty[L(\mc A(s), \mc A(s^*)) ]
    = \max_{\mc A}\Pr_{\substack{s\sim\mu\\ \mc A}} \qty[L(\mc A(s), \mc A(s^*)) = 1]\,,
\end{align*}
where $\Ex_\mc A$ denotes the expectation with respect to $\mc A$'s internal randomness and the loss $L(\mc A(s),\mc A(s^*))$ equals one, if at least one of $\mc A$'s $q$-many queries distinguishes $s$ from $s^*$ and zero else.
The final bound is then obtained via the union bound and bounding the maximal fraction of $\mc S$ with respect to $\mu$, which any query can distinguish from $s^*$.

The upper bound on the other side relies on von Neumann's minimax theorem.
In particular, by von Neumann's minimax theorem, we obtain a measure $\nu$ over $\mc X$ which, borrowing the term from Feldman \cite{feldman_general_2017}, defines a \emph{random cover} of the decision problem.

\subsubsection{Lower Bounds for Verifiable Learning Problems}\label{sec:intro-verifiable}
\addsectionheader{sec:intro-verifiable}

The lower bound for decision problems can be directly applied to prove lower bounds to \emph{verifiable learning problems}, which are defined as follows.
Let $\mc Z$ be a learning problem over $\mc S\subset\mc S'=M^\mc X$ and $\mc T$ and denote by $\mc Z_t=\set{s\in\mc S\midvert t\in\mc Z(s)}$.
We say that $\mc Z$ is $p$-verifiable with respect to $s^*\in\mc S'\setminus\mc S$ from access to $\eval_\tau$, if there exists a family of algorithms $\set{\mc A_t\midvert t\in\mc T}$ such that for every $t\in\mc T$ the algorithm $\mc A_t$, \
solves $\dec(\mc Z_t, s^*)$ from $p$ queries to $\eval_\tau$.

As such, we lower bound the \emph{average case} query complexity of learning $\mc Z$ with respect to the measure $s\sim\mu$ over instances by $q$, where 
\begin{align}
    q+p 
    \geq 2\cdot\qty(\alpha-\sfrac12)\cdot\beta\cdot\kfrac\qty(\mu,s^*,\tau)^{-1}\,.
\end{align}
Here $\alpha$ is the success probability with respect to the algorithm's internal randomness and $\beta$ is the success probability with respect to $s\sim\mu$.

The intuition behind this bound is the observation that any algorithm for learning $\mc S$ from $q$ queries implies, based on the assumption of $p$-verifiability, an algorithm for solving $\dec(\mc S, s^*)$ from $q+p$ queries. 
The actual bound then follows from applying this argument to an arbitrary $\beta$-fraction of $\mc S$ on which the average-case learner is successful.

Similarly, we can iterate the same argument in another way.
For some $s^*\in\mc S'$ let $\mc S^*$ be any set such that the restriction of $\mc Z$ to $\widetilde{\mc S}=\mc S\setminus\mc S^*$, is $p$-verifiable with respect to $s^*$.
Then, we find that any algorithm for learning $\mc Z$ with probability $\beta$ over $s\sim\mu$ and $\alpha$ over its internal randomness requires at least $q$ queries to $\eval_\tau$, with
\begin{align}\label{eq:intro-learning-by-verifiable-learning-bound}
    q+p 
    \geq 2\cdot\qty(\alpha-\sfrac12)\cdot\qty(\beta - \Pr_{s\sim\mu}\qty[s\in\mc S^*])\cdot\kfrac\qty(\mu,s^*,\tau)^{-1}\,.
\end{align}

\subsubsection{Lower Bounds for General Learning Problems}
\label{sec:intro-general-bounds}
\addsectionheader{sec:intro-general-bounds}

Let us now look at bounds for general learning problems. 
As before we set $\mc S'=M^\mc X$ and let $\mc S\subseteq\mc S'$. We consider the worst-case, as well as deterministic, joint and random average-case query complexity, definitions for which can be found in \Cref{sec:onComplexity}. 

\paragraph[Deterministic Algorithms]{Deterministic Algorithms:} The previous lower bound is similar to the deterministic average-case lower bound we obtain for general learning problems.
Any deterministic algorithm for learning $\mc Z:\mc S\to\mc P(\mc T)$ with probability $\beta$ over $s\sim\mu$, given access to $\eval_\tau$, must make at least $q$ queries, where for any $s^*\in\mc S'$ it holds 
\begin{align}\label{eq:intro-general-avg-lower}
    &q\geq \qty(\beta - \sup_{t\in\mc T}\qty[s\in\mc Z_t])\cdot\kfrac\qty(\mu,s^*,\tau)^{-1}\,,
    &&\text{which implies}
    &&& q\geq \qty(\beta-\triv[\mc Z,\mu])\cdot\qnt[\mu,\tau]\,.
\end{align}
The proof is almost identical to that of the deterministic bound of solving $\dec(\mc S, s^*)$ together with the ideas leading to \Cref{eq:intro-learning-by-verifiable-learning-bound}.

\paragraph[Random Algorithms]{Random Algorithms:} 
The previous lower bound immediately translates to a joint average-case lower bound for random algorithms, where $\beta$ denotes the joint success probability over $\mu$ and the algorithm's internal randomness.
The random average-case lower bound however is more involved. 
It follows the same ideas as Feldman \cite{feldman_general_2017}.
We first exclude all source objects $s\in\mc S$, for which there exists a trivial algorithm that, without making any query succeeds on $s$ with some probability at least $\gamma$.
In the next step, we apply a similar argument as for $\dec(\mc S,s^*)$ in the random setting, where we estimate the probability of correctly solving the problem due to a distinguishing query. 
Finally, applying the same arguments that lead to \Cref{eq:intro-learning-by-verifiable-learning-bound} we then obtain for any $s^*$
\begin{align}
    q\geq (\alpha-\gamma)\cdot\qty(\beta-\sup_\vartheta\Pr_{s\sim\mu}\qty[s\in\mc Z_{\vartheta}(\gamma)])\cdot\kfrac\qty(\mu,s^*,\tau)^{-1}\,.
\end{align}
Here, $\vartheta\in\Delta(\mc T)$ is any probability measure over $\mc T$ and $\mc Z_\vartheta(\gamma)=\set{s\in\mc S\midvert \vartheta(\mc Z(s))\geq\gamma}$ is the excluded set of ``trivially-to-solve'' instances: $\mc Z$ can be solved on this set with probability $\gamma$ over $\vartheta$ -- the algorithms internal randomness -- without queries, by simply returning a randomly sampled $t\sim\vartheta$.

\subsubsection{Characterizing Linear Learning Problems}
\label{sec:intro-linear}
\addsectionheader{sec:intro-linear}

Finally, we consider learning problems over \emph{linear function classes}. 
We will refer to learning problems over linear function classes as linear learning problems.
Simply put, a linear function class is a class of functions that is contained in the dual space of the function domain.
To be more precise, let $\mc V$ be a vector space over $\mb K$ (real or complex) with the metric on $\mb K$ given by the absolute value.
Further, let $\mc V$ be endowed with a norm $\norm{\cdotspace}$ and denote by $\mc V^\star$ the corresponding dual space endowed with norm $\dualnorm{\cdotspace}$.
A linear function class then is a subset $\mc S\subseteq\mc V^\star$. 
To have a well-defined tolerance, we identify the domain of the functions $f\in\mc S$ with the closed unit ball $\overline{B}_1(0)\subset\mc V$ with respect to $\norm{\cdotspace}$
\begin{align}
    f:\overline{B}_1(0)\to\mb K\doublecolon g\mapsto f(g)=\lrangle{f, g}\,,
\end{align}
where the last equality holds if $\mc V$ has an inner product $\lrangle{\cdotspace,\cdotspace}$ and $\mc V^\star=\mc V$.

Linear function classes are a powerful concept.
The statistical query framework can be seen as the prototypical linear learning setup.
In \Cref{tab:linearfunctions} we sketch some possible settings.
The oracle corresponds to the evaluation oracle for the given class with respect to $\overline{B}_1(0)$.

\renewcommand\arraystretch{1.65} 
\begin{table}[htbp]
    \begin{adjustwidth}{-0.2cm}{-0.3cm}
        \begin{center}
            \begin{tabular}{ |l|l|l|l|l|l| }
                \hline
                Oracle & $\mc V$ & $\lrangle{\cdotspace,\cdotspace}$ & $\dualnorm{\cdotspace}$ & $\norm{\cdotspace}$  & Details \\
                \hline\hline
                $\stat_\tau(P)$ & $\mb R^N$,  ($\Delta(N)$) & $\lrangle{P,\phi}_{\ell_2}=\Ex_{x\sim P}[\phi(x)]$ & $\norm{\cdotspace}_1$ & $\norm{\cdotspace}_\infty$  & \Cref{ex:distributionsandSQ}\\
                \hline
                $\qstat_\tau(\rho)$ & $\mb H^{N\times N}$, ($\mc S(N)$) & $\lrangle{\rho, O}_{HS}=\tr[\rho O]$ & $\norm{\cdotspace}_{\tr}$ & $\norm{\cdotspace}_\mrm{op}$ & \Cref{ex:statesandQSQ}\\
                \hline
                $\csq_\tau(f)$ & $[N]\to\mb R\; (= \mb R^N)$ & $\lrangle{f,g}_\lambda=\Ex_{x\sim\lambda}[f(x)g(x)]$ & $\norm{\cdotspace}_{L^2(\lambda)}$ & $\norm{\cdotspace}_{L^2(\lambda)}$ & \Cref{ex:functionalsandCSQ}\\
                \hline
                $\qcsq_\tau(M)$  & $\mb H^{N\times N}$, ($\norm{M}_\mrm{op}\leq1$) & $\lrangle{M,O}_\lambda=\Ex_{\rho\sim\lambda}\qty[\tr[\rho M]\tr[\rho O]]$ & $\norm{\cdotspace}_{L^2(\lambda)}$ & $\norm{\cdotspace}_{L^2(\lambda)}$ & \Cref{ex:qcsq}\\
                \hline
                $\qusq_\tau(U)$ & $\mb C^{N\times N}$, ($\U(N)$) & $\lrangle{U,V}_\lambda=\Ex_{\rho\sim\lambda}\qty[\tr[U^\dagger V \rho]]$ & $\norm{\cdotspace}_{L^2(\lambda)}$ & $\norm{\cdotspace}_{L^2(\lambda)}$ & \Cref{ex:qusq}\\
                \hline
                $\mcsq_\tau(A)$ & $\mb C^{N\times N}$ & $\lrangle{A,B}_\rho=\tr[A^\dagger \rho B]$ & $\norm{\cdotspace}_{L^2(\rho)}$ & $\norm{\cdotspace}_{L^2(\rho)}$ & \Cref{ex:matrixfunctionalsandMCSQ}\\
                \hline
            \end{tabular}
        \end{center}
    \end{adjustwidth}
    \caption{Some examples for linear functions. The oracle corresponds to the evaluation oracle $\eval_\tau(s)$ of the given class for the source object $s$ when the domain is restricted to $\overline{B}_1(0)$. 
    The set in the brackets next to the vector space is the set in which the function class is conventionally contained.
    This containing set in turn is itself contained in the dual unit sphere $\set{s\midvert\dualnorm{s}=1}$.  
    As throughout, $\Delta(N)$ denotes the simplex of $N$-dimensional distributions, $\mc S(N)$ denotes the set of $N$-dimensional quantum states and $\U(N)$ are the $N$-dimensional unitary matrices.
    The representation of the definition of the inner product in the first example only holds for $P\in\Delta(N)\subset\mb R^N$ and is $\sum_x P(x)\phi(x)$ more generally. By $\mb H^{N\times N}$ we denote the set of $N\times N$ dimensional Hermitian matrices.}
    \label{tab:linearfunctions}
\end{table}

\paragraph[Characterization of Linear Learning]{Characterization of Linear Learning:} We have two main findings regarding the query complexity of linear learning problems in the evaluation oracle framework.
First, we generalize Feldman's characterization of statistical query learning \cite{feldman_general_2017} to linear learning problems.
Let $\mc Z$ be a linear learning problem over $\mc S\subseteq \mb K^{\overline{B}_1(0)}$ and $\mc T$ and define for any $s^*\in \mb K^{\overline{B}_1(0)}$ the \emph{random dimension}
\begin{align}
    \rd(\mc S, s^*,\tau) = \sup_\mu\kfrac\qty(\mu,s^*,\tau)^{-1} = \sup_\mu\qty(\max_{x\in\overline{B}_1(0)}\Pr_{s\sim\mu}\qty[\abs{s(x)-s^*(x)}>\tau])^{-1}\,.
\end{align}
Note that $\rd(\mc S, s^*,\tau)$ characterizes the complexity of solving $\dec(\mc S, s^*)$ from $\eval_\tau$.
The \emph{random dimension of learning} for the problem $\mc Z$ is then defined via 
\begin{align}
    \rdl(\mc Z,\tau, \gamma) = \sup_{s^*}\inf_{\vartheta}\rd(\mc S\setminus\mc Z_\vartheta(\gamma), s^*, \tau)\,,
\end{align}
with $\mc Z_\vartheta(\gamma)$ as defined in the previous section.
By the previous section, $\rdl(\mc Z, \tau, \gamma)$ lower bounds the worst-case query complexity as $q\geq(\alpha-\gamma)\rdl(\mc Z, \tau, \gamma)$. 
In fact, under some assumptions (for details see \Cref{thm:randomupper}), this characterization is tight in the sense that there is a random algorithm for learning $\mc Z$ with success probability $\alpha-\delta$ from $q$ queries to $\eval_\tau$, with 
\begin{align}
    q = O\qty(\frac{r\cdot\rdl(\mc Z,\tau,\alpha)}{\zeta\cdot\tau^2}\cdot\log\qty(\frac{r}{\tau\cdot\delta}))\,.
\end{align}
Here we assumed that $\mc S$ as well as $\overline{B}_1(0)$ are compact and that we can set up a mirror descent (MD) scheme. 
The factor $\zeta$ then is due to the $\zeta$-\emph{regularizer} $R$ and $r$ is the radius of $\mc S$ as measured by the Bregman divergence $D_R$.
We note that all settings in \Cref{tab:linearfunctions} admit an MD scheme. 
In particular, in the first setting in \Cref{tab:linearfunctions}, when setting $R(x)=\sum_i x_i\ln(x_i)$ the negative entropy function, we obtain the standard multiplicative weights algorithm with $\zeta=1$ and obtain the same bounds as in \cite{feldman_general_2017}.
Alternatively, in the QSQ setting, when setting $R(X)=\tr[X\ln(X)]$, then $D_R$ becomes the quantum relative entropy with $\zeta=\frac{1}{\ln2}$ and MD corresponds to matrix multiplicative weights.
This setting is closely related to Aaronson, Chen, Hazan, Kale and Nayak's ``Online Learning of Quantum States'' \cite{aaronsonOnlineLearning2019}.
Similarly, interpreting the upper bound for the correlational statistical query setting $\csq_\tau$ with $R(x)=\norm{x}_2^2$, with $\zeta=1$, yields gradient descent-based learning scheme for functions that is closely related to Trevisan, Tulsiani and Vadhan's celebrated result \cite{trevisanRegularityBoosting2009}.

\paragraph[$\epsilon$-Learning]{$\epsilon$-Learning:} The lower bounds for random algorithms in the general framework are cumbersome to work with.
Our second result for linear learning thus concerns a more practical lower bound that is inspired by \cite{nietnerAveragecaseComplexityLearning2023}.
Denote by $\met_\star(x,y)=\frac12\dualnorm{x-y}$ the distance induced by the dual norm. 
For example, for SQ's $\dualnorm{\cdotspace}=\norm{\cdotspace}_1$, such that the induced distance equals the total variation distance $\met_\star=\tv$, or similarly, for QSQ's $\dualnorm{\cdotspace}=\norm{\cdotspace}_{\tr}$ and thus $\met_\star=\met_{\tr}$.
Then, for a linear function class $\mc S\subseteq\mc V^\star$ we define $\epsilon$-learning by the map 
\begin{align}
    \learn(\mc S,\epsilon):\mc S\to\mc P(\mc V^\star)\doublecolon s\mapsto B_{\epsilon,\met}^\star(s)=\set{y\in\mc V^\star\midvert \met_\star(x,y)<\epsilon}\,.
\end{align}
In \Cref{cor:linear-eps-learning-complexity} we show that any random algorithm for learning $\learn(\mc S,\epsilon)$ with probability at least $\beta$ over $s\sim\mu$ and $\alpha$ over its internal randomness needs to make at least $q$ queries, where, for any $s^*\in\mc V^\star$ it holds
\begin{align}\label{eq:intro-eps-learning-lower}
    q+1 \geq 2\cdot\qty(\alpha-\sfrac12)\cdot\qty(\beta - \Pr_{s\sim\mu}\qty[\met_\star(s,s^*)<2\epsilon+\tau])\cdot\kfrac\qty(\mu,s^*,\tau)^{-1}
    \,.
\end{align}

As a side result, we show a connection between \Cref{eq:intro-general-avg-lower,eq:intro-eps-learning-lower}.
Inspired by \cite{nietnerAveragecaseComplexityLearning2023}, we show in \Cref{cor:farfrom} that 
$\kfrac\qty(\mu,s^*,\tau)$
together with the mass of the $2\epsilon+\frac\tau2$ ball with respect to $\met_\star$ and $\mu$ around $s^*$ characterizes the mass of the largest $\epsilon$-ball with respect to $\mu$ in $\mc V^\star$:
\begin{align}\label{eq:intro-farfrom}
    \max_{r\in\mc V^\star}\Pr_{s\sim\mu}\qty[\met_\star(s,r)<\epsilon] \leq \max\set{\kfrac\qty(\mu,s^*,\tau), \Pr_{s\sim\mu}\qty[\met_\star(s, s^*)<2\epsilon+\frac\tau2]}\,.
\end{align}
For $\kfrac\qty(\mu,s^*,\tau)$ sufficiently small, plugging \Cref{eq:intro-farfrom} into \Cref{eq:intro-general-avg-lower} almost yields \Cref{eq:intro-eps-learning-lower}, up to the contribution of a $\frac\tau2$ shell around the $2\epsilon+\frac\tau2$-ball around $s^*$.

\subsection{Overview of Applications}\label{sec:applications-overview}
\addsectionheader{sec:applications-overview}

We give an overview about our framework applied to quantum statistical query learning in \Cref{sec:intro-qsq} and to parametrized learning problems in \Cref{sec:intro-parametrized}.

\subsubsection{Quantum Statistical Learning}\label{sec:intro-qsq}
\addsectionheader{sec:intro-qsq}

\paragraph[Related Work]{Related Work:} The statistical query model corresponds to empirical expectation values of bounded functions with respect to a probability distribution.
Analogously, the quantum statistical query (QSQ) model corresponds to empirical expectation values of bounded observables with respect to quantum states, see \Cref{tab:linearfunctions}. 
It was first studied by Arunachalam, Grilo and Yuen \cite{arunachalam2021quantum} and has since been studied various contexts such as quantum differential privacy \cite{angrisaniQLD2022} and classical verification of quantum learning \cite{caroVerification2023}.
We also refer to our Note added \Cref{sec:noteadded} for other more recent related work.

\paragraph[Positive Results]{Positive Results:} Before applying our formalism to show QSQ lower bounds we give an incomplete collection of positive results for QSQ learning for various classes in \Cref{sec:qsq-positive}. 
This includes learning parities with noise in the quantum PAC setting as shown by \cite{arunachalamQSQ2020}, a reinterpretation of matrix product state tomography in the QSQ framework based on \cite{cramerEfficient2011} and free fermion tomography via QSQ's inspired by \cite{banchiQuantumInfoGeom2014,swingleFermionic2019,gluzaFidelity2018}, the latter, together with \cite{nietnerFree2023}, implies a new separation between SQ and QSQ learning. 

\paragraph[Hardness Results]{Hardness Results:}
We exemplify our formalism in the quantum statistical query setting by analyzing the output states, and Born distributions, of brickwork random quantum circuits, random Clifford unitaries and Haar random unitaries.
As such, our work extends the studies by Hinsche et al. \cite{hinsche_single_2022} and Nietner et al. \cite{nietnerAveragecaseComplexityLearning2023}.
Nietner et al. investigate the average-case complexity of $\epsilon$-learning the Born distributions of random quantum circuits $P(x)=\abs{\mel{x}{U}{0^n}}$, with $U$ being a random quantum circuit from statistical queries. 
They focus on two parameters: The depth $d$ of the circuit and the fraction $\beta$ which the algorithm is required to learn.

We extend this analysis in two ways.
First, we provide the algorithm with the more powerful quantum statistical query oracle,
and second, add as additional parameter the \emph{number of copies} of the unknown state each query can access.
In particular, we introduce the $\mqstat{k}_\tau$ oracle 
\begin{align}
    \mqstat{k}_\tau(\rho) = \qstat_\tau(\rho^{\otimes k})\,.
\end{align}
This additional parameter tunes the allowed correlations and entanglement between individual copies, which the algorithm can exploit.

Through \Cref{sec:single-copy-qsq,sec:two-copy-qsq,sec:multicopy-qsq} we analyze the learnability of the random unitary and Clifford circuits at different depth and for different numbers of copies $k$. We summarize our results for learning the corresponding quantum states in \Cref{tab:qsq-hardnes}. 
Our \Cref{cor:single-copy-clifford-circuits,thm:single-stabilizer-hard,thm:two-copy-clifford} imply a novel separation between learning stabilizer states from quantum copies \cite{aaronson_gottesman_2004,montanaroLearningStabilizerStates2017} and multi-copy QSQs. 
While stabilizer states are known to be efficiently learnable from $O(n)$ entangled, respectively $O(n^2)$ unentangled copies, our results show that no such multi-copy QSQ algorithm can exist.
\setlength{\extrarowheight}{1pt}
\begin{table}[htbp]
    \begin{adjustwidth}{-0.1cm}{-0.15cm}
        \begin{center}
            \begin{tabular}{ 
                | >{\!\!\centering\arraybackslash}m{1.5cm}  
                | >{\!\!\!\centering\arraybackslash}m{2.2cm}
                | >{\!\!\!\centering\arraybackslash}m{2.4cm}
                | >{\!\!\!\centering\arraybackslash}m{2.4cm}
                | >{\!\!\!\centering\arraybackslash}m{3.0cm}
                | >{\!\!\!\centering\arraybackslash}m{2.4cm}| }
                \hline
                Depth $d$ &$O(\log(n))$ 
                & $\omega(\log(n))$ 
                & $\Omega(n)$ 
                & $\Omega(n\cdot\poly(k))$ & $d\to\infty$ 
                \\
                \hline\hline
                \multirow{2}{*}{BW-RQC's} 
                & \multirow{2}{*}{
                    $\begin{matrix}
                        O(n^3)\,,\; k=1\\
                        \text{\footnotesize{(\Cref{thm:qsq-mps})}}
                    \end{matrix}$}
                & \multirow{2}{*}{
                    $\begin{matrix}
                        2^{\omega(\log(n))}\,,\;k=1\\[-5pt]
                        \text{\footnotesize{(\Cref{cor:single-copy-clifford-circuits})}}
                    \end{matrix}$} 
                & $\Omega(2^n)\,,\;k=1$ \footnotesize{(\Cref{thm:single-coopy-rqc-hard})} 
                & \multirow{2}{*}{
                    $\begin{matrix}
                        \Omega\qty(\tfrac{2^n}{k^2}) \,,\; k=o(2^{\sfrac{n}{2}})\\
                        \text{\footnotesize{(\Cref{thm:t-design-multi-copy-hardness})}}
                    \end{matrix}$
                } 
                & \multirow{2}{*}{
                    $\begin{matrix}
                        \exp(\Omega\qty(\tfrac{2^n}{k^2}))\,,\\
                          k=o(2^{\sfrac{n}{2}})\\[-5pt]
                          \text{\footnotesize{(\Cref{thm:hardness-haar-random})}}
                    \end{matrix}$
                }
                \\
                \cline{4-4}
                &  
                &  
                & $\Omega(2^n)\,,\;k=2$ \footnotesize{(\Cref{cor:rqc-two-copy-hard})}
                & 
                &  
                \\
                \hline
                \multirow{3}{*}{$\Cl$ Circuits} 
                & \multirow{3}{*}{
                    $\begin{matrix}
                        O(n^3) \,,\; k=1\\
                        \text{\footnotesize{(\Cref{thm:qsq-mps})}}
                    \end{matrix}$}
                & \multirow{3}{*}{
                    $\begin{matrix}
                        2^{\omega(\log(n))}\,,\;k=1\\[-5pt]
                        \text{\footnotesize{(\Cref{cor:single-copy-clifford-circuits})}}
                    \end{matrix}$}  
                & \multicolumn{3}{c| }{$\Omega(2^n)\,,\;k=1$ \footnotesize{(\Cref{thm:single-stabilizer-hard})}} 
                \\
                \cline{4-6}
                &  
                &  
                & \multicolumn{3}{c| }{$\Omega(2^n)\,,\;k=2$ \footnotesize{(\Cref{thm:two-copy-clifford})}}
                \\
                \cline{4-6}
                &  
                &  
                & \multicolumn{3}{c| }{$O(n)\,,\;k=\Omega(n)$  \footnotesize{(\cite{montanaroLearningStabilizerStates2017} and \Cref{lem:multi-copy-to-qsamp})}}
                \\
                \hline
            \end{tabular}
        \end{center}
    \end{adjustwidth}
    \caption{Lower and upper bounds for $\epsilon$-learning the output states of $n$-qubit brickwork random quantum circuits (BW-RQC) and Clifford ($\Cl$) circuits in the $k$-copy quantum statistical query setting $\kqstat$. The same bounds apply to learning the Born distribution. All lower bounds hold for any $\tau=\Omega(\overpoly)$. While our work only proves the $d=\omega(\log(n))$ column in a worst-case sense, we prove all other lower bounds in the average-case setting for suitable $\beta=1-\Omega(\overpoly)$. For the detailed dependencies on $\tau$, $\beta$ and $\alpha$ we refer to the respective theorem statements. The $d\to\infty$ column corresponds to unitary Haar random (in the case of BW-RQC's) and uniform random Clifford unitaries. The last three Clifford columns are joined since any Clifford unitary can be generated by a $d=O(n)$ Clifford circuit.}
    \label{tab:qsq-hardnes}
\end{table}
Note that the upper bound for learning the output states of Clifford circuits, i.e. stabilizer states, in the $k=\Omega(n)$-copy setting is due to \Cref{lem:multi-copy-to-qsamp}.
Every QSQ algorithm with inverse polynomial tolerance can be efficiently simulated from access to copies of the unknown state. 
\Cref{lem:multi-copy-to-qsamp} then regards the opposite direction. 
It states that every algorithm that requires $k$-copies of an unknown quantum state can be simulated from $\log(\abs{\mc Z(\mc S)})$ many queries to $\kqstat_\tau$. 
As such, the lemma makes rigorous the above-mentioned interpolation between statistical learners on the one hand, and correlated and entangled learners on the other hand.

\subsubsection{Parametrized Learning}\label{sec:intro-parametrized}
\addsectionheader{sec:intro-parametrized}

Parametrized learning algorithms are defined with respect to a parametrized model.
Their aim is to find a parameter in parameter space $\vartheta\in\Theta\subset\mb R^m$ such that the model fits the data. 
This fit is captured by a loss function and an optimality criterion.
The loss function encodes the notion of approximation and can be evaluated up to some precision $\tau$.
The optimality criterion defines what loss is considered sufficiently good a fit.

\paragraph[Prior Work]{Prior Work:} The vast majority of practically used classical machine learning algorithms can be recast in the SQ framework.
However, the same ceases to be true for quantum machine learning algorithms and either statistical or quantum statistical queries.
Thus, to analyze the performance of QML algorithms the community borrows from the classical machine learning community another major framework: \emph{Loss Landscape Analysis} (LLA) of an individual loss function.

In LLA one investigates the graph of an individual loss function, which is referred to as the loss landscape.
The most popular obstructions to learning due to LLA are known under the name of barren plateau (BP) and narrow gorge (NG) \cite{mccleanBarren2018,Arrasmith_2022}, or cost concentration more generally \cite{nappQuantifyingBarrenPlateau2022a}.
Common to these obstructions is that the analysis is done with respect to a probability measure $\nu$ over parameter space $\Theta$.
If a loss function admits a BP, then the gradient exponentially vanishes with overwhelming probability.
If the loss function admits a NG, then the loss is exponentially close to the mean loss with overwhelming probability, while there exists a non-trivial optimal loss which is at least inverse polynomially bounded from the mean loss.
Cost concentration more generally refers to the concentration of the loss function, its gradient, or both around some value.
Cost concentration, BP and NG thus correspond to average-case statements about properties -- such as the gradient -- of the loss function with respect to $\nu$. 
These phenomena therefore imply hardness of learning by algorithms that rely on those properties and which explore the parameter space ``according to'' $\nu$.
However, these phenomena do not imply rigorous hardness results such as provided by the SQ framework.

Progress in this direction was made by Anschuetz and Kiani.
In \cite{anschuetz2022} they introduced the \emph{quantum correlational} and the \emph{quantum unitary statistical query} oracles $\qcsq$ and $\qusq$.
These oracles are designed to model the parameter shift routine used for evaluating the gradient in many QML models ($\qcsq$) and the evaluation of the average fidelity in unitary compiling ($\qusq$). 
Further, they generalized Blum et al.'s statistical dimension \cite{blum1994weakly} to their setting and apply this to show hardness for several learning problems.
The $\qcsq$ as well as the $\qusq$ setting assumes that the unknown object has to be approximated with respect to a distribution over quantum states.
This assumption is crucial for their analysis as it maps the problem into the statistical query framework.

\paragraph[Our Contribution]{Our Contribution:}
On a conceptual level, our approach can be seen as a combination of the SQ and the LLA philosophy. 
We improve the previously mentioned approaches in the following ways.
\begin{itemize}
    \item Due to its generality, our framework can be applied to virtually any parametrized learning problem. As such, we can drop the central assumptions in \cite{anschuetz2022} including the ability to map the problem to the SQ setting and that the optimizer uses the parameter shift rule ($\qcsq$) respectively the average fidelity ($\qusq$).
    \item Similar to the statistical dimension by Blum et al. the dimension due to Anschuetz and Kiani is of a worst-case nature. Our formalism in contrast is of an average-case nature and as such yields insights about the generality of the derived bounds. This is particularly appealing in the analysis of practical algorithms since for any sufficiently universal underlying model worst-case hardness is to be expected.
    \item Our framework can be applied both, either directly to the problem on the level of the loss function, or to the induced problem on the level of the actual access on which the optimization algorithm relies on (e.g. the parameter shift evaluation in case of $\qcsq$). This can be used in order to distinguish whether hardness is due to the structure of the analyzed class, or whether hardness is due to the implementation of the optimization algorithm.
    \item We can directly connect the existence of BP's and NG's with hardness results in our framework. 
    As such, our framework naturally points out the difference between inherently hard problems, and loss functions where hardness of optimization is due to a poorly chosen measure.
\end{itemize}

\paragraph[Set-up]{Set-up:}
A loss function is a map $l:\Theta\to[0,1]$ and for a class of loss functions $\mc L\subset[0,1]^\Theta$ an algorithm is given access to an unknown $l\in\mc L$ called the \emph{access loss function} via the oracle $\eval_\tau(l)$\footnotemark.
\footnotetext{For different access, such as gradient evaluations, one can simply modify the given oracle access.}
In general, we assume that to each loss function $l$ it exists an associated \emph{operational loss function} $l_\mrm{op}$ which, in principle, may be any function $\Theta\to\mb R$. 
The aim of introducing the operational loss function is to model settings in which an algorithm optimizes one loss function in order to achieve low risk with respect to another loss function.
An example would be $\epsilon$-learning in total variation distance $l_\mrm{op}(\vartheta)=\tv(P_\vartheta , P_\text{true})$ from maximum mean discrepancy access $l(\vartheta)=(P_\vartheta -P_\text{true})^\dagger\cdot K\cdot(P_\vartheta -P_\text{true})$.
The corresponding learning problem is then defined via the map 
\begin{align}
    \mc Z^\epsilon:\mc L\to\mc P(\Theta)\doublecolon 
    l\mapsto \set{\hat\vartheta\midvert l_\mrm{op}(\hat\vartheta)\leq \mrm{opt}(\Theta,l_\mrm{op},\epsilon)}\,,
\end{align}
where, if not stated differently, $\mrm{opt}(\Theta,l_\mrm{op},\epsilon)=\min_\vartheta l_\mrm{op}(\vartheta)+\epsilon$ is the optimality criterion defining the solution set $\mc Z^\epsilon(l)$.

Most optimization routines used in practice are either deterministic, or only exploit a mild degree of randomness.
This is, randomness is usually restricted to a random initialization of parameters and a probabilistic perturbation to the update rule, for example a stochastic evaluation of the loss function, or its derivatives. 
We exemplify in \Cref{sec:derandomize-random-init} how random initializations can be analyzed by our deterministic bounds, which also implies the conventional interpretation of barren plateaus as in \Cref{rem:random-init2}. 
Further, we assume that the parametrization and the initial parameters are not fine-tuned to the precise probabilistic update as in e.g. \cite{abbe_poly-time_2020}.
To be precise, we assume that the stochastic update is well modelled by the imprecise evaluation as measured by $\tau$. 

We now focus on the deterministic average-case bounds.
From \Cref{sec:formalism-overview} we immediately inherit the lower bound on the query complexity of solving $\mc Z^\epsilon$ from $q$ many queries to $\eval_\tau$. 
For any $g:\Theta\to[0,1]$ it holds
\begin{align}\label{eq:intro-loss-hardness}
    &q\geq \frac{\beta-\sup_{\vartheta\in\Theta}\Pr_{l\sim\mu}\qty[l_\mrm{op}(\vartheta)\leq\mrm{opt}(\Theta,l_\mrm{op},\epsilon)]}{\sup_{\vartheta\in\Theta}\Pr_{l\sim\mu}\qty[\abs{l(\vartheta) - g(\vartheta)}>\tau]}
    &&\text{or similarly}
    &&&q\geq (\beta-\triv[\mc Z^\epsilon,\mu])\cdot\qnt[\mu,\tau]\,.
\end{align}

Finally, note that the special case $l=l_\mrm{op}$ and $\min_\vartheta l(\vartheta)=0$ for all $l$ is $1$-verifiable. 
Thus, in this setting the above bounds hold for any probabilistic or quantum algorithm with $\triv[\mc Z^\epsilon,\mu]$ replaced by $\sup_{\vartheta}\Pr_{l\sim\mu}[l(\vartheta)\leq\epsilon]$.

\paragraph[Connection to BP and NG]{Connection to BP and NG:}
In the following, we assume that $l_\mrm{op}=l$ with $\min_\vartheta l(\vartheta)=0$.
Introducing the dual learning problem $\widehat{\mc Z}$ of a learning problem $\mc Z:\mc S\to \mc P(\mc T)$ as
\begin{align}
    \widehat{\mc Z}:\mc T\to\mc P(\mc S)\doublecolon t\mapsto\mc Z_t\,,
\end{align}
we can then connect BP's and NG's with hardness of learning as in \Cref{eq:intro-loss-hardness}.

We show in \Cref{cor:hardness-and-narrowgorge} that average-case hardness in the sense of $\triv[\mc Z^\epsilon,\mu]=1-\Omega(\overpoly)$ together with $\kfrac\qty(\mu,g,\tau)=\tau^{2} 2^{-\Omega(n)}$ implies a narrow gorge in an $\Omega(\overpoly)$ fraction of instances of the dual problem $\widehat{\mc Z}^\epsilon$. 
Here, the vanishing maximally distinguishable fraction translates to the sharp concentration of \emph{every} dual loss function. 
Similarly, the assumption on the probability of trivial learning translates to the existence of a non-trivial minimal loss for a non-negligible fraction of dual loss functions.
Using a trick from \cite{Arrasmith_2022}, namely that, for many quantum models one can compute each gradient component from two specific queries to the loss function, we can then relate cost concentration to barren plateaus.
Our \Cref{obs:hardness-implies-barren} assumes parametrized quantum models that admit an efficient parameter shift rule. 
As such, we find that hardness of non-trivial learning as measured by $\kfrac\qty(\mu,g,\tau)=\tau^{2} 2^{-\Omega(n)}$ implies the existence of a barren plateau in \emph{every} instance of the dual problem.
Here we stress that the same argument works for any model class for which one can calculate the gradient from few specific function evaluations. 
Moreover, in contrast to e.g. \cite{anschuetz2022}, we require the parameter shift rule only as a mathematical trick and do not rely on any particular implementation.

Based on the above observations we introduce the notion of \emph{self-learning} in \Cref{def:self-learning}, which leads to self-dual parametrized learning problems. 
Thus, hardness of self learning in the sense above is equivalent to the existence of a NG in many loss functions, respectively a BP in every loss function if there is an efficient parameter shift rule.

\paragraph[Explicit Results]{Explicit Results:}
We exemplify a prototypical hard problem for loss function learning in \Cref{sec:basis-states}.
In particular, we show the hardness of learning computational basis states $\ket{z}$ from the loss function $l^z(\ket{\psi})=1-\abs{\braket{z}{\psi}}$ for both, $\ket\psi$ being a computational basis state, corresponding to self-learning, and for general $\ket\psi$. 
Since computational basis states can be easily learned from QSQ's, it is apparent that the hardness of this task is due to the access given to the problem instance (c.f. \Cref{rem:basis-state-hardness}). 
This underlines the care that has to be taken in the design of loss functions. 

Identifying the local projector in terms of the Pauli $Z$ gate as $\ketbra{z_i}=\frac12(\id+(-1)^{z_i}Z)$ we can connect hardness of learning computational basis states with Cerezo et al.'s work on barren plateaus at any depth for global cost functions \cite{cerezoCostFunction2021a}. 
We make this connection precise in \Cref{thm:globalcostfunctionandhardness} and \Cref{rem:regardingPauliLearning} where we prove and discuss the hardness of learning global Pauli strings. 
We accompany this result with a positive result in \Cref{thm:learningZXStrings} where we give an efficient algorithm for learning $ZX$ strings in the same setting.
This highlights that non-commutativity is not sufficient for the hardness of learning global Pauli strings.

We show how Cerezo et al.'s result on barren plateaus is tightly connected with self-learning parametrized quantum states.
In \Cref{sec:PQCs} we prove hardness of the latter task for any parametrization that contains a $1$-design, which implies a barren plateau in every loss function of the problem at any depth.
As such, we see the above series of results as a generalization of \cite{cerezoCostFunction2021a} which offers an intuitive interpretation.

In \Cref{sec:data-re-up} we investigate self-learning of linear and data re-uploading QML models. 
Those correspond to models for learning functions $f:\mc X\to[-1,1]$ with respect to the square loss $l^f(g)=\Ex_x[(f(x)-g(x))^2]$.
As we show and explain in \Cref{lem:function-selflearning-variance} and \Cref{rem:function-learning-triviality}, hardness of non-trivial self-learning of function classes is implied by concentration of $f_\vartheta$ with respect to $\vartheta\sim\mu$.
The models considered here are parametrized by a PQC as $f(x)=\tr[M\mc U_{\vartheta,x}(\ketbra{0})]$ with $\mc U_{\vartheta,x}$ the unitary channel corresponding to the PQC $U(\vartheta, x)$ and $M$ a fixed bounded observable. 
Via the concentration argument, we prove the hardness of non-trivial learning for linear and data re-uploading models in various regimes, implying the existence of a barren plateau in all instances.
However, we also find that the learning problem is average-case trivial, again due to concentration.
Let us highlight the difference between \Cref{thm:bp-linear-models} and \Cref{cor:bp-linear-models-part-2}. 
While concentration in the first case is due to the parametrization -- for every $x\in\mc X$ most functions have the same value -- concentration in the latter case is due to the encoding -- every function has the same function value for most $x\in\mc X$.
The former resembles the hardness of self-learning parametrized quantum states while the latter resembles hardness induced by a random initialization as in \Cref{sec:derandomize-random-init}.

While the barren plateau in QML models, parametrized quantum states and VQAs is always due to the design properties of the parametrization, our formalism makes explicit the key differences between the different settings. 
The barren plateau in VQAs is first and foremost about the initialization of parameters.
In contrast, for parametrized quantum states, and similarly for a global cost function, the barren plateau requires only a 1-design, corresponds to an actual hardness lower bound due to a loss function that obfuscates any useful information and thus can not be alleviated.
The case of QML models is particularly interesting. 
Here, a design-induced barren plateau states that, with respect to that measure, most functions are the same.
As such, it is a statement about the initialization. 
The fact that those problems are average-case trivial may be an indication that a careful choice of parameters could lead to successful algorithms.
However, a barren plateau due to the design properties of the encoding gates immediately renders the function class impractical as all functions are identical for all but a negligible subset of the data. 

We provide an exemplary comparison of our formalism to the SQ formalism introduced by Anschuetz and Kiani \cite{anschuetz2022} in \Cref{sec:simple-hard-class}.
In analogy to their Proposition 4, we analyze self-learning of the respective function class which is related to optimizing a global cost function. 
Using the similarity to linear models from the preceding section we find that self-learning of the function class is average-case trivial while the complexity of non-trivial learning is exponential. 
However, in their particular setting we find a much stronger result:
The learning problem they consider is hard already for any non-trivial subset of functions from the class, i.e. for any subset of size at least two.
This result is beyond what can be obtained by analyzing the statistical dimension from \cite{anschuetz2022} and highlights the fact that hardness in this scenario 
is due to the loss function obfuscating the relevant information.

Finally, in \Cref{sec:remark-local-hamiltonians} we show that our framework is incapable of yielding super polynomial lower bounds on the query complexity of VQAs for local Hamiltonians. 
We contrast this with the results of \cite{anschuetz2022} about shallow loss functions in \Cref{sec:discussion}.

\subsection{Discussion and future work}\label{sec:discussion}
\addsectionheader{sec:discussion}
This work revisits the evaluation oracle in a fresh light and provides a new framework for unifying (quantum) statistical and parametrized (quantum) algorithms. 
Our framework brings new insights and results into quantum statistical learning and the generalizations thereof.
Crucially, we see the evaluation query toolbox provided as the natural continuation of the SQ toolbox and believe
that it
has the potential to be a core element
in the future analysis of
generic quantum learning algorithms.

Our proofs for the lower bounds in \Cref{sec:quantumlearning,sec:variational} heavily rely on concentration 
bounds for ensembles of unitaries. 
However, we are not aware of existing bounds similar to our multi-copy variance bounds as in \Cref{thm:two-copy-clifford,thm:4-design-two-copy-hardness,thm:hardness-haar-random,thm:t-design-multi-copy-hardness}.
Thus, we believe that these results will be of independent interest to related fields in quantum information such as metrology, tomography and benchmarking.

The simplest way to learn stabilizers is via Bell difference sampling \cite{grossSchurWeylDuality2021}, which reduces the problem to learning affine subspaces of $\mb F_2^{2n}$.
Almost the same can be done from two copy access \cite{montanaroLearningStabilizerStates2017}, with the only difference being an additional unknown shift of the affine subspace. 
We have shown in \Cref{thm:two-copy-clifford} that learning stabilizers from two-copy QSQ access is hard.
We believe that this persists for larger yet constant $k$ and conject that no efficient $k=O(1)$ $\kqstat$ algorithm for learning stabilizer states exists.

We note that our formalism can be directly applied to the quantum approximate optimization algorithm (QAOA)~\cite{farhiQAOA2014}.
As such
it is worth mentioning
recent results~\cite{bassoLimitationsQAOA2022} which prove concentration of the set of QAOA loss functions for a practically motivated class of problems in precisely the sense that $\kfrac=o(1)$. 
It remains open whether concentration is sufficiently pronounced to yield a rigorous hardness result or may even imply a rigorous triviality statement, the latter also crucially depending on the precise optimality criterion.

Note that the local VQE architecture studied in Corollary 2 \cite{anschuetz2022} does not satisfy our assumption made in \Cref{sec:remark-local-hamiltonians}.
Hence, it is natural to ask which rigorous hardness result one obtains by applying our framework to their setting. 
We believe that one can apply the techniques from their analysis to bound the expected variance in the setting from \Cref{sec:derandomize-random-init} yielding rigorous lower bounds. 
However, it is unclear whether their result, which is in expectation,  sufficiently concentrates to obtain a super polynomial lower bound both in our, as well, as in their framework.

Finally, we note that our results are information-theoretic and thus yield unconditional lower bounds.
Consequently, this has the drawback of not capturing the computational complexity. 
In particular, a learning problem may be computationally hard yet query efficient, as famously conjectured for LPN with respect to sample versus computational complexity.
In this regard, we point to important work by Bittel and Kliesch~\cite{bittelTrainingVariationalQuantum2021}, which investigates this precise question.

\subsection{Note Added}\label{sec:noteadded}

During the finalization of this manuscript, we became aware of a work by  Arunachalam, Havlicek and Schatz\-ki~\cite{arunachalamRoleEntanglementStatistics2023}, which analyzes the role of entanglement and statistics in learning. 
In particular, to compare entangled and statistical measurements they, too, adapt Feldman's~\cite{feldman_general_2017} to derive lower bounds on the single-copy quantum statistical query complexity of learning which, consequently, largely match the bounds presented here applied to single-copy QSQs. 
The general scope of the two papers however is much different.

Similarly, the recently introduced frameworks for learning quantum processes by quantum statistical queries \cite{angrisaniLearningUnitariesQuantum2023,wadhwaLearningQuantumProcesses2023} can be naturally analyzed in our formalism. 
In particular, the $\mathtt{QPStat}$ oracle from \cite{wadhwaLearningQuantumProcesses2023} corresponds to a linear evaluation oracle as in \Cref{sec:intro-linear}.
The $\mtt{QPStat}$ implements linear functions over the real vector space $\mb H^{N\times N}\otimes \mb H^{N\times N}$ spanned by pairs of states and observables, where the function can be written in terms of a channel $\mc E$ as $f_\mc E(O\otimes\rho)=\lrangle{\mc E, O\otimes\rho}_{\mc V}=\lrangle{\lrangle{\mc E\midvert O\otimes\rho}}=\tr[O\mc E(\rho)]$.

\subsection*{Acknowledgements}\label{sec:acknowledge}

I thank Hakop Pashayan for many insightful discussions and, in particular, for sharing an initial form of \Cref{lem:multi-copy-to-qsamp} and proof sketch.
Many thanks go to Richard K\"ung for valuable support and insights about the unitary and Clifford Haar measure which led to the proofs of \Cref{thm:two-copy-clifford,thm:t-design-multi-copy-hardness}.
Then I am grateful to Christian Bertoni for great support in dealing with random quantum circuits.
More generally, I thank Ryan Sweke, Johannes Jakob Meyer, Marcel Hinsche, Janek Denzler, Marios Ioannou, Elies Gil-Fuster, Marek Gluza, Nathan Walk, Jonas Haferkamp, Matthias Caro and Jens Eisert for great help, support, \href{https://en.wikipedia.org/wiki/Rubber_duck_debugging}{rubberducking} and many provoking, disillusioning, inspiring and fruitful discussions.
This work was supported by the BMBF (QPIC-1, Hybrid), DFG (CRC 183), the BMBK (EniQmA), and the Munich Quantum Valley (K-8).

\clearpage
\part{Formalism}\label{sec:formalism}
\section{General}\label{sec:general}
\addsectionfooter{sec:general}
\addsectionheader{sec:general}
\startcontents[general]
\vspace{1cm}
\printcontents[general]{}{1}{}
\vspace{1cm}

    \subsection{Set-Up}\label{sec:setup}
    \addsectionheader{sec:setup}
    We adapt the framework of \cite{feldman2017statistical} to formalize general learning tasks.

\begin{definition}[General Learning Problem $\mc Z$]\label{def:learnZ}
    Let $\mc S$ and $\mc T$ be sets. We refer to a map 
    \begin{align}
        \mc Z: \mc S\to \mc P(\mc T)
    \end{align} 
    from the source set $\mc S$ to the power set of the target set $\mc P(\mc T)$ as a \emph{Learning Problem}.
    For any source $s\in\mc S$ we refer to $\mc Z(s)$ as the solutions associated with $s$. 
    Moreover, for any $\mc S'\subseteq \mc S$ we denote by $\mc Z(\mc S)\subseteq\mc P(\mc T)$ the set of solution sets corresponding to $\mc S'$ under $\mc Z$ and use the shorthand 
    \begin{equation}
        \bigcup \mc Z(\mc S') = \bigcup_{s\in \mc S'} \mc Z(s)\,.
    \end{equation}
    For any target object $t\in\mc T$ we denote by 
    \begin{align}
        \mc Z_t=\set{s\in\mc S \midvert t\in\mc Z(s)}\subseteq \mc S\,,
    \end{align}
    the set of sources compatible with $t$, where $\mc Z_t=\varnothing$ if $t$ is no solution to any $s\in\mc S$.
    By $\mc Z_\mc T$ we then denote the set of sets of source objects with equivalent solutions
    \begin{align}
        \mc Z_\mc T=\set{\mc Z_t \midvert t\in\mc T}\subseteq \mc P(\mc S)\,,
    \end{align}
    and we say that the learning problem is \emph{surjective} if it holds
    \begin{equation}\label{eq:surjectivelearningproblem}
        \bigcup_{s\in\mc S}\mc Z(s) = \mc T\,.
    \end{equation}  
    For a subset $\mc S'\subset\mc S$ we denote by $\mc Z\vert_{\mc S'}$ the \emph{Restricted Problem} of $\mc Z$ to $\mc S'$, which is defined as 
    \begin{equation}
        \mc Z\vert_{\mc S'}:\mc S'\to\mc P(\mc T)\doublecolon s\mapsto \mc Z(s)\,.
    \end{equation}
\end{definition}

\begin{remark}[On Surjectivity]
    Note that the surjectivity condition in \Cref{eq:surjectivelearningproblem} is in general different from the requirement of surjectivity of the map $\mc Z$ and regards the question, whether for any $t\in\mc T$ there is an $s\in\mc S$ such that $t$ is contained in the solution set $\mc Z(s)$.
    As such, for a surjective learning problem, every target is eventually a possible solution. 
    In contrast, for a non-surjective learning problem, some targets are never a valid solution for no source object $s$.
    Moreover, note that any learning problem can be recast as a surjective learning problem by restricting to the relevant targets.
\end{remark}

Let us now define the computational task of learning as defined by a general learning problem $\mc Z$.
We assume that the potential algorithm is given access to instances $s\in\mc S$ by some oracle,
where we understand an oracle $\mc O$ as an abstract operational model for this access.
This induces a particular oracle $\mc O(s)$ to each $s\in\mc S$, which has to satisfy certain properties as defined by the model $\mc O$.

\begin{problem}[Learning $\mc Z$]\label{prob:learnZ}
Given a general learning problem $\mc Z:\mc S\to\mc T$ as in \Cref{def:learnZ}. The problem of \emph{learning} $\mc Z$ from oracle access $\O$ is defined as, given oracle access to $\O(s)$ for an unknown $s\in\mc S$, to return a description of some $t\in\mc Z(s)$.
\end{problem}

A natural example for a learning problem is  Valiant's theory of PAC learning \cite{valiant_theory_1984}.

\begin{myproblem}[PAC Learning]\label{prob:pac-learning}
    Let $n>0$, $\epsilon,\delta\in[0,1)$, let $\mc H$ be a class of boolean functions $f:\mb F_2^n\to\mb F_2$ and let $\mc D$ be a class of distributions over $\mb F_2^{n+1}$.
    The following task is referred to as \emph{Probably Approximately Correct} (PAC) learning of $\mc H$ with respect to $\mc D$.
    Given oracle access $\mc O$ to an unknown $D\in\mc D$ return some $h\in\mc H$, referred to as the \emph{Hypothesis}, such that with probability at least $1-\delta$ it holds 
    \begin{equation}
        \Ex_{(x,y)\sim D}\qty[y\neq h(x)]<\epsilon\,.
    \end{equation}
\end{myproblem}

The original setting, as introduced by Valiant and which is referred to as the \emph{Realizable} setting, is concerned with classes of distributions $\mc D(P,\mc H)=\set{P^f \midvert f\in\mc H\,,\; P^f(x,y)}$, where
\begin{align}
    P^f(x,y) = 
    \begin{cases}
        0,&\text{if}\;y\neq f(x)\\
        P(x),&\text{if}\;y=f(x)\,.
    \end{cases}
\end{align}

We say a class $\mc H$ \emph{can be efficiently PAC learned with respect to} $\mc D$ if there is an algorithm for PAC learning $\mc H$ with respect to $\mc D$ that runs in time $O(\poly(n,\sfrac{1}{\epsilon},\sfrac{1}{\delta}))$.
If $\mc H$ can be PAC learned with respect to the class of all $n+1$ bit distributions $\Delta(n+1)$, then we say that $\mc H$ \emph{is PAC learnable}.
The case $\epsilon=0$ is referred to as \emph{exact} PAC learning
It is a simple exercise to recast PAC learning in terms of a learning problem $\mc Z$:

\begin{example}[PAC Learning as Learning Problem]\label{ex:pac-as-learning}
    Let $\mc H$ be a class of boolean functions $f:\mb F_2^n\to\mb F_2$ and let $\mc D\subseteq\Delta(n+1)$.
    PAC learning of $\mc H$ with respect to $\mc D$ then corresponds to the learning problem 
    \begin{equation}
        \mc Z:\mc D \to \set{\mc Z(D)=\set{h\in\mc H\midvert \Ex_{(x,y)\sim D}\qty[y\neq h(x)]<\epsilon}\midvert D\in\mc D}\,.
    \end{equation}
    While $\epsilon$ translates to the learning criterion defining the sets $\mc Z(D)$, 
    $\delta$ translates to the success probability $\alpha=1-\delta$ defining the success statistics of random algorithms for learning $\mc Z$. Here, any learning algorithm in Valiant's original model would be considered a random algorithm due to the inherent randomness of the random example oracle. 
\end{example}

Let us also introduce the notions of dual learning problems, equivalence of learning problems and self-dual learning problems, which will turn out handy later on.

\begin{definition}[Duality and Equivalence]\label{def:duallearningproblem}
    Let $\mc Z:\mc S\to\mc P(\mc T)$ be a surjective learning problem.
    We denote by $\widehat{\mc Z}$ the dual learning problem which is defined as 
    \begin{equation}
        \widehat{\mc Z}:\mc T\to\mc P(\mc S)\doublecolon t\mapsto \mc Z_t\,.
    \end{equation}
    Further, We say that two learning problems $\mc Z:\mc S\to\mc P(\mc T)$ and $\mc Y:\mc S'\to\mc P(\mc T')$ are \emph{Equivalent}, if there exists bijections $\Phi:\mc S\to\mc S'$ and $\Psi:\mc T\to\mc T'$ such that for any $s\in\mc S, s'\in\mc S'$ and any $t\in\mc T, t'\in\mc T'$ it holds
    \begin{align}
        &\Psi(\mc Z(s)) = \mc Y(\Phi(s)) &&\text{and} &&&\Phi(\mc Z_t) = \mc Y_{\Psi(t)}\\[10pt]
        &\Psi^{-1}(\mc Y(s')) = \mc Y(\Phi^{-1}(s')) &&\text{and} &&&\Phi^{-1}(\mc Z_{t'}) = \mc Y_{\Psi^{-1}(t')}\,.
    \end{align}
    A surjective learning problem that is equivalent to its own dual learning problem is said to be \emph{Self-Dual}.
\end{definition}

In this work we focus on learning problems where the elements of the source set $\mc S$ admit the structure of functions and assume an oracle access that allows to query those functions up to some imprecision. This is formalized as follows.

\begin{definition}[Evaluation Oracle]\label{def:evaloracle}
    Let $(M, \met)$ be a metric space with metric $\met:M\times M\to\mb R^+_0$, let $\mc X$ be a set and let $\mc S\subseteq M^\mc X$. For any $s\in\mc S$ and $\tau>0$ we denote by $\eval_\tau(s)$ the \emph{evaluation oracle} of $s$ with tolerance $\tau$. When queried with any $x\in\mc X$ it returns some $v\in M$ such that it holds $\met(v,s(x))\leq\tau$. 
    We refer to a query with respect to $\eval_\tau(s)$ as a $\tau$ accurate evaluation query.
\end{definition}

In most cases we will consider $M\subseteq\mb C$ and consider the metric $\met(\cdotspace,\cdotspace)=\abs{\cdotspace-\cdotspace}$. 
However, $M$ does not need to be a set of scalars and may be any set endowed with a metric.

\subsubsection{Notions of Complexity}\label{sec:onComplexity}
\addsectionheader{sec:onComplexity}

Here we briefly state the notions of complexity we work with in this work: the deterministic and random worst- and average-case complexity. 

\begin{definition}[Worst Case Query Complexity]
    The \emph{deterministic worst-case query complexity} of the problem of learning $\mc Z$ is the smallest $q$ such that 
    there exists a deterministic algorithm $\A$ that
    solves \Cref{prob:learnZ} from at most $q$ many queries to $\O(s)$.

    The \emph{random worst-case query complexity} of the problem of learning $\mc Z$ with success probability $\alpha\in[0,1]$ is the smallest $q$ such that there 
    exists a random algorithm $\A$ that solves \Cref{prob:learnZ} 
    from $q$ queries to $\O$
    with probability at least $\alpha$ over $\A$'s internal randomness. 
\end{definition}

In this work we will deal with deterministic, classical probabilistic and quantum algorithms.  
We refer to classical probabilistic algorithms in short by \emph{probabilistic} algorithms. 
By the term \emph{random} algorithm we refor to both, quantum and probabilistic algorithms.

\begin{definition}[Average Case Query Complexities]
    Let $\mu$ be a measure over $\mc S$.
    The \emph{deterministic average-case query complexity} of the problem of learning $\mc Z$ with success probability $\beta\in[0,1]$ is the smallest $q$ such that 
    there exists a deterministic algorithm $\A$ that
    solves \Cref{prob:learnZ} on at least a $\beta$ fraction of instances from at most $q$ many queries to $\O(s)$. In particular, $\A^{\mc O(s)}$ has a deterministic output for every $s\in\mc S$, and it holds
    \begin{equation}
        \Pr_{s\sim\mu} \qty[\A^{\O(s)}\to\mc Z(s) \;\text{using at most $q$ queries}] \geq \beta\,.
    \end{equation}

    The \emph{joint average-case query complexity} of the problem of learning $\mc Z$ with success probability $\beta\in[0,1]$ is the smallest $q$ such that there 
    exists a random algorithm $\A$ such that it holds
    \begin{equation}
        \Pr_{s\sim\mu,\A} \qty[\A^{\O(s)}\to\mc Z(s) \;\text{using at most $q$ queries}] \geq \beta\,,
    \end{equation}
    where $\Pr_\A$ denotes the probability with respect to the internal randomness of $\A$ (and the oracle).

    The \emph{random average-case query complexity} of the problem of learning $\mc Z$ with success probabilities $\alpha\in[0,1]$ and $\beta\in[0,1]$ is the smallest $q$ such that there 
    exists a random algorithm $\A$ such that it holds
    \begin{equation}
        \Pr_{s\sim\mu} \qty[\Pr_\A\qty[\A^{\O(s)}\to\mc Z(s) \;\text{using at most $q$ queries}]\geq\alpha] \geq \beta\,.
    \end{equation}
    In particular, there is a set $\mc S'\subseteq\mc S$ with $\mu(\mc S')\geq\beta$ such that, with probability at least $\alpha$ over $\A$'s (and the oracle's) internal randomness, $\A$ solves $\mc Z$ restricted to $\mc S'$ from $q$ queries to $\O(s)$.
\end{definition}

    \subsection{Deterministic Average-Case Lower Bound}\label{sec:deterministiclowerbounds}
    \addsectionheader{sec:deterministiclowerbounds}
    
Throughout this and the following \Cref{sec:deterministiclowerbounds,sec:joinedlowerbounds,sec:randomlowerbounds,sec:simplified} we assume $\tau>0$ and $\mc S, M, \met$ and $\mc X$ as in \Cref{def:evaloracle}.

\begin{theorem}[Deterministic Average Case Lower Bound. Adapted from Lemma C.2 in \cite{feldman_general_2017}]\label{thm:deterministicavglowerbound}
    Suppose there is a deterministic algorithm $\A$ that learns $\mc Z:\mc S\to\mc P(\mc  T)$ with probability $\beta$ over the instance $s\sim\mu$ from $q$ many $\tau$ accurate evaluation queries. Then, for any $f\in M^\mc X$ it holds
    \begin{align}\label{eq:deterministic-avg-lower}
        q\geq\frac{\beta - \sup_{t\in\mc T}\Pr_{s\sim\mu}\qty[s\in\mc Z_t]}{\max_{x\in\mc X}\Pr_{s\sim\mu}\qty[\met\qty(s(x),f(x))>\tau]}\,.
    \end{align}
\end{theorem}

\begin{proof}
    Denote by $\mc S'\subseteq\mc S$ a set of size $\mu(\mc S')=\beta$ on which $\mc A$ successfully learns $\mc Z$ from $q$ many $\tau$ accurate evaluation queries.
    We run $\mc A$ and simulate every query response to $x\in\mc X$ by $f(x)$. By assumption the algorithm makes $q$ queries $x_1,\dots,x_q$ and, without loss of generality, we assume that $\A$ returns some $t\in\mc T$. 
    Let $s'\in\mc S'\setminus \mc Z_t$.
    Since $s'\in\mc S'$ the algorithm will be correct when given access to $\eval_\tau(s')$. Because $s'\not\in\mc Z_t$, 
    on the other hand, $t$ is not a valid solution to $s'$. 
    Hence, either at least one query must distinguish $s'$ from $f$ in the sense that for some $i\in[q]$ it holds 
    $\met\qty(s'(x_i), f(x_i))>\tau$, or we arrive at a contradiction, because all responses $f(x_i)$ are valid responses for $\eval_\tau(s')$. 
    Hence, it holds
    \begin{equation}\label{eq:mockdeterministicavg}
        \begin{split}
            \beta - \Pr_{s\sim\mu}\qty[s\in\mc Z_t] &\leq \Pr_{s\sim\mu} \qty[\exists i : \met\qty(s(x_i), f(x_i))>\tau]\\
            &\leq \sum_{i=1}^q \Pr_{s\sim\mu} \qty[\met\qty(s(x_i), f(x_i))>\tau]\\
            &\leq q \max_{x\in\mc X} \Pr_{s\sim\mu} \qty[ \met\qty(s(x), f(x))>\tau]\,.
        \end{split}
    \end{equation}
    Now assume that, after interacting with $f$ the algorithm did not return a $t\in\mc T$. then again by $\A$'s determinism, for \emph{any} $s'\in\mc S'$ there must exist a distinguishing query $x_i$ that distinguishes $s'$ from $f$. The claim then follows by taking the supremum over $t\in\mc T$ to bound the mass of the unknown $\mc Z_t$.
\end{proof}

    \subsection{Joint Average-Case Lower Bound}\label{sec:joinedlowerbounds}
    \addsectionheader{sec:joinedlowerbounds}
    
\begin{theorem}[Joint Average Case Lower Bound for Probabilistic Algorithms. Adapted from Lemma C.2 in \cite{feldman_general_2017}]\label{thm:probabilisticavglowerbound}
    Suppose there is a classical probabilistic algorithm $\A$ that learns $\mc Z:\mc S\to\mc P(\mc  T)$ with probability $\beta$ over the joint measure over the instance $s\sim\mu$ and $\A$'s internal randomnes, from $q$ many $\tau$ accurate evaluation queries. Then, for any $f\in M^\mc X$ it holds
    \begin{align}\label{eq:probabilisticlowerbound}
        q\geq\frac{\beta - \sup_{t\in\mc T}\Pr_{s\sim\mu}\qty[s\in\mc Z_t]}{\max_{x\in\mc X}\Pr_{s\sim\mu}\qty[\met\qty(s(x),f(x))>\tau]}\,.
    \end{align}
\end{theorem}

\begin{proof}
    We interpret any probabilistic algorithm $\A$ as an ensemble $\set{\A_r}$ of deterministic algorithms depending on the internal randomnes (as made explicit by $r\in R$ for a suitable set $R$). 
    Now suppose a probabilistic algorithm $\A$ that succeed with success rate at least $\beta$ over the joint measure of its internal randomnes, abusing notation $r\sim\A$, and the instance $s\sim\mu$. 
    Further denote by $P^+\subseteq R\times \mc S$ the subset such that for any $(r, s)\in P^+$ the algorithm $\A_r$ succeeds when given access to $s$, denote by $\mc S_r\subseteq\mc S$ the set on which $\A_r$ is succesfull and denote by $\A\times\mu$ the joint measure over $R\times\mc S$.
    Then, $\A\times\mu(P^+)\geq\beta$ implies that there exists some $r\in R$ such that $\mu(\mc S_r)\geq\beta$. 

    To arrive at the claim as in \Cref{eq:probabilisticlowerbound} recall that by 
    \Cref{thm:deterministicavglowerbound} the average-case problem with success probability $\beta$ over the instance requires 
    at least $q$ many queries in case of deterministic algorithms.
\end{proof}

The general statement for random algorithms now follows from the following simple observation.

\begin{observation}
    Any quantum algorithm with evaluation query access translates to a probabilistic algorithm with evaluation query access, at the price of an at most exponential overhead in computational time and hence random resources.
    The statement in \Cref{thm:probabilisticavglowerbound} on the other hand bounds the number of queries any probabilistic algorithm must make, independent of its computational resources. 
    Thus, as the just mentioned translation does not increase the number of queries made the probabilistic bound also holds for quantum algorithms yielding the following corollary.
\end{observation}

\begin{corollary}[Joint Average Case Lower Bound.]\label{cor:jointavglowerbound}
    Suppose there is a random algorithm $\A$ that learns $\mc Z:\mc S\to\mc P(\mc  T)$ with probability $\beta$ over the joint measure over the instance $s\sim\mu$ and $\A$'s internal randomnes, from $q$ many $\tau$ accurate evaluation queries. Then, for any $f\in M^\mc X$ it holds
    \begin{align}\label{eq:randomlowerbound}
        q\geq\frac{\beta - \sup_{t\in\mc T}\Pr_{s\sim\mu}\qty[s\in\mc Z_t]}{\max_{x\in\mc X}\Pr_{s\sim\mu}\qty[\met\qty(s(x),f(x))>\tau]}\,.
    \end{align}
\end{corollary}

    \subsection{Random Average-Case Lower Bound}\label{sec:randomlowerbounds}
    \addsectionheader{sec:randomlowerbounds}

In order to derive a general bound the random average-case complexity we continue to follow Feldmans work \cite{feldman_general_2017}.
Later we introduce simplified bounds in \Cref{sec:verifiable-lowerbounds} for verifyable learning problems.
To begin with, a crucial notion is that of the probability of guessing the correct solution. 
To this end introduce the following definition.

\begin{definition}\label{def:alphaprobablesolutions}
    Let $\mc Z:\mc S\to\mc P(\mc T)$ be a learning problem. For any distribution $\vartheta$ over $\mc T$ and $\gamma>0$ denote by 
    \begin{align}
        \mc Z_\vartheta(\gamma) = \set{s\in\mc S\midvert \vartheta\qty(\mc Z(s))\geq \gamma}
    \end{align}
    the set of sources whos solutions are at least $\gamma$ probable under $\vartheta$.
\end{definition}

\begin{theorem}[Random Average-Case Lower Bound]\label{thm:randomavglowerbound}
    Suppose there is a random algorithm $\A$ that learns $\mc Z:\mc S\to\mc P(\mc  T)$ with probability $\beta$ with respect to $s\sim\mu$ and with probability $\alpha$ with respect to $\A$'s internal randomness, from $q$ many $\tau$ accurate evaluation queries. Then, for any $f\in M^\mc X$ and any $\gamma\in(0,\alpha)$ it holds
    \begin{equation}
        q >  \frac{\qty(\alpha-\gamma)\cdot \qty(\beta - \sup_\vartheta\Pr_{s\sim\mu}\qty[s\in\mc Z_{\vartheta}(\gamma)])}{\max_{x\in\mc X}\Pr_{s\sim\mu} \qty[ \met\qty(s(x),f(x))>\tau]}\,,
    \end{equation}
    where the supremum is over all measures $\vartheta$ over $\mc T$.
\end{theorem}

\begin{proof}
    Let $\A$ be a random algorithm that successfuly learns $\mc Z$ with probability $\beta$ with respect to $\mu$ and probability $\alpha$ with respect to its own internal randomnes. 
    Similar as in the proof of \Cref{thm:deterministicavglowerbound} we run $\A$ and simulate every query $x\in\mc X$ by $f(x)$. Denote by $x_1,\dots,x_q$ the queries made, which are now random variables with respect to $\A$'s randomnes.
    Without loss of generality assume that after interaction $\A$ returns some $t\in\mc T$, which is now a random variable and we denote by $\vartheta(f)$ the corresponding distribution induced by $f$ and $\mc A$ over $\mc T$.
    Denote by $\mc S'\subseteq\mc S$ the set on which $\A$ is correct with probability at least $\alpha$ over its internal randomnes. 
    For any $s\in M^\mc X$ denote by
    \begin{equation}
        p(s) = \Pr_\A\qty[\exists i\in[q] : \met\qty(s(x_i), f(x_i))>\tau]
    \end{equation}
    the probability that one of the queries made by $\A$ distinguishes between $s$ and $f$. 

    Now, let $s'\in\mc S'\setminus\mc Z_{\vartheta(f)}(\gamma)$.
    By the correctness of the algorithm, it has to output some $t\in\mc Z(s')$ with probability at least $(1-p(s'))-(1-\alpha)$,
    where the first term corresponds to the probability that all query responses given are valid for $s'$ 
    and the second term corresponds to the failure probability.
    In contrast, the probability of $\A$ returning a $t\in\mc Z(s')$ after interacting with $f$ is bounded by $\vartheta(f)(\mc Z(s'))<\gamma$ due to \Cref{def:alphaprobablesolutions} .
    Thus, $p(s')>\alpha-\gamma$.
    
    This implies 
    \begin{equation}\label{eq:mockproofrandomaverage}
        \begin{split}
            \alpha-\gamma &< \Pr_{s\sim\mu, \A} \qty[ \exists i\in[q] : \met\qty(s(x_i),f(x_i))>\tau \midvert s\in\mc S'\setminus\mc Z_{\vartheta(f)}(\gamma) ]\\[8pt]
            &\leq \frac{\Pr_{s\sim\mu, \A} \qty[\exists i\in[q] : \met\qty(s(x_i),f(x_i))>\tau]}{\Pr_{s\sim\mu}\qty[s\in\mc S'\setminus\mc Z_{\vartheta(f)}(\gamma)]} 
            \leq \frac{\Pr_{s\sim\mu, \A} \qty[\exists i\in[q] : \met\qty(s(x_i),f(x_i))>\tau]}{\beta - \Pr_{s\sim\mu}\qty[s\in\mc Z_{\vartheta(f)}(\gamma)]} \\[8pt]
            &\leq \sum_{i=1}^q  \frac{\Pr_{s\sim\mu, \A} \qty[ \met\qty(s(x_i),f(x_i))>\tau]}{\beta - \Pr_{s\sim\mu}\qty[s\in\mc Z_{\vartheta(f)}(\gamma)]} \leq q\cdot \frac{\max_{x\in\mc X}\Pr_{s\sim\mu} \qty[ \met\qty(s(x),f(x))>\tau]}{\beta - \Pr_{s\sim\mu}\qty[s\in\mc Z_{\vartheta(f)}(\gamma)]} \,.
        \end{split}
    \end{equation}
    Reordering yields
    \begin{equation}
        q >  \frac{\qty(\alpha-\gamma)\cdot \qty(\beta - \Pr_{s\sim\mu}\qty[s\in\mc Z_{\vartheta(f)}(\gamma)])}{\max_{x\in\mc X}\Pr_{s\sim\mu} \qty[ \met\qty(s(x),f(x))>\tau]}\,.
    \end{equation}
    The claim then follows from taking the supremum in order to bound the contribution by $\vartheta(f)$.
\end{proof}

    \subsection{Summary}\label{sec:formalism-summary}
    \addsectionheader{sec:formalism-summary}
    
The structure of \Cref{eq:deterministic-avg-lower} is typical for the bounds presented in this work, and it is easy to see, that the bound is tight in a particular sense.
Let us restate \Cref{eq:deterministic-avg-lower}
\begin{align*}
    q\geq\frac{\beta - \sup_{t\in\mc T}\Pr_{s\sim\mu}\qty[s\in\mc Z_t]}{\max_{x\in\mc X}\Pr_{s\sim\mu}\qty[\met\qty(s(x),f(x))>\tau]}\,,
\end{align*}
and start with investigating the numerator.
To begin with, the numerator is bounded by $\leq1$ and can therefore only weaken the bound on $q$.
Interestingly, the numerator is related to what we call the probability of trivial learning.
Let $\triv=\sup_{t\in\mc T}\Pr_{s\sim\mu}\qty[s\in\mc Z_t]$ and denote by  $t^*=\argmax_{t\in\mc T}\Pr_{s\sim\mu}\qty[s\in\mc Z_t]$.
Further, define $\mc A^*$ to be the trivial algorithm which, when run makes not even a single query and always outputs $t^*$.
Then, $\triv$ is the probability of success of $\A^*$ with respect to the instances drawn according to $s\sim\mu$.
Consequently, $q=0$ must hold for every $\beta\leq \triv$, which implies that \Cref{eq:deterministic-avg-lower} is tight with respect to $\beta$. 
We refer to $\triv$ as the \emph{Probability of Trivial Learning}.

The probability $\triv$ and thus the numerator depend on $\mc T$.
The denominator in contrast is independent of $\mc T$ and only depends on $\mc S$ (and thus $\mc X$) and $\mu$. 
This implies that the denominator is identical for every learning problem over $\mc S$. 
With the previous paragraph in mind, we find that the denominator of \Cref{eq:deterministic-avg-lower} asymptotically lower bounds \emph{every} sufficiently non-trivial learning problem over $\mc S$.
We thus call the denominator the \emph{Query Complexity of Non-Trivial Learning}.
Let us collect this in the following definition.

\begin{definition}[Trivial and Non-Trivial Learning]\label{def:trivial-and-non-trivial}
    Let $\mc Z:\mc S\to\mc P(\mc T)$ be a learning problem over $\mc S\subseteq M^\mc X$ and let $\mu$ be a measure over $\mc S$. We then make the following definitions.
    \begin{itemize}
        \item \textbf{(Non-)Trivial Learning:} We denote by 
        \begin{align}
            \triv[\mc Z,\mu]=\sup_{t\in\mc T}\;\Pr_{s\sim\mu}\qty[s\in\mc Z_t]
        \end{align}
        the \emph{Probability of Trivial Learning} $\mc Z$ with respect to $\mu$. 
        \item \textbf{Complexity of Learning:} For any $f\in M^\mc X$ we denote by 
        \begin{align}
            \kfrac\qty(\mu, f, \tau) = \max_{x\in\mc X}\Pr_{s\sim\mu}\qty[\met(s(x),f(x))>\tau]\,,
        \end{align} 
        the \emph{Maximally Distinguishable Fraction} from $f$ with respect to $\mu$. 
        The \emph{Query Complexity of Non-Trivial Learning} with respect to $\mu$ and tolerance $\tau$ is then defined as
        \begin{align}
            \qnt[\mu, \tau] = \max_{f\in M^\mc X}\kfrac\qty(\mu, f, \tau)^{-1} \,.
        \end{align}
    \end{itemize} 
\end{definition}

Intuitively, we say that a learning problem is non-trivial with respect to a measure $\mu$ over $\mc S$ if for every trivial algorithm $\mc A$ there is a realistic chance to observe an $s\sim\mu$ which cannot be solved by $\mc A$.
This is formalized as follows.

\begin{definition}[Non-Trivial Learnign Problems]\label{def:non-trivial-learning-problem}
    With respect to a scaling parameter $n\in\mb N$ we say that a family of learning problems $\{\mc Z_n\}$ is \emph{Non-Trivial} with respect to $\mu$ if it holds $\triv[\mc Z_n,\mu]=1-\Omega(\overpoly)$ as $n\to\infty$.
\end{definition}

In many cases, one can directly estimate the concentration that defines $\kfrac\qty(\mu, f, \tau)$.
However, sometimes it is more practical to apply either Markov's or Chebyshev's inequality.
For the sake of completeness, we state the corresponding bounds as follows.

\begin{corollary}[Variance and Average Discrimination Bound for $q_\mtt{nt}$]\label{cor:variance-bound-qnt}
    The complexity of non-trivial learning is bounded from below by the \emph{Average Discrimination} as
    \begin{align}
        \kfrac\qty(\mu, f, \tau) \leq \tau^{-1}\cdot\max_{x\in\mc X} \Ex_{s\sim\mu}[\met(s(x),f(x))]\,.
    \end{align}
    Moreover, if the variance is well-defined, then it holds
    \begin{align}
        \qnt[\mu, \tau] \geq \tau^2\cdot\Var_{s\sim\mu}[s(x)]^{-1}\,.
    \end{align}
\end{corollary}

\begin{proof}
    The first statement is an immediate consequence of Markov's inequality. The second follows from Chebyshev's inequality.
\end{proof}

While the average discrimination will in general be well-defined, the variance may be not defined in some metric spaces. We refer to 
\Cref{app:metric-variance}
for the more general definition of the metric variance covering those cases.

\begin{envbox}[opacityfill=.11]{Red}\begin{remark}
    We summarize this section in the following equation bounding the average-case query complexity for deterministic algorithms
    \begin{align}
        q\geq (\beta-\triv[\mc Z, \mu])\cdot\qnt[\mu,\tau] \geq (\beta-\triv[\mc Z, \mu]) \cdot\tau^2 \cdot\max_{x\in\mc X}\Var_{s\sim\mu}[s(x)]^{-1}\,,
    \end{align}
    with $\beta$ the success probability over $\mu$.
\end{remark}\end{envbox}

\stopcontents[general]


\section{Overview of Learning Problems}\label{sec:encyclopedia}
\addsectionfooter{sec:encyclopedia}
\addsectionheader{sec:encyclopedia}
\startcontents[encyclopedia]
\vspace{1cm}
\printcontents[encyclopedia]{}{1}{}
\vspace{1cm}

This section contains definitions and examples for particular types of learning problems, which we will analyze in more detail in this work.

\subsection{Decision Problems}\label{sec:encyclopedia-decission}
\addsectionheader{sec:encyclopedia-decission}

The following many-to-one decision problem is the paradigmatic learning problem to gain intuition about the query complexity of learning problems.

\begin{example}[Decission Problem $\dec$]\label{ex:decissionproblem}
    Let $\mc S'$ be a set, $s^*\in\mc S'$ and $\mc S\subseteq\mc S'$. We define the \emph{Decision Problem} of distinguishing between $\mc S$ and $s^*$ as 
    \begin{equation}
        \text{$\dec$}(\mc S, s^*):\mc S\cup\set{s^*}\to\set{\set{\mc S}, \set{\set{s^*}}}\subseteq\mc P(\set{\mc S, \set{s^*}})
    \doublecolon s\mapsto
    \begin{cases}
        \set{\mc S} &,\quad s\in\mc S\\
        \set{\set{s^*}} &,\quad s\notin\mc S\,.
    \end{cases}
    \end{equation}
    In words, given access to some unknown $s\in\mc S\cup\set{s^*}$, the task is to decide whether $s\in\mc S$ or $s\not\in\mc S$.
    If $s^*\in\mc S$ we say the decision problem is \emph{trivial} and else we say it is \emph{non-trivial}.
\end{example}

Typical instances of $\dec$ would be for example uniformity testing.

Similarly to the many-to-one decision problem just introduced is the many-to-many decision problem which, abusing notation, will also be denoted by $\dec$.

\begin{example}[Many-To-Many Decision Problem]
    Let $\mc S'$ be a set and for $i=0,1$ let $\mc S_i\subset\mc S'$.
    And assume that $\mc S_0\cap\mc S_1=\varnothing$.
    Define the \emph{Many-To-Many Decision Problem} $\dec(\mc S_0, \mc S_1)$ as 
    \begin{align}
        \text{$\dec$}(\mc S_0, \mc S_1):\mc S_0\cup\mc S_1\to\set{\set{\mc S_0}, \set{\mc S_1}}\subset\mc P(\set{\mc S_0, \mc S_1})\doublecolon
        s\mapsto
        \begin{cases}
            \set{\mc S_0}\,,&s\in\mc S_0\,,\\
            \set{\mc S_1}\,,&s\in\mc S_1\,.
        \end{cases}
    \end{align}
\end{example}

\subsection{Verifiable Learning Problems}\label{sec:encyclopedia-verifialbe}
\addsectionheader{sec:encyclopedia-verifialbe}

Verifiable learning problems are, in short, problems, for which there exists an efficient test to verify a solution. 
This is formalized as follows.

\begin{definition}[Verifyable Learning Problem]\label{def:verifyablelearning}
    Let $q\in\mb N$, 
    $\mc S\subseteq\widetilde{\mc S}$ 
    and let 
    $f\in\widetilde{\mc S}$. 
    We refer to a learning problem 
    $\mc Z:\mc S\to\mc P(\mc T)$ 
    as \emph{deterministic}
    $q$-\emph{verifiable} with respect to $f$ from oracle access $\O$, if there exists a family of deterministic
    algorithms $\set{\A_t\midvert t\in\mc T}$, such that, for any $t\in\mc T$, the algorithm $\A_t$ can solve $\dec(\mc Z_t, f)$ from $q$ many queries to $\O$.

    Similarly, we refer to the learning problem as \emph{random} $q$-\emph{verifiable} with respect to $f$ from oracle access $\O$ with success probability $\alpha$, if there exists a family of random algorithms $\set{\A_t}$, such that for any $t\in\mc T$ the algorithm $\A_t$ solves $\dec(\mc Z_t,f)$ with success probability at least $\alpha$ from at most $q$ queries to $\O$.
\end{definition}

Note, that this notion of verification is different from the notion of verification as introduced by Goldwasser et al. \cite{goldwasserInteractiveProofsVerifying2021}.
\Cref{def:verifyablelearning} is motivated by the necessities 
for the reduction in \Cref{sec:verifiable-lowerbounds}.
However, note, that problems that Goldwasser et al. refer to as trivially verifiable will in general translate to $p$-verifiable problems with respect to any $f$.
According to their notion, a trivially verifiable problem comes along with a simple objective that assures that the actual source object $s$ the verifier has access to is indeed contained in $\mc Z_t$. 
For example, PAC learning in the realizable setting is trivially verifiable, since every concept $g$ satisfying $\Ex_{x}[f(x)\neq g(x)]<\epsilon$ is a valid solution to $f$.
Hence, a correlational query with $g$ to $\csq_\tau[f]$ will suffice to estimate whether $g\in\mc Z_g$.
In particular, realizable PAC learning is $1$-verifiable from correlational statistical queries if $\tau\leq\epsilon$ against any $f$ and for any concept class.

\subsection{Linear Learning Problems}\label{sec:encyclopedia-linear}
\addsectionheader{sec:encyclopedia-linear}

Intuitively, a linear learning problem is a learning problem over a linear function class, where a linear function class is a class of linear functions over a linear space $\mc V$. 
In particular, a linear function class is a subset of the dual space of the linear space $\mc V$.
Here, we focus on linear spaces over $\mb R$ and $\mb C$.

\begin{definition}[Linear Function Class and Linear Learning Problems]\label{def:linearfunctionclass}
    Let $\mc V$ be a  linear space over $\mb K$ with $\mb K=\mb R$ or $\mb C$ equipped with the absolute value as metric, let $\norm{\cdotspace}$ be a norm on $\mc V$ and denote by $\overline{B}_r$ the closed ball with radius $r>0$ around zero in $\mc V$ with respect to $\norm{\cdotspace}$. 
    Denote by $\mc V^\star$ the dual space of $\mc V$
    and by $\dualnorm{\cdotspace}$ the dual norm
    \begin{equation}
        \dualnorm{f} = \sup\set{\abs{f(x)}\midvert \norm{x}\leq1} = \sup_{x\in\mc V}\frac{\abs{f(x)}}{\norm{x}}\,.
    \end{equation}
    We identify a subset $\mc S'\subseteq\mc V^\star$  with the set
    \begin{equation}
        \mc S = \set{f\vert_{\overline{B}_1}\midvert f\in\mc S'}=\mc S'\vert_{\overline{B}_1}\,,
    \end{equation}
    and refer to $\mc S$ as a \emph{Linear Function Class}. 
    We refer to a learning problem $\mc Z:\mc S\to\mc P(\mc T)$ over a linear function class $\mc S$ as a \emph{Linear Learning Problem}.
\end{definition}

The equivalence of $\mc S'$ and $\mc S$ follows immediately by linearity.
The restriction of functions in $\mc S$ to the unit ball has a simple reason.
The evaluation oracle for a function comes with a tolerance parameter $\tau$. 
Without some restriction of this kind one could easily boost the oracles' accuracy by scaling the function argument and exploiting the linearity.

We now list some linear function classes together with the respective evaluation oracles. 
We first consider the standard SQ oracle in terms of linear functions in \Cref{ex:distributionsandSQ} before doing the same for the QSQ oracle in \Cref{ex:statesandQSQ} and for functionals and the CSQ oracle in \Cref{ex:functionalsandCSQ}.
Then, we will consider the less known quantum correlational and quantum unitary statistical oracles in \Cref{ex:qcsq,ex:qusq} before introducing a generalization of the CSQ oracle based on the Gelfand–Naimark–Segal inner product in \Cref{ex:matrixfunctionalsandMCSQ}.

\begin{example}[Probability Distributions and Statistical Queries]\label{ex:distributionsandSQ}
    Let $\Delta(N)\subset\mb R^N$ be the $N$-dimensional sim\-plex, or likewise the set of $N$-dimensional probability distributions. 
    A subset $\D\subseteq\Delta(N)$ is referred to as a \emph{distribution class}.
    A distribution class $\D$ is identified with a linear function class as follows:
    Let $\mc V$ be $\mb R^N$ equipped with the infinity norm $\norm{\cdotspace}_{\ell_\infty}$. 
    The unit ball $\overline{B}_1$ then is the set of vectors with entries bounded by $1$, or equivalently, the set of functions $\phi:[N]\to[-1,1]$.
    Naturally $\mc V^\star=\mc V=\mb R^N$, the dual norm is the $\ell_1$-norm $\dualnorm{\cdotspace}=\norm{\cdotspace}_1$
    and
    $\D\subset\mc V^\star$ can be identified with the linear function class $\D\vert_{\overline{B}_1}$.
    In particular, for $P,Q\in\mc D$ it holds 
    \begin{align}
        \dualnorm{P-Q} = \norm{P-Q}_1 = 2\tv(P,Q)\,.
    \end{align}

    The evaluation oracle $\eval_\tau$ with respect to $\D\vert_{\overline{B}_1}$ is identical to the \emph{statistical query} (SQ) oracle~\cite{kearns_efficient_1998,feldman_complexity_2018} $\stat_\tau$ for $\D$ which, for every $P\in\D$ when queried with some function $\overline{B}_1\ni\phi:[N]\to[-1,1]$ returns a value $v\in \Ex_{i\sim P}[\phi(i)]+[-\tau,\tau]$.
\end{example}

\begin{example}[Quantum States and Quantum Statistical Queries]\label{ex:statesandQSQ}
    Let $\mb H^{N\times N}\subset\mb C^{N\times N}$ be the real vector space of $N\times N$ dimensional Hermitian matrices and denote by $\norm{\cdotspace}_{\tr}$ the trace norm on $\mb C^{N\times N}$ defined as
    \begin{equation}
        \norm{M}_{\tr} = \tr\qty[\sqrt{M^\dagger\cdot M}]\,.
    \end{equation}
    Denote by
    \begin{equation}
         \mc S(N) = \set{\rho\midvert \rho\in\mb H^{N\times N}\;\text{and}\; \norm{\rho}_{\tr}=1}
    \end{equation}
    the set of $N$-dimensional quantum states.
    We refer to a set $\mc Q\subseteq\mc S(N)$ as a \emph{class of quantum states}. 
    Similar to probability distributions we can associate $\mc Q$ to a linear function class as follows:
    Let $\mc V=\mb H^{N\times N}$ be equipped with the Schatten $\infty$-norm, or equivalently the $\ell_2$-induced operator norm $\norm{\cdotspace}_\mrm{op}$. 
    The closed unit ball $\overline{B}_1\subset\mc V$ corresponds to the set of Hermitian $N\times N$-dimensional operators with spectrum bounded by $1$. 
    It holds $\mc V=\mc V^\star=\mb H^{N\times N}$,
    the dual norm is the trace norm $\dualnorm{\cdotspace}=\norm{\cdotspace}_{\tr}$
    and $\mc Q\subset\mc V^\star$ can be identified with the linear function class
    $\mc Q\vert_{\overline{B}_1}$.
    
    The evaluation oracle $\eval_\tau$ with respect to $\mc D\vert_{\overline{B}_1}$ is identical to the \emph{quantum statistical query} (QSQ) oracle~\cite{arunachalamQSQ2020} $\qstat_\tau$ for $\mc Q$ which, for every $\rho\in\mc Q$  returns a value
    \begin{equation}
        v \in \tr\qty[\rho O] + [-\tau,\tau]\,,
    \end{equation} 
    when queried with an operator $O\in \overline{B}_1$.
\end{example}

\begin{example}[Functionals and Correlational Statistical Queries]\label{ex:functionalsandCSQ}
    Let $N\in\mb N$, $\lambda$ be a probability measure on $[N]$, let $\lrangle{\cdotspace,\cdotspace}_\lambda$ be the $L^2(\lambda)$ inner product on $\mb R^N$ given by 
    \begin{align}
        &\lrangle{f,g}_\lambda = \Ex_{i\sim\lambda}[f(i)g(i)]\,,
        &&\text{and let}
        &&&\norm{f}_{L^2(\lambda)} = \sqrt{\lrangle{f,f}_\lambda} = \sqrt{\Ex_{i\sim\lambda}\qty[\abs{f(i)}^2]}
    \end{align}
    be the induced norm on $L^2(\lambda)$.
    Then, let $\overline{B}_r\subset\mb R^N$ be the closed ball with radius $r>0$ with respect to the $L^2(\lambda)$-norm.
    Since the Hilbert space $L^2(\lambda)$ is self-dual it holds 
    \begin{align}
        &\mc V=L^2(\lambda)=\mc V^\star
        &&\text{and}
        &&&\norm{\cdotspace}=\norm{\cdotspace}_{L^2(\lambda)}=\dualnorm{\cdotspace}\,.
    \end{align} 

    We refer to a set $\mc F\subseteq \mb R^N$ as a \emph{functional class}
    and identify $\mc F$ with the linear function class $\mc F\vert_{\overline{B}_1}$.
    The evaluation oracle $\eval_\tau$ with respect to $\mc F\vert_{\overline{B}_1}$ is identical to the \emph{correlational statistical query} (CSQ) oracle~\cite{blum1994weakly,bshoutyUsingExtendedStatistical2001} $\csq_\tau$ with respect to $\mc F$ and $\lambda$ which, for every $f\in\mc F$ returns a value
    \begin{equation}
        v\in \Ex_{i\sim\lambda}\qty[g(i)f(i)] + [-\tau,\tau]\,,
    \end{equation}
    when queried with a function $\overline{B}_1\ni g:[N]\to[-1,1]$.
\end{example}

\begin{example}[Quantum Correlational Statistical Queries]\label{ex:qcsq}
    A special case of \Cref{ex:functionalsandCSQ} is due to \cite{anschuetz2022}. 
    Let $\mc Q\subseteq\mc S(N)$ be a class of quantum states, $\mc V$ be the real linear space of functions consisting of elements
    \begin{align}
        &f_M:\mc Q\to \mb R\doublecolon \rho\mapsto f_M(\rho)=\tr\qty[\rho M]
        &&\text{for}
        &&&M\in\mb H^{N\times N}\,.
    \end{align}
    This is well-defined by the linearity of the trace and since the Hermitian operators are a real linear space.
    In particular, $\mc V$ and $\mb H^{N\times N}$ are isomorphic. 
    For a distribution $\lambda$ over $\mc Q$ we define the $L^2(\lambda)$ inner product on $\mc V$ as
    \begin{equation}
        \lrangle{f, g}_\lambda = \Ex_{\rho\sim\lambda}\qty[f(\rho)g(\rho)]
    \end{equation}
    and let $\norm{\cdotspace}_{L^2(\lambda)}$ be the corresponding norm.
    Again, $L^2(\lambda)$ is a Hilbert space, is self-dual
    and for any $M\in\mb H^{N\times N}$ it holds
    \begin{equation}
        \norm{\tr\qty[\cdotspace M]}_{L^2(\lambda)}\leq \norm{M}_\infty\,.
    \end{equation}
    We identify any subset $\mc F\subset L^2(\lambda)$ with the linear function class $\mc F\vert_{\overline{B}_1}$.
    
    The evaluation oracle is then by \Cref{ex:functionalsandCSQ} equivalent to the correlational statistical query oracle which,
    in the special case of this example, is equivalent to the \emph{quantum correlational statistical query} (QCSQ) oracle $\qcsq_\tau$ as defined in \cite{anschuetz2022}.
    The QCSQ oracle for a function $f_M$ defined by the 
    Hermitian matrix $M$ denoted by $\qcsq_\tau(M)$ returns a value 
    \begin{equation}
        v \in \Ex_{\rho\sim\lambda}\qty[\tr[\rho\cdot O]\tr[\rho\cdot M]] + [-\tau,\tau]\,,
    \end{equation}
    when queried with another Hermitian matrix $O$ with $\norm{O}_\infty\leq1$.
\end{example}

\begin{example}[Quantum Unitary Statistical Query]\label{ex:qusq}
    Another special case of CSQ's due to \cite{anschuetz2022} is the following. 
    For a class of quantum states $\mc Q$ and $\lambda$ a measure over $\mc Q$ we can define an inner product on $\mb C^{N\times N}$ by
    \begin{equation}
        \lrangle{M,K}_{\lambda} = \Ex_{\rho\sim\lambda} \qty[\tr[M^\dagger \cdot K \cdot\rho]]\,,
    \end{equation}
    and denote by $\norm{\cdotspace}_{L^2(\lambda)}$ the induced norm. 
    Again, $L^2(\lambda)$ is a Hilbert space, is self-dual and it holds $\norm{M}_{L^2(\lambda)}\leq\norm{M}_\mrm{op}$.
    We can identify any subset $\mc F\subset \mb C^{N\times N}$ with the corresponding linear function class $\mc F\vert_{\overline{B}_1}$ containing elements 
    \begin{align}
        f_M:\overline{B}_1\to\mb C\doublecolon O\mapsto f_M(O)=\Ex_{\rho\sim\lambda}\qty[\tr[M^\dagger\cdot O\cdot\rho]]\,.
    \end{align}
    The corresponding evaluation oracle for $\mc F\subseteq\U(N)$ when restricted to unitary queries $Q\in\U(N)\subset\overline{B}_1$ corresponds to the \emph{quantum unitary statistical query} oracle $\qusq_\tau(U)[Q]$ from \cite{anschuetz2022} which returns a value 
    \begin{equation}
        v \in \Ex_{\rho\sim\lambda} \qty[\tr[U^\dagger \cdot Q\cdot\rho]] + \overline{B}^{\mb C}_\tau(0)\,,
    \end{equation} 
    when queried with unitary $Q$
    and
    with $\overline{B}^{\mb C}_{\tau}(0)$ is the complex closed $\tau$ ball around 0.
\end{example}

\begin{example}[Matrix Functionals and Matrix Correlational Statistical Queries]\label{ex:matrixfunctionalsandMCSQ}
    This example gives a generalization of \Cref{ex:functionalsandCSQ} to the quantum setting.
    We refer to matrix functionals and matrix correlational statistical queries in order to avoid confusion with \Cref{ex:qcsq}.
    Let $N>0$, $\rho\in\mc S(N)$ a quantum state and define the GNS-inner product on $\mb C^{N\times N}$ with respect to $\rho$ as
    \begin{align}
        &\lrangle{A,B}_\rho = \tr[A^\dagger\cdot\rho\cdot B]
        &&\text{with induced norm}
        &&&\norm{A}_{L^2(\rho)}=\sqrt{\tr[A^\dagger\cdot\rho\cdot A]}\,.
    \end{align}
    Again, $L^2(\rho)$ is a Hilbert space, self-dual, and it holds $\norm{M}_{L^2(\rho)}\leq\norm{M}_\mrm{op}$.

    We refer to a set $\mc F\subseteq \mb C^{N\times N}$ as a \emph{matrix functional class} and identify it with the linear function class $\mc F\vert_{\overline{B}_1}$ containing the elements
    \begin{align}
        f_A:\overline{B}_1\to\mb C\doublecolon B\mapsto f_A(B)=\lrangle{A,B}_\rho\,.
    \end{align}
    We then denote evaluation oracle $\eval_\tau$ with respect to $\mc F\vert_{\overline{B}_1}$ as the \emph{Matrix Correlational Statistical Query} (MCSQ) oracle $\mcsq_\tau$ for $\mc F$ and $\rho$.
    For every matrix $A\in\mc F$ the oracle $\mcsq_\tau(M)$ returns a value $v$ which is promised to be in the interval  
    \begin{equation}
        v\in \lrangle{M,O}_\rho + [-\tau,\tau]\,,
    \end{equation}
    when queried with a matrix $O\in \overline{B}_1$.
\end{example}
\stopcontents[encyclopedia]


\section{Simplified Lower Bounds}\label{sec:simplified}
\addsectionfooter{sec:simplified}
\addsectionheader{sec:simplified}
\startcontents[simplified]
\vspace{1cm}
\printcontents[simplified]{}{1}{}
\vspace{1cm}

    \subsection{Decision Problems}\label{sec:decissionproblembounds}
    \addsectionheader{sec:decissionproblembounds}
    
In this section we give worst-case bounds for the particular learning problem $\dec$ as in \Cref{ex:decissionproblem}.
Throughout this section we assume $\tau>0$ and $\mc S, M, \met, \mc X$ as in \Cref{def:evaloracle}.

\begin{theorem}[Deterministic Query Complexity of $\dec$]\label{thm:deterministichardnessofdeciding}
    Let $\mc S\subseteq M^\mc X$ and $f\in M^\mc X\setminus\mc S$. 
    Let $\A$ be a deterministic algorithm that solves $\dec(\mc S,f)$ from $q$ many $\tau$ accurate evaluation queries. 
    Then for any measure $\mu$ over $\mc S$ it holds  
    \begin{equation}
        q\geq \qty(\max_{x\in\mc X} \Pr_{s\sim\mu}\qty[\met\qty(s(x), f(x))>\tau] )^{-1}\,.
    \end{equation}
\end{theorem}

\begin{proof}
    We run $\A$ and answer every query $x$ by $f(x)$. Let $x_1,\dots,x_q$ be the queries made by $\A$. 
    For the sake of a contradiction assume the existence of some $s\in\mc S$ such that there is no distinguishing query. 
    In particular, assume that for all $i\in[q]$ it holds $\met\qty(s(x_i),f(x_i)\leq\tau)$. 
    Thus, all responses $f(x_i)$ made would correspond to valid evaluation queries with tolerance $\tau$ to $s$, so $\A$ can not distinguish $s$ from $f$ contradicting the correctness of $\A$.
    Thus, for any $s\in\mc S$ there must exist at least one distinguishing query.
    In particular, it must hold
    \begin{align}\label{eq:mockunionbound}
        1=\Pr_{s\sim\mu} \qty[\exists i\in[q]:\; \met\qty(s(x_i),f(x(i)))>\tau]
        \leq\sum_{i=1}^q \Pr_{s\sim\mu} \qty[ \met\qty(s(x_i),f(x_i))>\tau]\\[8pt]
        \leq q\max_{x\in\mc X} \Pr_{s\sim\mu} \qty[ \met\qty(s(x),f(x))>\tau]\,,
    \end{align}
    which completes the proof.
\end{proof}

\begin{theorem}[Random Query Complexity of $\dec$]\label{thm:randomhardnessofdeciding}
    Let $\alpha\in(0,1]$, $\mc S\subseteq M^\mc X$ and $f\in M^\mc X\setminus\mc S$. 
    Let $\A$ be a random algorithm that solves $\dec(\mc S,f)$ from $q$ many $\tau$ accurate evaluation queries with success probability $\alpha$ over $\A$'s internal randomness. 
    Then for any measure $\mu$ over $\mc S$ it holds  
    \begin{equation}
        q\geq \frac{2\cdot\qty(\alpha - \sfrac12)}{\max_{x\in\mc X} \Pr_{s\sim\mu}\qty[\met\qty(s(x), f(x))>\tau]}\,.
    \end{equation}
\end{theorem}

\begin{proof}
    We run $\A$ and answer any query $x$ by $f(x)$. 
    Denote by $x_1,\dots,x_q$ the queries made, which are now random variables with respect to $\A$'s randomness.  
    Assume the output is $X\in\set{\mc S,\set{f}}$.
    Let $s\in\mc S$ and denote by 
    \begin{equation}
        p(s) = \Pr_\A\qty[\exists i\in[q]:\; \met\qty(s(x_i),f(x_i))>\tau]
    \end{equation}
    the probability of the existence of a distinguishing query between $s$ and $f$.
    We can lower bound $p(s)$ following \cite{feldman_general_2017}. 
    By the correctness of $\A$ the output $X$ will be $\mc S$ with probability at most $1-\alpha$. 
    Now assume $p(s)<2(\alpha-\sfrac12)$.
    Thus, for some valid responses of $\eval_\tau(s)$ the algorithm $\A$ will output $\set{f}$ with probability at least $1-p(s)-(1-\alpha)>1-\alpha$, 
    contradicting the correctness assumption on $\A$.
    Hence, $p(s)\geq 2(\alpha-\sfrac12)$.

    The remainder of the proof follows as in the proof of \Cref{thm:deterministichardnessofdeciding} with the left-hand side of \Cref{eq:mockunionbound} replaced by $2(\alpha-\sfrac12)\leq p(s)$
    \begin{align}\label{eq:random-mockunionbound}
        2\cdot\qty(\alpha-\sfrac12)=\Pr_{\substack{s\sim\mu\\ \mc A}} \qty[\exists i\in[q]:\; \met\qty(s(x_i),f(x(i)))>\tau]
        \leq\sum_{i=1}^q \Pr_{\substack{s\sim\mu\\ \mc A}} \qty[ \met\qty(s(x_i),f(x_i))>\tau]\\[8pt]
        \leq q\max_{x\in\mc X} \Pr_{s\sim\mu} \qty[ \met\qty(s(x),f(x))>\tau]\,.
    \end{align}
\end{proof}

It is worth mentioning that, similar to the bounds in \cite{feldman_general_2017}, the previous bound is tight in the following sense.

\begin{theorem}[Random Query Complexity Upper Bound]
    Let $\alpha=1-\delta\in(0,1)$, $\mc X$ and $\mc S\subseteq M^\mc X$ compact sets and let $f\in M^\mc X\setminus\mc S$.
    Then, there exists a random algorithm that decides $\dec(\mc S, f)$ from $q$ many queries to $\eval_\tau$ and with success probability $\alpha$ over its internal randomness, with
    \begin{align}
        q \leq q_{\max} \coloneq \frac{\ln\qty(\frac{1}{\delta})}{\min_{\mu}\max_{x\in\mc X}\Pr_{s\sim\mu}\qty[\met(s(x), f(x))>\tau]}\,.
    \end{align}
\end{theorem}

\begin{proof}
    Since $\mc X$ and $\mc S$ are compact, the space of probability distributions $\nu$ over $\mc X$, respectively $\mu$ over $\mc S$ is convex and compact with respect to the weak-* topology \cite[Propositions 7.2.3 and 7.2.4]{walkdenErgodicTheory2018}. 
    Thus, for any decision problem $\dec(\mc S, f)$ we find 
    \begin{align}
        \min_\mu \max_{x\in\mc X} \Pr_{s\sim\mu}\qty[\met\qty(f(x), s(x))>\tau]
        = \min_\mu \max_{\nu} \Pr_{s\sim\mu, x\sim\nu}\qty[\met\qty(f(x), s(x))>\tau]&\\
        = \max_{\nu} \min_\mu \Pr_{s\sim\mu, x\sim\nu}\qty[\met\qty(f(x), s(x))>\tau]
        &\,,\label{eq:minimax-decision}
    \end{align}
    where in the third equality we have used von Neumann's minimax theorem.
    
    Let $\nu$ be the measure over $\mc X$ that maximizes \Cref{eq:minimax-decision}.
    The probability that none of $q_{\max}$ samples $x_1,\dots,x_{q_{\max}}\sim\nu$ solves the decision problem can then be bounded by \Cref{eq:minimax-decision} and the definition of $q_{\max}$ as
    \begin{align}
        \Pr_{x_1,\dots,x_{q_{\max}}\sim\nu}\qty[\abs{s(x_i)-f(x_i)}\leq\tau\midvert \forall i\in[q_{\max}]] \leq \qty(1 - \frac{\ln(\sfrac{1}{\delta})}{q_{\max}})^{q_{\max}}
        \leq \delta\,,
    \end{align}
    for any $s\in\mc S$.
    Thus, $q_{\max}$ non-adaptive queries selected according to $\nu$ suffice to solve $\dec(\mc S, f)$, which completes the proof.
\end{proof}

    \subsection{Verifiable Learning Problems}\label{sec:verifiable-lowerbounds}
    \addsectionheader{sec:verifiable-lowerbounds}

The previous section yields lower bounds on the query complexity of the decision problem $\dec(\mc S, s^*)$ from \Cref{sec:encyclopedia-decission}. 
While those bounds are important in themselves, they also deal as a prominent tool for obtaining simpler lower bounds for more general learning problems, 
namely verifiable learning problems as introduced in \Cref{sec:encyclopedia-verifialbe}.
This goes via the well known learning-to-deciding reduction which is formalized as follows.

\begin{lemma}[Learning is as Hard as Deciding]\label{lem:learning-is-as-hard-as-deciding}
    Let $p_v\geq0$, $\mc S\subseteq\widetilde{\mc S}$ and let $f\in\widetilde{\mc S}$. Further, let $\mc Z:\mc S\to\mc P(\mc T)$ 
    be a deterministic $p_v$-verifiable learning problem with respect to $f$ from oracle access $\O$ and denote 
    by $p_d\geq0$ the deterministic query complexity of $\dec(\mc S, f)$ with respect to $\O$. 
    Let $\A$ be a deterministic algorithm for learning $\mc Z$ from oracle access $\O$ using $q$ many queries. Then it must hold 
    \begin{equation}\label{eq:learningisashardasdeciding}
        q\geq p_d-p_v\,.
    \end{equation}
    Similarly, let $\alpha_v,\alpha_l\in(0,1]$ and define $\alpha_d=\alpha_l\cdot\alpha_v$. Let $\mc Z$ be random $p_v$-verifiable with respect to $f$ from oracle access $\O$ with success probability $\alpha_v$ and let $p_d\geq0$ be the random query complexity of $\dec(\mc S, f)$ with success probability $\alpha_d$ and with respect to $\O$. 
    Then again, for any random algorithm $\A$ for learning $\mc Z$ from oracle access $\O$ with probability at least $\alpha_l$ using $q$ many queries it must hold \Cref{eq:learningisashardasdeciding}.
\end{lemma}

\begin{proof}
    We begin with the deterministic case. 
    Let $\A$ be a deterministic algorithm for learning $\mc Z$ and, for any $t\in\mc T$ , let $\mc B_t$ be the of deterministic algorithm that decides $\dec(\mc Z_t, f)$ from at most $p_v$ queries to $\O$. 
    Assume we are given oracle access $\O(s)$ to an unknown $s\in\mc S\cup\set{f}$. 
    We run $\A$ on $\O(s)$ and without loss of generality obtain a solution $t\in\mc T$ (otherwise, by the correctnes of $\A$ we already know $s=f$). 
    Finally, we run $\mc B_t$ on $s$ and receive either ``$\mc Z_t$'' or ``$\set{f}$''. In the latter case return ``$\set{f}$'', in the former return ``$\mc S$''. 
    By the correctness of $\A$ and $\mc B_t$ we have thus constructed an algorithm for solving $\dec(\mc S, f)$ that makes at most $q+p_v$ queries.
    Thus, by assumption, $q+p_v\geq p_d$ which implies the claim. 

    The proof of the random case uses the identical construction. 
    The only difference is in the analysis. 
    The constructed algorithm for deciding between $\mc S$ and $f$ is certainly correct when both, $\A$ and $\mc B_t$ are correct. 
    Thus, we find it is correct with probability at least $\alpha_d\geq \alpha_l\cdot\alpha_v$.
\end{proof}

To connect to the discussion after \Cref{def:verifyablelearning}, we note that it is unclear whether a claim similar to \Cref{lem:learning-is-as-hard-as-deciding} can be made for verifiable problems in the sense of Goldwasser et al. \cite{goldwasserInteractiveProofsVerifying2021}, too.
In particular, the notion of verifiable learning by Goldwasser et al. leaves the case open that both, the verifier and the prover access the reference object $f$ which is not contained in the class to be learned, the verifier in their setting always is given access to some $s\in\mc S$.

We now combine \Cref{thm:deterministichardnessofdeciding,lem:learning-is-as-hard-as-deciding} to obtain the following bound.

\begin{theorem}[Deterministic Average-Case Lower Bound for Verifyable Learning Problems]\label{thm:deterministic-lower-verifyable}
    Let $p_v>0, \mc S\subseteq M^{\mc X}$ and let $f:\mc X\to M$ with $f\not\in\mc S$. 
    Let $\mc Z:\mc S\to\mc P(\mc T)$ be random $p_v$-verifiable with respect to $f$ from $\tau$ accurate evaluation queries.
    Further, let $\mu$ be a measure over $\mc S$.
    Then, any algorithm $\A$ that learns $\mc Z$ with probability at least $\beta$ over $s\sim\mu$ requires at least $q$-many $\tau$ accurate evaluation queries, with
    \begin{equation}\label{eq:deterministic-lower-verifyable}
        q+p_v \geq
        \frac{\beta}{\max_{x\in\mc X}\Pr_{s\sim\mu}\qty[\met\qty(s(x), f(x))>\tau]}\,.
    \end{equation}
\end{theorem}

\begin{proof}
    Let $\mc S^+\subseteq\mc S$ be a subset on which $\A$ is successful and for which it holds $\mu(\mc S^+)=\beta$.
    Let $\mu_+$ be the measure $\mu$ conditioned on $\mc S^+$, such that for any $s\in\mc S$ it holds $\mu_+(s)=\mu(s\mid s\in\mc S^+)$.
    Then, due to the definition of conditional probability we find, for any $x\in\mc X$
    \begin{equation}
        \Pr_{s\sim\mu_+}\qty[\met(s(x),f(x))>\tau]
        =
        \Pr_{s\sim\mu}\qty[\met(s(x),f(x))>\tau\midvert s\in\mc S^+] 
        \leq
        \beta^{-1}\cdot\Pr_{s\sim\mu}\qty[\met(s(x),f(x))>\tau]\,.
    \end{equation}
    Next we observe that a $\beta$-average-case learner $\A$ of $\mc Z$ implies a worst-case learner for $\mc Z^+=\mc Z\vert_{\mc S^+}$.
    The complexity of learning $\mc Z^+$ in turn can be bounded by \Cref{lem:learning-is-as-hard-as-deciding}. The bound as in \Cref{eq:deterministic-lower-verifyable} then follows from applying \Cref{thm:deterministichardnessofdeciding} to $\dec(\mc S^+,f)$ with respect to the measure $\mu_+$.
\end{proof}

Similarly we obtain a corresponding bound for the random average-case query complexity as follows.

\begin{theorem}[Random Average-Case Lower Bound for Verifyable Learning Problems]\label{thm:random-lower-verifyable}
    Let $p_v>0, \alpha_v\in(0,1], \mc S\subseteq M^{\mc X}$ and let $f:\mc X\to M$ with $f\not\in\mc S$. 
    Let $\mc Z:\mc S\to\mc P(\mc T)$, with probability $\alpha_v$, be random $p_v$-verifiable with respect to $f$ from $\tau$ accurate evaluation queries.
    Further, let $\mu$ be a measure over $\mc S$.
    Then, any algorithm $\A$ that learns $\mc Z$ with probability at least $\beta$ over $s\sim\mu$ and success probability $\alpha$ over the internal randomness of $\A$ requires at least $q$-many $\tau$ accurate evaluation queries, with
    \begin{equation}\label{eq:random-lower-verifyable}
        q+p_v \geq 
        \frac{2\cdot\qty(\alpha\cdot\alpha_v-\sfrac12)\cdot\beta}{\max_{x\in\mc X}\Pr_{s\sim\mu}\qty[\met\qty(s(x), f(x))>\tau]}\,.
    \end{equation}
\end{theorem}

\begin{proof}
    The proof is identical to that of \Cref{thm:deterministic-lower-verifyable} applying the second instead of the first part of \Cref{lem:learning-is-as-hard-as-deciding} and invoking \Cref{thm:randomhardnessofdeciding} instead of \Cref{thm:deterministichardnessofdeciding}.
\end{proof}

\Cref{def:verifyablelearning} and \Cref{thm:deterministic-lower-verifyable,thm:random-lower-verifyable}
place some constraints on the choice of $f$ in order to find non-trivial lower bounds.
On the one hand, $s\sim\mu$ must concentrate around $f$. 
On the other hand for any $t\in\mc T$ there must exist an algorithm to distinguish $\mc Z_t$ from $f$ with few queries.

Thus, the given bounds can not be directly used if $s\sim\mu$ concentrates within $\mc S$, or when $f$, around which $s\sim\mu$ concentrates, can not be distinguished query efficiently from a subset $\mc S^f\subset\mc S$. 
In these cases it is practical to apply \Cref{thm:deterministic-lower-verifyable,thm:random-lower-verifyable}
to a restricted problem $\mc Z'=\mc Z\vert_{\mc S'}$.
Lower bounds on learning $\mc Z'$ then imply the corresponding lower bounds on learning $\mc Z$.
Here, $\mc S'\subset \mc S$ is any suitable subset that ensures verifiability.
In particular:
\begin{itemize}
    \item For any $t\in\bigcup_{s\in\mc S'}\mc Z(s)$ it must hold that $\dec(\mc Z_t,f)$ can be solved using at most $p_v$ queries.
\end{itemize}
We write $\mc S'=\mc S\setminus\mc S^{f}$ with $\mc S^f$ the subset of sources, which cannot be distinguished from $f$ using $p_v$ queries.
Replacing the set $\mc S^+$ in the proofs of \Cref{thm:deterministic-lower-verifyable,thm:random-lower-verifyable}
by $\mc S^+\setminus\mc S^{f}$ and letting $\mu_+$ in the proof denote the measure $\mu$ restricted on the corresponding subset then yields the following corollary.

\begin{corollary}\label{rem:verifiable-bound}
    Let $\mc Z:\mc S\to\mc P(\mc T)$ be a learning problem such that $\mc Z'=\mc Z\vert_{\mc S'}$ for some $\mc S'\subset\mc S$ is, with probability $\alpha_v$ random $p_v$-verifiable with respect to $f$ and $\eval_\tau$.
    Let $\mu$ be a measure over $\mc S$. 
    Then, for any algorithm $\A$ that learns $\mc Z$ with probability at least $\beta$ over $s\sim\mu$ and $\alpha$ over $\A$'s internal randomness requires at least $q$ many queries to $\eval_\tau$, with 
    \begin{align}
        &q+p_v 
        \geq
        2\cdot\qty(\alpha-\sfrac12)\cdot\frac{\beta-\Pr_{s\sim\mu}\qty[s\in\mc S^f]}{\max_{x\in\mc X}\Pr_{s\sim\mu}\qty[\met(s(x), f(x))>\tau]}\,.
    \end{align}
    The deterministic bound corresponds to setting $\alpha_v=\alpha=1$.
\end{corollary}

\stopcontents[simplified]


\section{Special Case: Linear Function Classes}\label{sec:specialcase-distributions}
\addsectionfooter{sec:specialcase-distributions}
\addsectionheader{sec:specialcase-distributions}
\startcontents[linear]
\vspace{1cm}
\printcontents[linear]{}{1}{}
\vspace{1cm}

    \subsection{Characterization of Linear Learning}\label{sec:linearlearningupper}
    \addsectionheader{sec:linearlearningupper}
    
In this section we will characterize the query complexity of linear learning problems in the evaluation-oracle model.
In contrast to our original definition, we will identify linear function classes with sets in $\mc V$ rather than the dual space.
Our bounds will require that both $\mc S\subset\mc V$ and the closed dual unit ball $\overline{B}_1^\star(0)$ are compact.
Since this can be simplified to requiring $\mc V$ to be reflexive with respect to $\norm{\cdotspace}$, we are free to swap the roles of $\mc V$ and $\mc V^\star$ as long as we are consistent.

Before state upper bounds on the query complexity for learning linear function families, let us first introduce a corresponding dimension for learning with evaluation queries.
We will then prove, in analogy to Feldman's work in \cite{feldman_general_2017} that this dimension characterizes the query complexity of linear learning problems.

\begin{definition}[Random Dimension]\label{def:dimension}
    Let $\mc S\subseteq M^\mc X$ and $r\in M^\mc X$.
    We define the \emph{Random Dimension} of a decision problem $\dec(\mc S, r)$ with tolerance $\tau$ as
    \begin{equation}
        \rd(\mc S, r, \tau) =  \sup_\mu\qty(\max_{x\in\mc X} \Pr_{s\sim\mu}\qty[\met(r(x),s(x))>\tau])^{-1}\,,
    \end{equation}
    where the $\sup_\mu$ is over all probability measures over $\mc S$.

    For a learning problem $\mc Z:\mc S\to\mc P(\mc T)$ with $\mc S\subseteq M^\mc X$ we define the \emph{Dimension of Learning} and the \emph{Random Dimension of Learning} respectively as follows
    \begin{align}
        \dl(\mc Z, \tau) &= \sup_{f\in M^\mc X} \inf_{t\in\mc T} \rd\qty(\mc S\setminus \mc Z_t, f, \tau)\\
        \rdl(\mc Z,\tau, \gamma) &= \sup_{f\in M^\mc X} \inf_\vartheta \rd\qty(\mc S\setminus \mc Z_\vartheta(\gamma), f, \tau)\,,
    \end{align}
    for any $\gamma>0$, where $\mc Z_\vartheta(\gamma)$ is as defined in \Cref{def:alphaprobablesolutions} and where the infimum is over all probability measures $\vartheta$ over $\mc S$.
\end{definition}

As an immediate consequence we can adapt 
\Cref{thm:deterministicavglowerbound,thm:randomavglowerbound} 
to obtain worst-case lower bound for learning problems in terms of the (random) dimension of learning.

\begin{corollary}[Deterministic Worst Case Lower Bound of Learning]
    Let $\mc Z:\mc S\to\mc P(\mc T)$ be a learning problem with $\mc S\subseteq M^\mc X$ and let $\alpha,\tau>0$.
    Let $\A$ be any deterministic algorithm that learns $\mc Z$ from $q$-many $\tau$-accurate evaluation queries.
    Then it must hold
    \begin{equation}
        q\geq \dl(\mc Z, \tau)\,.
    \end{equation}
\end{corollary}

\begin{proof}
    The proof is identical to that of \Cref{thm:deterministicavglowerbound} with $\mc S'=\mc S$ and where the 
    left-hand side of \Cref{eq:mockdeterministicavg} is replaced by $\Pr_{s\sim\mu}[s\not\in \mc Z_t]$.
\end{proof}

\begin{corollary}[Random Worst Case Lower Bound of Learning]
    Let $\mc Z:\mc S\to\mc P(\mc T)$ be a learning problem with $\mc S\subseteq M^\mc X$ and let $\alpha,\tau>0$.
    Let $\A$ be any random algorithm that learns $\mc Z$ with success probability at least $\alpha$ over its internal randomness from $q$-many $\tau$-accurate evaluation queries.
    Then, for any $\alpha>\gamma>0$ it must hold
    \begin{equation}
        q\geq (\alpha-\gamma) \cdot\rdl(\mc Z, \tau, \gamma)\,.
    \end{equation}
\end{corollary}

\begin{proof}
    The proof is identical to that of \Cref{thm:randomavglowerbound} with $\mc S'=\mc S$ and where the conditional probability in  \Cref{eq:mockproofrandomaverage} is kept throughout.
\end{proof}

The above lower bounds hold for any learning problem in the evaluation-oracle model.
Proofs for upper bounds, however, require additional structure.
To this end, in this work and regarding upper bounds we focus on linear learning problems.

Interestingly, one can use the tools from convex optimization, namely the mirror descent algorithm,
in order to prove query complexity upper bounds for linear learning problems.
Throughout this section we denote by $\mc V$ a linear space over $\mb R$
and denote by $\lrangle{\cdotspace, \cdotspace}$ the inner product in $\mc V$ with the convention of anti-linearity in the first argument.
Let $\norm{\cdotspace}$ be a norm on $\mc V$, $\dualnorm{\cdotspace}$ its dual norm and
let $\mc W\subseteq\mc V$ be some open convex set. 
We refer to $\mc W$ as the primal space.
We refer to \Cref{app:convex} for the technical definitions that we skip for now.

\begin{definition}[Mirror Descent]\label{def:mirrordescent}
    The \emph{Mirror Descent} (MD) algorithm with $\zeta$-regular mirror map $R$ over primal space $\mc W$, constraint set $\mc K$, step size $\eta>0$, initial guess $f_1\in\mc K\cap\mc W$ and linear advice is defined as follows.
    In each update step $t$, the algorithm receives a linear function $\lrangle{g_t,\cdotspace}$ and updates according to
    \begin{equation}
        f_{t+1} = \Pi^R_{\mc K\cap\mc W}\qty((\nabla R)^{-1}(\nabla R(f_t)-\eta g_t))\,.
    \end{equation}
\end{definition}

\begin{theorem}[Analysis of Online Mirror Descent]\label{thm:mirrordescentanalysis}
   The mirror descent algorithm from \Cref{def:mirrordescent}, when given access to a list of linear functions $g_1,\dots,g_T$, will produce a list of vectors $f_1,\dots,f_T$ such that for any $f\in\mc K$ it holds
   \begin{equation}
        \sum_{t=1}^T\qty(\lrangle{g_t,f_t}-\lrangle{g_t,f}) \leq \frac{D_R(f,f_1)}{\eta} + \frac{\eta}{2\zeta} \sum_{t=1}^T\dualnorm{g_t}\,.
   \end{equation} 
   In particular, assuming $\dualnorm{g_t}=1$ for all $t$, the average regret is upper bounded by
   \begin{equation}
        \frac1T \sum_{t=1}^T\qty(\lrangle{g_t,f_t}-\lrangle{g_t,f}) \leq \frac{\eta}{\zeta} \,,
   \end{equation}
   for all $T\geq\frac{2\zeta D_R(f,f_1)}{\eta^2}$.
\end{theorem}

The proof can be found in e.g. \cite{harvey2018}.
We are now able to prove the upper bounds.

\begin{theorem}[Deterministic Upper Bound for Linear Learning]\label{thm:deterministicupper}
    Let $\mc Z:\mc S\to\mc P(\mc T)$ be a linear learning problem with $\mc S\subseteq\mc V$.
    Let $\mc W\subseteq\mc V$ be a primal space, $\mc K$ a constraint set such that $\mc S\subseteq\mc K$ and let $R$ be an $\zeta$-regularizer on $\mc W$ with Bregman divergence $D_R$. 
    Define $r=\min_{v\in\mc V}\max_{s\in\mc S}D_R(s,v)$.
    For every $\tau>0$ there exists a deterministic algorithm $\A$ that learns $\mc Z$ from $q$-many $\sfrac\tau3$-accurate evaluation queries, where
    \begin{equation}
        q\leq O\qty(\frac{r\cdot \dl(\mc Z,\tau)}{\zeta\cdot\tau^2}\cdot\ln(\abs{\mc S}))\,.
    \end{equation} 
\end{theorem}

\begin{proof}
    Assume we run the algorithm $\A$ with access to some unknown $\tilde{f}\in\mc S$.
    Let $f_1\in\mc S$ be a function that minimizes $r$. 
    At step $i$ denote by $f_i$ the current hypothesis of the algorithm. 
    Denote by $d=\rd(\mc S\setminus\mc Z_t, f_i,\tau)$.
    By \Cref{def:dimension} there exists a $t\in\mc T$ such that $d\leq\dl(\mc Z,\tau)$.
    Similar as in Lemma C.1 from \cite{feldman_general_2017} for some
    $p\leq d\ln(\abs{\mc S\setminus\mc Z_t})$ there exist functions $g_1,\dots, g_p$ with $\dualnorm{g_j}\leq1$ such that for every $s\in\mc S$ there exists a $j\in[p]$ with $\abs{\tilde f(g_j)- f_i(g_j)}>\tau$. 
    In particular, if no such $j$ exists, then $f_i\in\mc Z_t$ must hold.

    The algorithm now proceeds as follows. At each step $i$ let $t\in\mc T$ be the target maximizing $\rd(\mc S\setminus\mc Z_t, f_i, \tau)$, let $g_1,\dots,g_p$ be a cover as above 
    and denote by $v_1,\dots,v_p$ the responses of $\eval_{\frac\tau3}(\tilde f)$ to the queries $g_j$. 
    If there exists a $j$ such that $\abs{\tilde f(g_j)- f_i(g_j)}>\frac{2\tau}{3}$ define $h_i$ as $h_i=g_j$ if $f_i(g_j)>v_j$ and $h_i=-g_j$ if $f_i(g_j)\leq v_j$.
    Then, make a mirror descent update as in \Cref{def:mirrordescent} with the linear function $h_i$ and step size $\eta=\frac{\tau\zeta}{3}$.
    In contrast, if no such $j$ exists, return $t$.

    To establish the correctness of $\A$ note that the algorithm returns some $t\in\mc T$ only when it cannot find a distinguishing query $g_j$ at a given step. 
    However, by the discussion in the first paragraph,
    this only happens if $\tilde f\in\mc Z_t$, which proves the correctness.

    Let us now analyze the complexity of $\A$.
    In each step the algorithm makes $p\leq \dl(\mc Z,\tau)\ln(\abs{\mc S})$ queries.
    In case of an update we know that $\abs{\tilde{f}(g_j)-f_i(g_j)}>\frac{2\tau}{3}-\frac\tau3=\frac\tau3$.
    Thus, by the definition of $h_i$ it holds $\tilde f(h_i)-f_i(h_i)=\lrangle{h_i,\tilde f}-\lrangle{h_i,f_i}>\frac\tau3$.
    Finally, by \Cref{thm:mirrordescentanalysis} we know that there can be at most 
    $T=\frac{18 D_R(\tilde f, f_1)}{\zeta\tau^2}\leq\frac{18 r}{\zeta\tau^2}$ such steps completing the proof.
\end{proof}

In analogy, we can prove the same statement for random algorithms.

\begin{theorem}[Random Upper Bound for Linear Learning]\label{thm:randomupper}
    Let $\mc Z:\mc S\to\mc P(\mc T)$ be a linear learning problem with $\mc S\subseteq\mc V$ compact and compact closed unit ball $\overline{B}_1^\star(0)\subset\mc V^\star$.
    Let $\mc W\subseteq\mc V$ be a primal space, $\mc K$ a constraint set such that $\mc S\subseteq\mc K$ and let $R$ be an $\zeta$-regularizer on $\mc W$ with Bregman divergence $D_R$. 
    Define $r=\min_{v\in\mc V}\max_{s\in\mc S}D_R(s,v)$.
    For every $\tau>0$ and $0<\delta<\alpha\leq1$ there exists a random algorithm $\A$ that, with probability $\alpha-\delta$ learns $\mc Z$ from $q$-many $\frac\tau3$-accurate evaluation queries, where
    \begin{equation}\label{eq:randomupper}
        q= O\qty(\frac{r\cdot \rdl(\mc Z,\tau,\alpha)}{\zeta\cdot\tau^2}\cdot\log(\frac{r}{\tau\cdot\delta}))\,.
    \end{equation} 
\end{theorem}

\begin{proof}
    Set $T=\frac{18 r}{\zeta\tau^2}$, $\delta'=\frac\delta T$ and $p=d\ln\qty(\frac{1}{\delta'})$. 

    The requirements $\mc S$ and $\overline{B}_1^\star(0)$ compact ensure that the spaces of probability distributions $\mu$ over $\mc S$ and $\nu$ over $\overline{B}_1(0)^\star$ are convex and compact with respect to the weak-* topology \cite[Propositions 7.2.3 and 7.2.4]{walkdenErgodicTheory2018}. 
    Thus, for any decision problem $\dec(\mc S, r)$ we find for the random dimension
    \begin{equation}\label{eq:rd=rcvr}
        \begin{split}
            \rd(\mc S, r, \tau) &= \qty(\min_\mu \max_{x\in\mc X} \Pr_{s\sim\mu}\qty[\met\qty(r(x), s(x))>\tau])^{-1}\\
            &= \qty(\min_\mu \max_{\nu} \Pr_{s\sim\mu, x\sim\nu}\qty[\met\qty(r(x), s(x))>\tau])^{-1}\\
            &= \qty( \max_{\nu} \min_\mu \Pr_{s\sim\mu, x\sim\nu}\qty[\met\qty(r(x), s(x))>\tau])^{-1}\\
            &= \min_\nu\qty( \min_\mu \Pr_{s\sim\mu, x\sim\nu}\qty[\met\qty(r(x), s(x))>\tau])^{-1}\,,
        \end{split}
    \end{equation}
    where in the third equality we have used von Neumann's Minimax theorem.
    Now, set $d=\rdl(\mc Z)$. 

    The algorithm is now almost exactly as that from the proof of \Cref{thm:deterministicupper}. 
    The adaptations are as follows.
    Recall, that at any given step $i$ the algorithm has a current hypothesis $f_i$.
    At each step, instead of working with a deterministic target $t$ that maximizes $\rd(\mc S\setminus\mc Z_t,f_i,\tau)$ we work with a ``random target'', i.e. any measure $\vartheta$ over $\mc T$ such that $\rd(\mc S\setminus\mc Z_\vartheta(\alpha), f_i, \tau)\leq d$. 
    Instead of the deterministic cover we also use a ``random cover'', this is, any measure $\nu$ over $\overline{B}_1^\star(0)$ that satisfies, for all $f\in\mc S\setminus \mc Z_\vartheta(\alpha)$
    \begin{align}
        \Pr_{g\sim\nu}\qty[\abs{f(g)-f_i(g)}\geq\tau]\geq\frac{1}{d}\,.
    \end{align}
    The existence of which is guaranteed by \Cref{eq:rd=rcvr}. 

    The remainder is now as in the proof of \Cref{thm:deterministicupper}. 
    Sample $p$ many functions $g_1,\dots,g_p$ and obtain the values $v_j=\eval_{\frac{\tau}{3}}[g_j]$. Then, find a distinguishing query $g_j$ and depending on the sign of $f_i(g(j))-v_j$ define $h_i$ and apply a mirror descent update with $h_i$ and step size $\frac{\tau\zeta}{3}$.
    Else, if no distinguishing query exists return a random target $t\sim\vartheta$.

    We start by bounding the complexity of the algorithm.
    Again, by \Cref{thm:mirrordescentanalysis} we find that at most $T=\frac{18r}{\zeta\tau^2}$ updates will be performed. 
    Thus, the query complexity is obtained from the query complexity of each individual round times $T$ giving rise to \Cref{eq:randomupper}.

    Let us now analyze the correctness of the algorithm. 
    During the last round of the algorithm, either $\tilde f\in\mc Z_\vartheta(\alpha)$, or the random functions $g_j\sim\nu$ did not admit a cover, such that the algorithm wrongly did not update $f_i$.
    In the former case the algorithm will sample some valid solution $\mc Z(\tilde f)\ni t\sim\vartheta$ with probability at least $\alpha$.
    Let us now bound the probability of the latter case. 
    In each round $i$ the algorithm will not find a distinguishing query with probability 
    \begin{align}
        \Pr_{(g_1,\dots,g_p)\sim\nu}\qty[\forall j\in[p]\,:\;\abs{\tilde f(g_j)-f_i(g_j)}<\tau ]\leq \qty(1-\frac1d)^{-p} 
        \leq e^{-\frac{p}{d}} = \delta'\,.
    \end{align}
    Thus, the probability that the algorithm does not find a distinguishing query in one of the at most $T$ consecutive rounds is lower bounded by 
    \begin{align}
        \Pr\qty[\exists i\in[T]:\; \forall j\in[p]:\;  \abs{\tilde f(g_j) - f_i(g_j)}<\tau]\leq T\delta'=\delta\,,
    \end{align}
    where we have used the union bound.
    Thus, the probability of success is at least $\alpha-\delta$ proving the claim.
\end{proof}

    \subsection{Further Implications}\label{sec:linear-further-implications}
    \addsectionheader{sec:linear-further-implications}
    
This section contains further results about linear learning problems. 
Since linear learning problems can be seen as the prototype for many learning problems, we believe they may be of independent interest.

In contrast to the previous section, we return to our original identification of linear function classes with sets in the dual space.
Throughout, let $\mc V$ be a vector space over the field $\mb K=\mb R$ or $\mb C$ with norm $\norm{\cdotspace}$, dual norm $\norm{\cdotspace}_\star$ and denote by $\mc V^\star$ the dual space.
By $\met_\star(x,y)=\frac12\norm{x-y}_\star$ for any $x,y\in\mc V^\star$ we denote the distance induced by the dual norm.
As in \Cref{def:linearfunctionclass} we identify any $x\in\mc V^\star$ with a linear function 
\begin{align}
    x:\overline{B}_1(0)\to\mb K\doublecolon v\mapsto x(v)\,,
\end{align}
where $\overline{B}_1(0)$ denotes the closed ball around $0\in\mc V$ with radius $1$ with respect to $\norm{\cdotspace}$.
Similarly, we denote by $B_\epsilon^\star(x)$ the open $\epsilon$-ball around $x\in\mc V^\star$ with respect to $\norm{\cdotspace}_\star$.
Moreover, we denote by $B_{\epsilon,\met}^\star(x)$ the open $\epsilon$-ball around $x$ with respect to $\met_\star$.
Thus, $B_\epsilon^\star(x)=B_{2\epsilon,\met}^\star(x)$. 
For the sake of ease we assume $\mc V$ (and thus $\mc V^\star$) to be finite dimensional. 
The reason for this assumption is that it implies that $\overline{B}_1(0)$ is compact and hence we can replace any $\sup_{w\in\overline{B}_1(0)}$ by a $\max_{w\in\overline{B}_1(0)}$ and thus there exists a (non-unique) maximizer $v\in\argmax_{w\in\overline{B}_1(0)}$.
We note that our results carry over to the infinite dimensional case at the price of a more technical presentation.

\subsubsection{Learning Linear Functions as Verifiable Problem}\label{sec:learning-linear-as-verifiable}
\addsectionheader{sec:learning-linear-as-verifiable}

A prototypical learning problem that often arises is the following.

\begin{definition}[$\epsilon$-Learning]\label{def:epsilon-learning}
    Let $0<\tau\leq\epsilon$ and $\mc S\subseteq\mc V^\star$. 
    We denote by 
    \begin{align}
        \learn\qty(\mc S, \epsilon):\mc S\to\mc P(\mc V^\star)\doublecolon x\mapsto 
        B_{\epsilon,\met}^\star(x)=\set{y\in\mc V^\star\midvert \met_\star(x,y)<\epsilon}
    \end{align}
    the problem of \emph{$\epsilon$-Learning $\mc S$} with respect to $\met_\star$.
\end{definition}

In analogy to Lemma~28 \cite{nietnerAveragecaseComplexityLearning2023} we now modify $\learn(\mc S, \epsilon)$ to a deterministically 1-verifiable learning problem.

\begin{lemma}\label{lem:linear-1-verifiable}
    Let $0<\tau\leq\epsilon$ and $z\in\mc V^\star$.
    Define $\mc S'=\mc S\setminus B_{2\epsilon+\tau,\met}^\star(z)$.
    Then, $\learn(\mc S', \epsilon)$ is deterministically $1$-verifiable with respect to $z$ from $\eval_\tau$.
\end{lemma}

\begin{proof}
    By \Cref{def:verifyablelearning} we need to construct a family of algorithms $\set{\mc A_y\midvert y\in\mc V^\star}$ such that for every $y$, algorithm $\mc A_y$ solves 
    \begin{align}
        &\dec(\mc Z_y, z)
        &&\text{with}
        &&&\mc Z_y = \learn(\mc S',\epsilon)_y\,,
    \end{align}
    from a single query to $\eval_\tau(\tilde{x})$ to the unknown $\tilde{x}\in\mc Z_y\cup\set{z}$.

    Without loss of generality, let $y$ be such that $\mc Z_y\neq\varnothing$. 
    By definition, it holds
    \begin{align}
        \mc Z_y= B_{\epsilon,\met}^\star(y)\cap\mc S'\,.
    \end{align}
    Now, let $x\in\mc Z_y$ and denote by $v\in \mc V$ a vector that maximizes the dual distance
    \begin{align}
        v \in \argmax_{w\in B_1(0)}\frac12\abs{x(w)-z(w)}\,.
    \end{align}
    In particular, $\abs{x(v)-z(v)}=2\met_\star(x,z)\geq 4\epsilon+2\tau$.

    Then, we define $\mc A_y$ as follows: Query $\eval_\tau(\tilde{z})$ with $v$ and receive a value $r$. 
    If $\abs{r-z(v)}>\tau$ return $\set{\mc Z_y}$. 
    Else, if $\abs{r-z(v)}<\tau$ return $\set{\set{z}}$.

    If $\tilde{x}=z$ the algorithm will be correct by definition.
    Now assume $\tilde{x}=x'\in\mc Z_y$.
    Then, since $\met_\star(x,x')\leq 2\epsilon$ we find 
    \begin{align}\label{eq:verifer-linear-learning}
        \abs{x'(v)-z(v)}
        \geq \abs{\smash{\underbrace{\abs{z(v)-x(v)}}_{=2\met_\star(x,z)}} - \vphantom{\abs{x}}\smash{\underbrace{\abs{x(v)-x'(v)}}_{\leq2\met_\star(x,x')}}} 
        \geq 2\met_\star(x,z) - 2\met_\star(x,x')
        > 4\epsilon+2\tau - 4\epsilon
        =2\tau\,, \vphantom{\abs{\underbrace{\abs{x}}_{=X}}}
    \end{align}
    where we have used $\met_\star(x,z)\geq2\epsilon+\tau>2\epsilon>\met_\star(x,x')$.
    Thus, $\abs{r-z(v)}> \tau$,
    which completes the proof.
\end{proof}

Thus, we obtain the corollary.

\begin{corollary}[Complexity of $\epsilon$-Learning]\label{cor:linear-eps-learning-complexity}
    Let $0<\tau\leq\epsilon$, $\mc S\subseteq\mc V^\star$ and $\mu$ some measure over $\mc S$.
    Assume there exists an algorithm that learns $\learn(\mc S, \epsilon)$ with probability $\beta$ over $x\sim\mu$ and probability $\alpha$ over its internal randomness from $q$ queries to $\eval_\tau$. 
    Then, for every $z\in\mc V^\star$, it holds
    \begin{align}
        q+1\geq
        2\cdot(\alpha-\sfrac12)\cdot\frac{\beta-\Pr_{x\sim\mu}\qty[\met_\star(x,z)<2\epsilon+\tau]}{\max_{v\in \overline{B}_1(0)}\Pr_{x\sim\mu}\qty[\abs{x(v)-z(v)}>\tau]}\,.
    \end{align}
\end{corollary}

Similarly, for any closed subset $\mc M\subset \overline{B}_1(0)$ one can define the pseudo distance on $\mc V^\star$ 
\begin{align}
    \met_\mc M:\mc V^\star\times \mc V^\star \to\mb R\doublecolon (x,z)\mapsto \max_{v\in\mc M}\frac12\abs{x(v)-z(v)}\,.
\end{align}
Note that $\met_\mc M$ is symmetric and satisfies the triangle inequality.
However, depending on $\mc M$ it is not point separating.
We then define.

\begin{definition}
    Let $0<\tau\leq\epsilon$, $\mc M\subseteq \overline{B}_1(0)$ and $\mc S\subset\mc V^\star$. 
    Then, we denote by 
    \begin{align}
        \learn\qty(\mc S, \epsilon, \mc M):\mc S\to\mc P(\mc V^\star)\doublecolon x\mapsto B_{\epsilon,\mc M}^\star(x)\coloneq \set{y\in\mc V^\star\midvert \met_\mc M(x,y)<\epsilon}
    \end{align}
    the problem of \emph{$\mc M$-restricted $\epsilon$-Learning }$\mc S$ with respect to $\mc M$.
\end{definition}

Redoing the arguments in \Cref{lem:linear-1-verifiable} with $\met_\mc M$ instead of $\met_\star$ then yields.

\begin{corollary}[Complexity of $\mc M$-Restrivted $\epsilon$-Learning]\label{cor:complexity-restricted-linear-eps-learning}
    Let $0<\tau\leq\epsilon$, $\mc S\subseteq\mc V^\star$, $\mc M\subseteq \overline{B}_1(0)$ and $\mu$ some measure over $\mc S$.
    Assume there exists an algorithm that learns $\learn(\mc S, \epsilon, \mc M)$ with probability $\beta$ over $x\sim\mu$ and probability $\alpha$ over its internal randomness from $q$ queries to $\eval_\tau$. 
    Then, for every $z\in\mc V^\star$, it holds
    \begin{align}
        q+1\geq
        2\cdot(\alpha-\sfrac12)\cdot\frac{\beta-\Pr_{x\sim\mu}\qty[\met_\mc M(x,z)<2\epsilon+\tau]}{\max_{v\in \overline{B}_1(0)}\Pr_{x\sim\mu}\qty[\abs{x(v)-z(v)}>\tau]}\,.
    \end{align}
\end{corollary}

\subsubsection{Far From Any Fixed Function}\label{sec:farfromeverything}
\addsectionheader{sec:farfromeverything}

A simple yet important observation from \cite{nietnerAveragecaseComplexityLearning2023} is their ``far from any fixed distribution'' observation.

\begin{mylemma}[Lemma 39 from \cite{nietnerAveragecaseComplexityLearning2023}]
    Let $\epsilon,\tau>0$ and let $\mc D\subseteq\Delta(\mc X)$ be a class of distributions over domain $\mc X$.
    Let $\mu$ be a probability measure over $\mc D$ and let $Q\in\mc D$ be an arbitrary distribution.
    Then, for any $D\in\mc D$ it holds 
    \begin{align}
        \Pr_{P\sim\mu}\qty[\tv(P,D)<\epsilon] \leq \max\set{\kfrac\qty(\mu, Q, \tau), \Pr_{P\sim\mu}\qty[\tv(P,Q)\leq 2\epsilon+\tfrac\tau2]}\,.
    \end{align}
\end{mylemma}

Where we slightly adapted the $\tau$ dependence.

This lemma is interesting for two reasons.
First, it characterizes the maximal weight of any $\epsilon$-ball in terms of the maximally distinguishable fraction against $Q$ and the weight of the $2\epsilon+\tau$-ball around $Q$. 

Secondly, we note that the Lemma can be seen as a connection between the general deterministic bound for $\epsilon$-learning linear function classes as by \Cref{thm:deterministicavglowerbound} and the lower bound via verifiable learning as in \Cref{cor:linear-eps-learning-complexity}.
In particular, while the former translates to a bound in terms of the largest $\epsilon$-ball, the latter works in terms of just a single $2\epsilon+\tau$-ball.
In the following we show that this is true in general for learning linear function classes. 

\begin{lemma}\label{lem:far-from}
    Let $\epsilon,\tau>0$ and let $z,x\in\mc V^\star$ be such that 
    \begin{align}
        \met_\star(x,z)\geq\epsilon+\frac\tau2\,.
    \end{align}
    Then, for any measure $\mu$ over $\mc V^\star$ it holds 
    \begin{align}
        \Pr_{y\sim\mu}\qty[\met_\star(x,y)<\epsilon]
        \leq \kfrac\qty(\mu, z, \tau)
        =\sup_{v\in \overline{B}_1(0)}\Pr_{y\sim\mu}\qty[\abs{y(v)-z(v)}>\tau]\,.
    \end{align}
\end{lemma}

\begin{proof}
    Recall that by the definition of the dual norm it holds
    \begin{align}
        \met_\star(x,y)=\sup_{v\in \overline{B}_1(0)} \frac{1}{2}\abs{x(v)-y(v)}\,.
    \end{align}
    Thus, for every $y\in B_\epsilon^\star(x)$ and any $v\in\overline{B}_1(0)$ we find
    \begin{align}
        \abs{y(v) - z(v)} 
        \geq \abs{ \abs{y(v) - x(v)} - \abs{x(v)-z(v)} }
        \geq \abs{2\met_\star(x,y) - 2\met_\star(x,z)}
        > \abs{2\epsilon +\tau - 2\epsilon} = 2\tau\,,
    \end{align}
    where we used $\met_\star(x,z)\geq\epsilon+\frac\tau2>\epsilon>\met_\star(x,y)$.
    Thus, for any $v\in \overline{B}_1(0)$ it holds
    \begin{align}
        \Pr_{y\sim\mu}\qty[\met_\star(x,y)<\epsilon]
        \leq\Pr_{y\sim\mu}\qty[\abs{y(v)-z(v)}>\tau] 
        \leq \kfrac\qty(\mu, z, \tau)\,.
    \end{align}
\end{proof}

Similar to \cite[Lemma 39]{nietnerAveragecaseComplexityLearning2023} we conclude with.

\begin{corollary}\label{cor:farfrom}
    Let $\epsilon,\tau>0$ and let $z\in\mc V^\star$ and let $\mu$ be a measure over $\mc V^\star$. 
    Then for every $x\in\mc V^\star$ it holds 
    \begin{align}
        \Pr_{y\sim\mu}\qty[\met_\star(x,y)<\epsilon]
        \leq \max\set{\kfrac\qty(\mu, z, \tau), \Pr_{y\sim\mu}\qty[\met_\star(y,z)<2\epsilon+\tfrac{\tau}{2}]}\,.
    \end{align}
\end{corollary}

\begin{proof}
    If $\met_\star(x,z)<\epsilon+\frac{\tau}{2}$, then $\Pr_{y\sim\mu}\qty[\met_\star(x,y)<\epsilon]\leq \Pr_{y\sim\mu}\qty[\met_\star(y,z)<2\epsilon+\frac{\tau}{2}]$.
    Otherwise, if $\met_\star(x,z)\geq\epsilon+\frac\tau2$ it holds by \Cref{lem:far-from} that $\Pr_{y\sim\mu}\qty[\met_\star(x,y)<\epsilon] \leq \kfrac\qty(\mu, z, \tau)$.
\end{proof}

\stopcontents[linear]

\clearpage
\part{Applications}\label{sec:applications}

\section{Preliminaries}\label{sec:applications-preliminaries}
\addsectionfooter{sec:applications-preliminaries}
\addsectionheader{sec:applications-preliminaries}
\startcontents[app-prelims]
\vspace{1cm}
\printcontents[app-prelims]{}{1}{}
\vspace{1cm}

We denote by $\Delta(N)\subset\mb R^N$ the set of $N$-dimensional probability distributions and similarly by $\Delta(\mc X)$ the set of probability distributions over the set $\mc X$\footnotemark.
\footnotetext{Throughout this work we assume to work over ``nice'' subsets of $\mb C^N$, such that without loss of generality, we assume that the event space, or $\sigma$-algebra, corresponds to the induced Lebesgue measurable sets. Similarly, by $X\subseteq\mc X$ we implicitly assume that $X$ is also contained in the respective $\sigma$-algebra.} 
For two probability distributions $P,Q\in\Delta(\mc X)$ we write $\tv(P,Q)=\max_{X\subseteq\mc X}\abs{P(X)-Q(X)}=\frac12\sum_{x\in\mc X}\abs{P(x)-Q(x)}$, where abusing notation we write $P(x)=P(\{x\})$ in the general case and $P(X)=\sum_{i\in\mc X}P_i$ for discrete probability distributions. 
We denote by $\mb H^{N\times N}\subset\mb C^{N\times N}$ the set of $N$-dimensional Hermitian operators and by $\mc S(N)\subset\mb H^{N\times N}$ the set of $N$-dimensional quantum states $0\leq\rho=\rho^\dagger$, $\tr[\rho]=1$.
Further, for any $M\in\mb C^{N\times N}$ we denote by 
\begin{align}
    &\norm{M}_{\tr}=\tr[\sqrt{M^\dagger M}]
    &&\text{the trace norm and by}
    &&&\norm{M}_\mrm{op}=\sup_x\frac{\norm{Mx}_2}{\norm{x}_2}
\end{align}
the operator norm,
with $\norm{x}_2$ the $\ell_2$ norm of the vector $x\in\mb C^N$.
We use the standard braket notation for pure state vectors $\rho=\ketbra{\psi}$ with $\ket\psi\in\mb C^N$ and $\bra\psi$ its dual. 
It holds $\sqrt{\braket{\psi}}=\norm{\ket\psi}_2$ as and denote by $\braket{\phi}{\psi}$ the inner product between two state vectors.
If not mentioned differently, we assume $\ket\psi$ to be normalized $\norm{\ket\psi}_2=1$.

As already introduced before, we denote by $\Delta(N)\subset\mb R^N$ the simplex of $N$ dimensional probability distributions. 
By $\norm{x}_1=\sum_i\abs{x_i}$ we denote the $1$-norm and define the total variation distance between two probability distributions $P,Q\in\Delta(N)$ via $\tv(P,Q)=\frac12\norm{P-Q}_1$.
Similarly, we define the trace distance between quantum states by $\met_{\tr}(\rho,\sigma)=\frac12\norm{\rho-\sigma}_{\tr}$. 
The Fidelity between quantum states is defined as 
\begin{align}
    F(\rho,\sigma)=\qty(\tr[\sqrt{\sqrt{\sigma}\rho\sqrt{\sigma}}])^2\,.
\end{align} 
By the Fuchs-van de Graaf inequality it holds 
\begin{align}\label{eq:fuchs-van-de-graaf}
    1-\sqrt{F(\rho,\sigma)} \leq \met_{\tr}(\rho,\sigma) \leq \sqrt{1-F(\rho,\sigma)}\,.
\end{align}
In particular, for $\rho=\ketbra\psi$ pure it holds
\begin{align}\label{eq:trace-dist-fidelity-pure}
    \frac{1}{2}\norm{\rho-\sigma}_{\tr} = \sqrt{1-F(\rho,\sigma)} = \sqrt{1-\mel{\psi}{\sigma}{\psi}}\,.
\end{align}
We denote by $\id_N$ the $N$ dimensional identity matrix and, if $N$ is clear from context simply write $N$. Similarly, it will be made clear if the index has a different meaning.

\subsection{Random Quantum Circuits}\label{sec:prelimRQC}
\addsectionheader{sec:prelimRQC}

Most of our applications will consider quantum circuits and ensembles of random quantum circuits. Those are defined as follows.
Important objects we are interested in then are the output states of quantum circuits and the corresponding Born distributions.

\begin{definition}[Brickwork Quantum Circuits]\label{def:brickwork-classes}
    For a gate set $\mc G\subseteq \U(4)$ and $n,d\in\mb N$ denote by $\Cbrqc[,n,d](\mc G)\subseteq\U\qty(2^n)$ the set of \emph{Brickwork Quantum Circuits} on $n$ qubits of depth $d$. 
    $\Cbrqc$ consists of all unitaries that can be decomposed as 
    \begin{equation}
        U 
        = U^{(d)}\cdots U^{(2)}\cdot U^{(1)}
        =\qty(\mds 1_2\otimes U_{23}^{(d)}\otimes U_{45}^{(d)}\otimes\cdots)\cdots\qty(\mds 1_2\otimes U_{23}^{(2)}\otimes U_{45}^{(2)}\otimes\cdots)\cdot\qty(U_{12}^{(1)}\otimes U_{34}^{(1)}\otimes\cdots)\,,
    \end{equation} 
    where each $U_{ij}^{(l)}\in\mc G$ acts on qubits $i$ and $j$ only and where for simplicity we have assumed $d$ even. 

    Similarly, we denote by $\Qbrqc[,n,d](\mc G)$ the class of states 
    \begin{equation}
        \Qbrqc[,n,d]=\set{\rho = \rho(U)\midvert \rho(U) = U\ketbra{0} U^\dagger\,,\; U\in\Cbrqc[,n,d](\mc G)}\,,
    \end{equation} 
    which we call \emph{Brickwork Quantum States}.

    For any quantum state $\rho$ we define its \emph{Born Distribution} as the probability distribution $P_\rho(x)=\mel{x}{\rho}{x}$. Denote by
    \begin{equation}
        \Dbrqc[,n,d]=\set{P_\rho\in\Delta\qty(2^n)\midvert \rho\in\Qbrqc}\,,
    \end{equation}
    the set of \emph{Brickwork Quantum Circuit Born Distributions}.
    Similarly, for a unitary (quantum circuit) $U$ we refer to $P_U\coloneq P_{\rho(U)}$ the \emph{Born Distribution} or \emph{Output Distribution} of the unitary (quantum circuit) $U$.
\end{definition}

Random quantum circuits are then defined via the following process.

\begin{definition}[Brickwork Random Quantum Circuits]\label{def:rqc}
    Let $\mubrqc[,n,d]$ be the probability distribution over $\U\qty(2^n)$ that corresponds to drawing a unitary $U\in\Cbrqc[,n,d]$ by drawing each $U_{ij}^{(l)}$ i.i.d. from the Haar measure over $\U(4)$.
    We refer to the distribution $\mubrqc[,n,d]$ as \emph{Brickwork Random Quantum Circuits} on $n$ qubits of depth $d$.
\end{definition}

\subsection{Unitary Designs and the Haar Measure}\label{sec:haar-preliminaries}
\addsectionheader{sec:haar-preliminaries}

One of the key features behind that makes random quantum circuits interesting is that they approximate the Haar measure on the unitary group $\U(N)$ in a specific sense.

\begin{definition}[Unitary $t$-Designs]\label{def:unitary-t-design}
    For any measure $\mu$ over $\U(N)$ we define the $t$\emph{'th Moment Operator} as 
    \begin{align}
        \Phi^{(t)}_{\mu}:\mb C^{N^t\times N^t}\to\mb C^{N^t\times N^t}\doublecolon A\mapsto \Ex_{U\sim\mu}\qty[U^{\otimes t}\cdot A\cdot (U^\dagger)^{\otimes t}]\,.
    \end{align}
    We denote by $U\sim\U(N)$ the \emph{Haar measure} (or similarly, the uniform measure) over $\U(N)$. 
    
    We say that a measure $\mu$ over $\U(N)$ forms a \emph{Unitary $t$-Design} if it holds
    \begin{align}
        \Phi^{(t)}_{\U(N)} = \Phi^{(t)}_{\mu}\,.
    \end{align}
    By $U\stacksim{t}\U(N)$ we denote the process of sampling from a unitary $t$ design.
\end{definition}

Note that any $t$-design forms a $t-1$-design since 
\begin{align}
    \Ex_{U\sim\mu}\qty[\tr_{t}\qty[U^{\otimes t}(A\otimes \tfrac{1}{N} \id_N)(U^\dagger)^{\otimes t}]] 
    = \Ex_{U\sim\mu}\qty[U^{\otimes t-1}A(U^\dagger)^{\otimes t-1}]\,, 
\end{align}
where $\tr_{t}[\dots]$ denotes the partial trace on the $t$'th copy and $N^{-1}\id_N$ denotes the maximally mixed state in $N$ dimensions.

\begin{lemma}[Second Moments \cite{klieschTheory2021}]\label{lem:unitary-second-moments}
    Let $A\in\mb C^{N^2\times N^2}$ be an operator with $N\geq2$, then 
    \begin{align*}
        \Phi_{\U(N)}^{(2)} (A) &= \frac{2\tr[A P_\vee]}{N(N+1)}P_\vee + \frac{2\tr[A P_\wedge]}{N(N-1)} P_\wedge \\[5pt]
        &= \frac{2}{(N-1)(N+1)}\qty(\tr[A]\id + \tr[A\mb F]\mb F) 
        + \frac{2}{(N-1)N(N+1)} \qty(\tr[A]\mb F + \tr[A\mb F]\id)\,,
    \end{align*}
    where $P_\vee=\frac12(\id+\mb F)$ and $P_\wedge=\frac12(\id-\mb F)$.
\end{lemma}

The typical examples for designs are the Pauli group, which forms a $1$-design, and the Clifford group.

\begin{lemma}[The Clifford Group Forms a $3$-Design \cite{zhuClifford3Design2017, zhuCliffordFails2016}]\label{lem:clifford-3-design}
    The set $\Cl(2,n)$ of $n$-qubit Clifford unitaries forms a unitary $3$-design $\Phi_{\Cl(2,n)}^{(3)}=\Phi_{\U(2^n)}^{(3)}$. However, The Clifford group fails "gracefully" to be a unitary $4$-design $\Phi_{\Cl(2,n)}^{(4)}\neq\Phi_{\U(2^n)}^{(4)}$.
\end{lemma}

Approximations of $t$-designs exist for various norms. 
The most common definition is with respect to the diamond norm $\norm{\cdotspace}_\diamond$ \cite{haferkamp2022random}. 
For a measure $\mu$ over $\U(N)$ it requires 
\begin{align}\label{eq:eps-approx-design-conventional}
    \norm{\Phi^{(t)}_{\U(N)} - \Phi^{(t)}_{\mu}}_\diamond 
    \leq \frac{\epsilon}{N^{t}}\,,
\end{align}
to hold for $\mu$ to form an $\epsilon$-approximate unitary $t$-design.
The dimensional factor is connected to the definition of a \emph{relative} $\epsilon$-approximate unitary $t$-design, which requires \cite{brandao2016local}
\begin{align}\label{eq:relative-eps-approx-design}
    (1-\epsilon)\qty(\Phi_{\U(N)}^{(t)}\otimes\id^{\otimes t})\qty(\Omega_N^{\otimes t})
    \leq \qty(\Phi_{\mu}^{(t)}\otimes\id^{\otimes t})\qty(\Omega_N^{\otimes t})
    \leq (1+\epsilon)\qty(\Phi_{\U(N)}^{(t)}\otimes\id^{\otimes t})\qty(\Omega_N^{\otimes t})
\end{align}
to hold, with $\ket{\Omega_N}=\sum_{i=1}^N\ket{ii}$ the maximally entangled state.
In particular, while \Cref{eq:eps-approx-design-conventional} refers to an additive error between the maps $\Phi_{\U(N)}^{(t)}$ and $\Phi_\mu^{(t)}$, \Cref{eq:relative-eps-approx-design} enforces a small a relative error. 

Another definition is that of an $\epsilon$-approximate \emph{Tensor Product Expander} (TPE). 
This is, $\mu$ forms an $\epsilon$-approximate TPE if it holds \cite{brandao2016local}
\begin{align}
    g(\mu,t)\coloneq\norm{\Phi^{(t)}_{\U(N)} - \Phi^{(t)}_{\mu}}_\mrm{op} 
    \leq \eps\,.
\end{align}
Since the operator norm and the diamond norm are only related by a dimensional factor $\norm{\cdotspace}_\diamond\leq N \norm{\cdotspace}_\mrm{op}$ (in dimension $N$) the TPE condition is much weaker.
Yet, the TPE condition can be easily boosted as $g(\mu^{*r},t)\leq g(\mu,t)^r$.
Here $\mu^{*r}$ denotes the measure that samples $U_1,\dots,U_r\sim\mu$ identically and independently and then returns $U=U_r\cdots U_2\cdot U_1$.
Thus, most techniques for proving approximate unitary designs go via first proving a statement on $g(\mu,t)$ and then boosting to the required accuracy.

In this work we are mainly interested in approximate designs with a small additive error. 
We thus define.

\begin{definition}[Additive $\epsilon$-Approximate Unitary $t$-Design]
    A measure $\mu$ over $\U(N)$ is said to form an \emph{Additive $\epsilon$-Approximate Unitary $t$-Design} if it holds
    \begin{align}
        \norm{\Phi^{(t)}_{\U(N)} - \Phi^{(t)}_{\mu}}_\diamond 
        \leq \epsilon\,,
    \end{align}
    We write $U\stacksim{\epsilon,t}\U(N)$ if we sample from an additive $\epsilon$-approximate unitary $t$-design.
\end{definition}

Importantly, as shown by Brand\~ao, Harrow and Horodecki and improved upon by Haferkamp, brickwork random quantum circuits form approximate designs \cite{brandao2016local,haferkamp2022random}.

\begin{lemma}[Random Quantum Circuits Form a $t$-Design \cite{haferkamp2022random}]\label{lem:rqc-design}
    Let $n\geq\ceil*{2\log(4t) +1.5\sqrt{\log(4t)}}$, then $\mubrqc[,n,d]$ forms an additive $\epsilon$-approximate unitary $t$-design if 
    \begin{align}
        d\geq C\ln(t)^5 t^{4+3\frac{1}{\log(t)}} \qty(nt+\log\qty(\epsilon^{-1}))\,,
    \end{align}
    where $C$ is a constant that can be taken to be $C=2^{44}$.
\end{lemma}

Moreover, for small $t$ those bounds were improved by Haferkamp and Hunter-Jones.

\begin{lemma}[RQC Design Depth for Small $t$ \cite{haferkamp2021improved}]
    Let $t\in\set{2,4}$ and $n$ as before. 
    Then $\mubrqc[,n,d]$ forms an additive $\epsilon$ approximate $t$-design if 
    \begin{align}
        &d\geq 6.4\qty(n+\log(n)+\log\qty(\epsilon^{-1})) 
        &&\text{for} 
        &&&t=2
        \intertext{and}
        &d\geq 38\qty(4n+\log\qty(\epsilon^{-1})) 
        &&\text{for} 
        &&&t=4\,.
    \end{align}
\end{lemma}

Note that we adapted the prefactor of the $O(n)$ term by $\sfrac12$ compared to the values in \cite[Table~I]{haferkamp2021improved}.
This is because we state the values for additive approximate designs, whereas the numbers in their Table~I are for approximate designs.

\subsection{Spherical and Projective Designs}\label{sec:sphericaldesigns}
\addsectionheader{sec:sphericaldesigns}

The main use case of unitary designs in this work is to define complex spherical and complex projective designs.
Those correspond to measures over the complex unit sphere $\mb S^{N-1}\subset\mb C^N$. 
We identify the set of $N$ dimensional pure state vectors with the complex unit sphere in $N$-dimensions $\ket\psi\in\mb S^{N-1}$ and use the standard braket notation for dual vectors $\bra\psi$.
We call a finite subset $T\subset\mb N^2$ a \emph{Lower Set} if for any $(k,l)\in T$ it holds that every $(m,n)\in T$ with $m\leq k$, $m\leq l$. 

\begin{definition}[Complex Spherical $T$-Designs \cite{roySpherical2011}]\label{def:spherical-design}
    For any $(s,t)\in\mb N$ and any measure $\mu$ over $\mb S^{N-1}\subset\mb C^N$ denote by 
    \begin{align}
        K_\mu^{(s,t)} = \Ex_{\psi\sim\mu}\qty[\ket\psi^{\otimes s}\bra\psi^{\otimes t}]\in\mb C^{N^s\times N^t}\,,
    \end{align}
    the $(s,t)$'th moment operator. 
    We use the shorthand $K_\mu^{(t)}=K_\mu^{(t,t)}$.
    We denote by $\psi\sim\mb S^{N-1}$ the \emph{Haar Measure} (or similarly the uniform measure) over $\mb S^{N-1}$.
    For a lower set $T\subset\mb N^2$ we say that $\mu$ forms a \emph{Complex Spherical $T$-Design} if for all $(s,t)\in T$ it holds 
    \begin{align}
        K_\mu^{(s,t)} = K_{\mb S^{N-1}}^{(s,t)}\,.
    \end{align}
    Similarly, we say that $\mu$ forms a \emph{Complex Projective $t$-Design} if for every $s\leq t$ it holds
    \begin{align}
        K_\mu^{(s)} = K_{\mb S^{N-1}}^{(s)}\,.
    \end{align}
    By $\psi\stacksim{(t,t)}\mb S^{N-1}$ ($\psi\stacksim{t}\mb S^{N-1}$) we denote the process of sampling from a complex spherical $(t,t)$-design (projective $t$-design).
    We use the shorthand $(s,t)$-design for a $T=\set{(k,l)\midvert k\leq s, l\leq l}\subset\mb N^2$-design.
\end{definition}

The definition of projective and spherical designs is very similar.
The practical difference, why in some cases it is handy to require the stronger notion of a spherical $t$-design, is that in principle nothing is known about the ``off-diagonal'' moments of a projective design. 
To see this first note that for a $T$-design it holds
\begin{align}
    &K_\mu^{(k,l)} = 0
    &&\text{for all}
    &&&(k,l)\in T
    &&&&\text{with}
    &&&&&k\neq l\,,
\end{align}
see for example Lemma~3.3 in \cite{roySpherical2011}.
Now, as in  \cite[Example~3.5]{roySpherical2011},
let $L$ be the standard basis in $\mb C^N$.
$L$ forms a projective $1$-design, but
$K_L^{(0,1)}=\Ex_{\psi\sim L}[\ket{\psi}]=\frac{1}{N}(1,\dots,1)\neq 0$.

Explicit formula for projective designs can be found in e.g. \cite{klieschTheory2021}.

\begin{lemma}[Projective Moment Operators: Lemma~34 from \cite{klieschTheory2021}]\label{lem:moments-lemma34}
    It holds 
    \begin{align}
        K_{\mb S^{N-1}}^{(t)} = \binom{N+t-1}{t}^{-1} P_{\vee^t}\,,
    \end{align}
    with $P_{\vee^t}$ the projector on the symmetric subspace with respect to the permutation group $S_t$ on the $t$ copies of Hilbert space.
    In particular, for $t=1,2$ this implies 
    \begin{align}
        &K_{\mb S^{N-1}}^{(1)} = \frac{\id}{N}
        &&\text{and}
        &&&K_{\mb S^{N-1}}^{(2)} = \frac{\id\otimes\id + \mb F}{N(N+1)}
    \end{align}
    with $\mb F$ the operator that flips the two copies of Hilbert space.
\end{lemma}

Similar to the unitary case we now define approximate designs.
In going from channels to states, the diamond norm naturally translates to the trace norm.

\begin{definition}[Additive $\epsilon$-Approximate Spherical $T$-Design]
    Let $T\subset\mb N^2$ be a lower set. A measure $\mu$ over $\mb S^{N-1}$ forms an \emph{Additive $\epsilon$-Approximate  Complex Spherical $T$-Design} if for all $(s,t)\in T$ it holds
    \begin{align}
        \norm{K_{\mb S^{N-1}}^{(s,t)}-K_{\mu}^{(s,t)}}_{\tr} \leq \epsilon\,.
    \end{align}
    Again, we use the shorthand $(t,t)$-design as in \Cref{def:spherical-design} and define \emph{Additive $\epsilon$-Approximate Complex Projective $t$-Designs} accordingly.
    We write $\psi\stacksim{\epsilon,(t,t)}\mb S^{N-1}$ ($\psi\stacksim{\epsilon,t}\mb S^{N-1}$) if we sample from an additive $\epsilon$-approximate complex spherical $(t,t)$-design (projective $t$-design).
\end{definition}

The main reason, why we are interested in additive approximate designs is the following lemma.

\begin{lemma}\label{lem:additive-design-variance-implication}
    Let $\mu,\nu$ be measures over $\mb S^{N-1}$ and let $O\in\mb C^{N\times N}$.
    Then,
    \begin{align}
        &\norm{K_{\nu}^{(t)}-K_{\mu}^{(t)}}_{\tr} \leq \epsilon\,,
        &&\text{for $t=1,2$ implies}
        &&&\Var_{\psi\sim\mu}\qty[\mel{\psi}{O}{\psi}]\leq\Var_{\psi\sim\nu}\qty[\mel{\psi}{O}{\psi}] + \epsilon + \epsilon^2\,.
    \end{align}
\end{lemma}

\begin{proof}
    The claim follows from direct calculation 
    \begin{align}
        \Var_{\psi\sim\mu}\qty[\mel{\psi}{O}{\psi}]
        =\Ex_{\psi\sim\mu}\qty[(O\otimes O)K_\mu^{(2)}] - \Ex_{\psi\sim\mu}\qty[OK_\mu^{(1)}]^2
        \leq \qty(\Ex_{\psi\sim\nu}\qty[(O\otimes O)K_\nu^{(2)}]+\epsilon) - \qty(\Ex_{\psi\sim\nu}\qty[OK_\nu^{(1)}] - \epsilon)^2\\
        = \Var_{\psi\sim\nu}\qty[\mel{\psi}{O}{\psi}] + \epsilon + \epsilon^2 - 2\epsilon\Ex_{\psi\sim\nu}\qty[OK_\nu^{(1)}] 
        \leq\Var_{\psi\sim\nu}\qty[\mel{\psi}{O}{\psi}] + \epsilon + \epsilon^2\,.
    \end{align}
\end{proof}

A direct consequence of \Cref{lem:rqc-design} is the following corollary.

\begin{corollary}[Random Quantum Circuits Form a Projective $t$-Design]\label{cor:circuits-projective-design}
    Let $n\geq\ceil*{2\log(4t) +1.5\sqrt{\log(4t)}}$, $N=2^n$ and denote by $\mubrqc[,n,d]^\star$ the measure on $\mb S^{N-1}$ induced by $\mubrqc[,n,d]$ which corresponds to first sampling $U\sim\mubrqc[,n,d]$ and then outputting $U\ket0$.
    Then, for any $t\in\mb N$ it holds that $\mubrqc[,n,d]^\star$ forms an additive $\epsilon$-approximate complex projective $t$-design if 
    \begin{align}
        d\geq C\ln(t)^5 t^{4+3\frac{1}{\log(t)}} \qty(nt + \log\qty(\epsilon^{-1}))\,.
    \end{align}
    Moreover, $\mubrqc[,n,d]^\star$ forms an additive $\epsilon$-approximate $t$-design if 
    \begin{align}
        &d\geq 6.4\qty(n+\log(n)+\log\qty(\epsilon^{-1})) 
        &&\text{for} 
        &&&t=2
        \intertext{and}
        &d\geq 38\qty(4n+\log\qty(\epsilon^{-1})) 
        &&\text{for} 
        &&&t=4\,.
    \end{align}
\end{corollary}

\stopcontents[app-prelims]

\section{Quantum Statistical Query Learning}\label{sec:quantumlearning}
\addsectionfooter{sec:quantumlearning}
\addsectionheader{sec:quantumlearning}
\startcontents[quantum]
\vspace{1cm}
\printcontents[quantum]{}{1}{}
\vspace{1cm}

The quantum statistical query model has been studied in a number of works \cite{arunachalamQSQ2020,angrisaniQLD2022,caroVerification2023}.
Recall the definition of the quantum statistical query oracle as introduced in \Cref{ex:statesandQSQ}.

\begin{definition}[Quantum Statistical Query Oracle $\qstat$]\label{def:qsq}
    Let $N>0$, $\tau>0$ and let $\rho\in\mc S(N)\subset\mb H^{N\times N}$ with $0<\rho<\id$ be an $N$ dimensional quantum state. 
    We denote by $\qstat_\tau(\rho)$ the \emph{Quantum Statistical Query} (QSQ) oracle of $\rho$ with tolerance $\tau$. 
    When queried with an observable $O\in\mb H^{N\times N}$ with $\norm{O}_\mrm{op}\leq1$ it returns a value $v$, which is promised to lie in the interval
    \begin{equation}
        v\in\qty[\tr[\rho \cdot O]-\tau, \tr[\rho \cdot O]+\tau]\,.
    \end{equation}
\end{definition}

As mentioned previously, the $\qstat_\tau$ oracle implements the $\eval_\tau$ oracle with respect to the absolute value $\met(v,w)=\abs{v-w}$ in the following way:
We identify a quantum states $\rho$ with the corresponding functions over bounded operators $\norm{O}_\mrm{op}\leq 1$, which maps $O\mapsto\tr[\rho O]$.
Thus, the $\qstat$ query complexity of non-trivial learning for any problem $\mc Z:\mc Q\to\mc P(\mc T)$ over a set of quantum states $\mc Q\subset\mc S(N)$ and measure $\mu$ over $\mc Q$ is lower bounded by  
\begin{align}
    \qnt[\mu, \tau] \geq  \tau^2\cdot\min_O\Var_{\rho\sim\mu}\qty[\tr[\rho O]]^{-1}\,.
\end{align}

\subsection{Positive Results}\label{sec:qsq-positive}
\addsectionheader{sec:qsq-positive}

We begin by collecting a few positive results about QSQ based learning algorithms.

\subsubsection{State Testing}\label{sec:statetesting}
\addsectionheader{sec:statetesting}

It is easy to see that for any pure state $\rho=\ketbra\psi$ one can test from a single query to $\qstat_\tau$ whether an unknown pure state $\sigma$ equals $\rho$ or is at least $\epsilon$ far, for $\tau<\epsilon^2$. 
To this end, denote by $\mc S_\mrm{pure}(N)=\set{\rho\in\mc S(N)\midvert \tr[\rho^2]=1}$ the set of $N$ dimensional pure states.

\begin{theorem}[$\epsilon$-Testing Pure States]
    Let $N\in\mb N$, $0<\epsilon\leq1$ and $0\leq\tau<\sfrac{\epsilon^2}{2}$, $\rho=\ketbra\psi\in\mc S(N)$ and denote by $\mc Q=\mc S_\mrm{pure}(N)\setminus B_\epsilon(\rho)$, where $B_\epsilon$ is the open $\epsilon$-ball with respect to the trace distance. Then $\dec(\mc Q, \rho)$ can be solved from a single query to $\qstat_\tau$.
\end{theorem}

\begin{proof}
    Let $\sigma=\ketbra\varphi\in\mc Q\cup\set{\rho}$. Then, querying $\qstat_\tau(\sigma)$ with $\rho$ returns some $v\in\abs{\braket{\varphi}{\psi}}^2+[-\tau,\tau]$.
    For the trace distance of pure states it holds
    \begin{align}
        \met_{\tr}(\rho,\sigma)=\frac12\norm{\rho-\sigma}_{\tr} = \sqrt{1-\abs{\braket{\varphi}{\psi}}^2}\,.
    \end{align}
    Thus, by assumption $\abs{\braket{\varphi}{\psi}}^2$ is either $1$ or at most $1-\epsilon^2$.
    Therefore, for any $\tau<\sfrac{\epsilon^2}{2}$ the response $v$ can distinguish between $\abs{\braket{\varphi}{\psi}}^2=1$ and $\abs{\braket{\varphi}{\psi}}^2\leq 1-\epsilon^2$.
\end{proof}

\subsubsection{PAC Learning}\label{sec:QSQPAClearning}
\addsectionheader{sec:QSQPAClearning}

Let us now consider PAC learning 
as in \Cref{prob:pac-learning}
in terms of the QSQ oracle.

\begin{definition}[Quantum PAC and QSQ]\label{def:qpac-state}
    Let $n\in\mb N$, let $f:\mb F_2^n\to\mb F_2$ be a boolean function and let $\mc D$ be a distribution over $\mb F_2^n$.
    We refer to the $n+1$-qubit state $\rho_f=\ketbra{\psi_f}$ with
    \begin{equation}
        \ket{\psi_f} = \sum_{x\sim\mc D} \sqrt{\mc D(x)} \ket{x, f(x)}
    \end{equation}
    as the \emph{Quantum Probably Approximately Correct} (qPAC) state with respect to $f$ and $\mc D$. 
    A \emph{qPAC Oracle} then is any oracle that provides copies of the qPAC state $\rho_f$.
    Similarly, the \emph{Noisy qPAC Oracle} with noise rate $\eta\in[0,1]$ is defined similarly with respect to $\rho_f^\eta=\ketbra{\psi_{f,b}}$, where 
    \begin{align}
        &\rho_f^\eta = \Ex_{b}\qty[\ketbra{\psi_{f,b}}] &&\text{with} &&&\ket{\psi_{f,b}} = \sum_{x\sim\mc D} \sqrt{\mc D(x)} \ket{x, f(x)\oplus b_x}\,,
    \end{align}
    with $b$ a length $2^n$ bit string where each $b_i$ is sampled i.i.d. according to a biased coin flip $b_i\sim \qty(1-\eta,\;\eta)$\footnotemark.
    For a class $\mc C$ of boolean functions denote by $\mc Q_\mc C$ the set of qPAC states and by $\mc Q_\mc C^\eta$ the set of noisy qPAC states.
    The oracle that provides copies of $\rho_f$ is referred to as qPAC oracle and similarly in case of noisy qPAC.
    We say we are given QSQ access to the (noisy) qPAC oracle if we are given access to a QSQ oracle $\qstat_\tau(\rho_f)$  (respectively $\qstat_\tau(\rho_f^\eta)$).
\end{definition}
\footnotetext{
        Each $2^n$-bit string $b$ encodes a specific noise pattern. The state $\rho_f^\eta$ then is the mixture over all noise configurations.
    }

The statistical query model was introduced by Kearns \cite{kearns_efficient_1998} in order to model learning algorithms in the presence of noise.
Thus, one of the core contributions was a result showing that SQ based PAC learners imply PAC learners, which are robust to random classification noise.
In \cite{arunachalamQSQ2020} the corresponding result is shown for QSQ based learners.

\begin{theorem}[QSQ, PAC and noisy PAC, Theorem 3.4 from \cite{arunachalamQSQ2020}]\label{thm:QSQ-to-PAC}
    Let $\tau,\delta\in(0,1)$ and let $\eta<\max\set{\sfrac12, 2\tau^2}$. Further, let $\mc C$ be a class of boolean functions and
    suppose there exists an algorithm for $\epsilon$-learning $\mc Q_\mc C$ from $q$ many queries to $\qstat_\tau$.
    Then it holds:
    \begin{enumerate}
        \item There exists a quantum algorithm for $(\epsilon,\delta)$-PAC learning $\mc C$ from $O(\tau^{-2}q\log(\sfrac{q}{\delta}))$ many copies of the qPAC state.
        \item There exists a quantum algorithm for $(\epsilon,\delta)$-PAC learning $\mc C$ from $O((\tau-\sqrt{\eta})^{-2}q\log(\sfrac{q}{\delta}))$ many copies of the noisy qPAC state with noise rate $\eta$.
    \end{enumerate}
\end{theorem}

Moreover, Arunachalam, Grilo and Yuen have shown in \cite{arunachalamQSQ2020} that, in contrast to learning from statistical queries, parities can be learned from quantum statistical queries.

\begin{theorem}[Learning Parities, Lemma 4.2 in\cite{arunachalamQSQ2020}]\label{thm:qsq-parity}
    Let $n\in\mb N$ and let $\mc C$ be the class of parity functions 
    \begin{equation}
        \mc C = \set{\chi_s:\mb F_2^n\to\mb F_2\doublecolon x\mapsto x\cdot s\midvert s\in\mb F_2^n}\,.
    \end{equation}
    Then for any $\tau<\sfrac12-\Omega(\overpoly)$ the class $\mc C$ can be learned exactly from $O(n)$  QSQ queries to the qPAC oracle with respect to $\mc D$ the uniform distribution.
\end{theorem}

Similarly, in \cite{arunachalamQSQ2020} it is shown that the class of $k$-juntas can be $\epsilon$-learned efficiently from $O(n+2^{O(k)})$ many quantum statistical queries of tolerance $\tau=O(\epsilon 2^{-\sfrac{k}{2}})$ and that 
the class of $\poly(n)$ sized DNF's can be learned from $O(\poly(n))$ many $\overpoly$-accurate quantum statistical queries (Lemmas 4.3 and 4.5).

Combining \Cref{thm:QSQ-to-PAC,thm:qsq-parity} yields then the following corollary.

\begin{corollary}[Solving LPN From QSQ's \cite{arunachalamQSQ2020}]\label{cor:qsq-lpn}
    Let $\tau>0$. The learning parities with noise problem can be solved efficiently from $\tau$-accurate QSQ access to the noisy qPAC state for any noise rate $\eta < \max\set{1/2, 2\tau^2}$.
\end{corollary}

\subsubsection{Matrix Product States}\label{sec:QSQMPSlearning}
\addsectionheader{sec:QSQMPSlearning}

Matrix product states are a prominent class of states. 

\begin{definition}
    Let $n,D\in\mb N$. An $n$-qubit pure state $\rho=\ketbra{\psi}$ is referred to as \emph{Matrix Product State} (MPS) with \emph{Bond Dimension} $D$ if it can be decomposed as 
    \begin{equation}
        \ket\psi = \ket{\mrm{MPS}(A_1,\dots,A_n)}
        = \sum_{x\in\mb F_2^n} A_1^{(x_1)}\cdot A_2^{(x_2)}\cdots A_{n-1}^{(x_{n-1})}\cdot A_n^{(x_n)} \ket{x}\,,
    \end{equation}
    where the $A_l^{(i)}$ are complex $D_{l,1}\times D_{l,2}$ matrices with $D_{1,1}=D_{n,2}=1$ and all other $D_{l,1}=D_{l,2}=D$.
\end{definition}

It is easy to see, that the MPS tomography schemes as introduced in \cite{cramerEfficient2011} can in fact be implemented by quantum statistical queries.
Thus, it holds.

\begin{theorem}[Learning MPS]\label{thm:qsq-mps}
    Let $n,D\in\mb N$ and let $\mc Q$ be the set of matrix product states on $n$-qubits of bond dimension $D$ endowed with the trace distance. 
    Then, $\mc Q$ can be efficiently $\epsilon$-learned from $O(nD^2)$ many $O(\sfrac{\epsilon}{D^2})$-accurate quantum statistical queries.   
\end{theorem}

\begin{proof}
    Note that the sequential scheme of unitary transformations in \cite{cramerEfficient2011} only requires the $\epsilon$-ap\-pro\-ximate knowledge of $\kappa$-local reduced density matrices of unitary transformations of $\rho$, 
    with $\kappa=\ceil{\log(D)}+1$.
    The claim then follows from the fact that any $\kappa$-qubit state can be measured to accuracy $\epsilon$ from $2^{2\kappa}$ many $(\epsilon 2^{-2\kappa})$-accurate statistical queries. 
\end{proof}

\subsubsection{Free Fermions and Free Bosons}\label{QSQfreefermions}
\addsectionheader{QSQfreefermions}

Another example of quantum systems that can be learned in the quantum statistical query framework are free fermions \cite{bravyiFermionic2004,gluzaFidelity2018,swingleFermionic2019,aaronsonNonInteracting2023,nietnerFree2023}.
In short, due to Wick's theorem this class of states can be learned efficiently from knowledge about all $O(n^2)$ many two-point functions, which can be evaluated from quantum statistical queries.

\begin{definition}[Fermionic Gaussian Operators and Pure States]\label{def:gaussian-operators-and-pure-state}
    For $l\in\mb N$ denote for $i\in[l]$ by $f_i^\dagger$ and $f_i$ the fermionic creation and annihilation operators which satisfy the anti-commutation relations 
    \begin{align}
        &\set{f_i,f_j^\dagger}=\delta_{ij}\,,
        &&\text{and}
        &&&\set{f_i,f_j} = \set{f_i^\dagger,f_j^\dagger} =0\,.
    \end{align}
    We denote by $n_i=f^\dagger_if_i$ the number operator of the $i$'th mode and by $n=\sum_in_i$ the total number operator.
    The self adjoint Majorana operators are then defined via
    \begin{align}
        &m_{2i-1}=f_i+f_i^\dagger
        &&\text{and}
        &&&m_{2i}=-i(f_i-f_i^\dagger)\,,
        &&&&\text{and they satisfy}
        &&&&&\set{m_i,m_j}=2\delta{ij}\,.
    \end{align}
    A Hermitian operator $H\in\mb C^{N\times N}$ with $N=2^l$ is called Gaussian (or free) if there exists a matrix $\mbf{A}=-\mbf{A}^T\in\mb R^{2l\times2l}$ such that it holds 
    \begin{align}
        &H = \frac{i}{4} \mbf{m}^T\cdot \mbf{A}\cdot\mbf{m}\,,
        &&\text{with}
        &&&\mbf{m}=(m_1,\dots,m_{2l})\,.
    \end{align}
    Similarly, we call a unitary $U\in\U(N)$ Gaussian if it can be written in terms of a Gaussian Hamiltonian as $U(t)=e^{-itH}$. 
    A pure Gaussian state $\rho=\ketbra\psi$ can then be defined by the state vector
    \begin{align}
        &\ket{\psi} = U\ket{\mbf{n}}\,,
        &&\text{with}
        &&&\ket{\mbf{n}}=\ket{(\mbf{n}_1,\dots,\mbf{n}_{n})} = (f^\dagger_n)^{\mbf{n}_n}\cdots (f^\dagger_1)^{\mbf{n}_1}\ket{0^n}\,,
    \end{align}
    where $U$ is a Gaussian unitary and $n_i\ket{\mbf{n}}=\mbf{n}_i\ket{\mbf{n}}$.
\end{definition}

Then we find.

\begin{theorem}[Learning Pure Gaussian States]\label{thm:learning-pure-gaussian-states}
    Let $l\in\mb N$, $N=2^l$, $\epsilon\in(0,1)$ and denote by $\mc G\subset\mc S(N)$ the class of pure Gaussian states. 
    Then, for every $\tau<\frac{\epsilon}{2l^2}$, the class $\mc G$ can be learned to accuracy $\epsilon$ in fidelity from $l(l-1)$ many queries to $\qstat_\tau$.
\end{theorem}

We introduce the covariance matrix before proving \Cref{thm:learning-pure-gaussian-states} and restate a basic lemma about Gaussian unitaries.

\begin{definition}[Covariance Matrix]\label{def:covariance-matrix}
    Let $\rho$ be an arbitrary fermionic state on $l$ modes and let $m_i$ denote the Majorana operators.  
    The covariance matrix (CM) of $\rho$ is defined as
    \begin{align}
        \mbf{M}(\rho)_{ij} = \frac{i}{2}\tr[\rho\qty[m_i,m_j]]\,.
    \end{align}
    It holds that $\mbf{M}(\rho)$ is real and anti-symmetric.
\end{definition}

\begin{lemma}[From \cite{gluzaFidelity2018}]\label{lem:gaussian-unitaries}
    Let $U(t)$ be a Gaussian unitary generated by $H$ which is defined by $\mbf{A}\in\mb R^{2l\times 2l}$ and let $\mbf{m}=(m_1,\dots,m_{2l})$.
    Then it holds 
    \begin{align}
        &\mbf{m}(t) = \qty(U^\dagger(t)m_1 U(t),\dots,U^\dagger(t)m_{2l}U(t)) 
        = \mbf{Q}(t)\cdot\mbf{m}\,,
        &&\text{with}
        &&&\mbf{Q}(t)=e^{t\mbf{A}}\in\mb{SO}(2l)\,.
    \end{align}
\end{lemma}

Note, that \Cref{lem:gaussian-unitaries} implies that a pure Gaussian state $U\ket{\mbf{n}}$ in the basis, which diagonalizes $U$, has  a block diagonal CM 
\begin{align}
    \mbf{M} = \bigoplus_{i=1}^l
    \begin{pmatrix}
        0 & \lambda_i\\
        -\lambda_i & 0
    \end{pmatrix}
    \eqcolon\bigoplus_{i=1}^{l} B(\lambda_i)
\end{align}
with $\lambda_i=\pm1$ for all $i$.
To see this we note $\sfrac{i}{2}\tr[U\ketbra{\mbf{n}}U^\dagger[m_i,m_j]]=\sfrac{i}{2}\tr[\ketbra{\mbf{n}}[U^\dagger m_i U, U^\dagger m_j U]]$.

\begin{proofof}[\Cref{thm:learning-pure-gaussian-states}]
    We follow the standard approach of learning the CM which is what is also performed in fermionic quantum optics, see for example \cite{gluzaFidelity2018}.

    Let $\rho\in\mc G$ be the unknown state and let $\mbf M$ be the corresponding CM and let $\Gamma_{ij}=\frac12[m_i,m_j]$.
    Querying $\qstat_\tau(\rho)$ with $i\Gamma_{ij}$ for $0\leq j<i\leq l$ we obtain values $v_{ij}$ and thus an estimate $\mbf{M}'=-\mbf{M}'^T$ of $\mbf M$ with 
    \begin{align}
        \mbf M'_{ab} = 
        \begin{cases}
            v_{ij}\,,&j<i\,,\\
            v_{ji}\,,&i<j\,.
        \end{cases}
    \end{align}
    Note that the $\tau$-guarantee translates to a bound in $\ell_\infty$ norm which in turn implies an $\ell_2$-norm bound $\norm{A}_2\leq k\norm{A}_\infty$ for $A$ a $k\times k$ dimensional matrix.

    Since $\rho$ is a pure Gaussian state, $\mbf M$ has eigenvalues in $\pm1$ as argued after \Cref{lem:gaussian-unitaries}. 
    Therefore, we modify $\mbf{M}'$ as follows.
    Let $\mbf{Q}'$ be the matrix that block diagonalizes $\mbf{M}'=\mbf{Q}'^T\cdot\mbf{D}'\cdot\mbf{Q}'$ with $\mbf{D}'=\bigoplus_i B(\lambda_i')$.
    Since $\mbf{M}'$ has dimension $2l$ and is $\tau$-close in $\ell_\infty$-norm to $\mbf M$ it follows that
    each $\lambda_i'$ is at least $2l\tau$-close to $\pm1$. 
    Thus, setting $\widehat{\mbf{M}}=\mbf{Q}'^T\widehat{\mbf{D}}\mbf{Q}'$ with $\widehat{\mbf{D}}$ being $\mbf{D}'$ with the $\lambda_i'$ set to $\pm1$ results in a valid pure Gaussian state CM.

    We now use Equation~(8) from Gluza et al.~\cite{gluzaFidelity2018} to estimate the fidelity between our estimate and the true state
    \begin{align}
        F\qty(\rho,\widehat{\rho})\geq 1-\frac14\tr[\qty(\mbf{M}-\widehat{\mbf{M}})^T\cdot\mbf{M}]
        \geq 1-\frac14 \norm{\mbf{M}-\widehat{\mbf{M}}}_2 \norm{\mbf{M}}_2
        \geq 1-\frac{2l}{4}\norm{\mbf{M}-\widehat{\mbf{M}}}_2\,,
    \end{align}
    in terms of the $\ell_2$-norm (which for matrices is the Frobenius norm) between the true and the observed CM.

    By construction,
    \begin{align}
        \norm{\mbf{M}-\widehat{\mbf{M}}}_2
        \leq\norm{\mbf{M}'-\widehat{\mbf{M}}}_2+\norm{\mbf{M}-\mbf{M}'}_2'
        \leq2l\tau+2l\tau=4l\tau\,.
    \end{align}
    This implies $F(\rho,\widehat\rho)\geq 1-2l^2\tau$ or equivalently $F(\rho,\widehat\rho)\geq1-\epsilon$ for $\epsilon>2l^2\tau$, which proves the claim.
\end{proofof}

\subsection{Single Copy QSQ's}\label{sec:single-copy-qsq}
\addsectionheader{sec:single-copy-qsq}

As a consequence of our formalism, we will now show the extension of the results from \cite{nietnerAveragecaseComplexityLearning2023} from the statistical query to the quantum statistical query setting.
In particular, we derive bounds on learning brickwork random quantum circuit Born distributions in \Cref{sec:single-copy-rqc}. 
Then, we give the corresponding results for Clifford unitaries in \Cref{sec:single-copy-clifford}.

\subsubsection{Random Quantum Circuits: Part I}\label{sec:single-copy-rqc}
\addsectionheader{sec:single-copy-rqc}

To begin, recall the task of $\epsilon$-learning of distributions as analyzed in \cite{nietnerAveragecaseComplexityLearning2023} which is an instance of \Cref{def:epsilon-learning}.

\begin{definition}[$\epsilon$-Leraning of Distributions]\label{def:distribution-epsilon-learning}
    Let $N\in\mb N$, $\epsilon>0$, $\mc D\subseteq\Delta(N)$ be a class of distributions and denote by $\tv$ the total variation distance.
    We define $\epsilon$-\emph{Learning} of distributions with respect to the total variation distance as the learning problem
    \begin{equation}
        \learn(\mc D,\epsilon):\mc D\to\set{B_\epsilon(D)\subset\Delta(N)\midvert D\in\mc D}
        \doublecolon D\mapsto B_\epsilon(D)\,,
    \end{equation}
    with $B_\epsilon(D)$ the open $\epsilon$-ball with respect to $\tv$ in $\Delta(N)$ around $D$.
\end{definition}

In words, the task of $\epsilon$-learning of a class of distributions $\D$ with respect to the total variation distance is, when given access to an unknown $P\in\D$, to learn the description of a distribution $Q$ that is $\tv(P, Q)<\epsilon$ close.
Recall the definition of the  statistical query oracle, which 
maps a class of distributions $\mc D$ to a class of linear functions as in \Cref{ex:distributionsandSQ}.

\begin{definition}[Statistical Query Oracle $\stat$]\label{def:sq}
    Let $N>0$, $\tau>0$ and let $P\in\Delta(N)$ be an $N$-dimensional probability distribution. 
    We denote by $\stat_\tau(P)$ the \emph{Statistical Query} (SQ) oracle of $P$ with tolerance $\tau$. 
    When queried with a function $\phi:[N]\to[-1,1]$ it returns a value $v$, which is promised to lie in the interval
    \begin{align}
        &v\in\qty[P[\phi]-\tau, P[\phi]+\tau]\,,
        &&\text{where}
        &&&P[\phi] = \Ex_{i\sim P}\qty[\phi(i)]\,.
    \end{align}
\end{definition}

As shown in \cite{nietnerAveragecaseComplexityLearning2023} and similarly, by \Cref{cor:linear-eps-learning-complexity}, the query complexity of $\epsilon$-learning as by \Cref{def:distribution-epsilon-learning} from $\stat_\tau$ is lower bounded by 
\begin{equation}\label{eq:sq-lower-bound-eps-learning}
    q 
    \geq \frac{\beta - \Pr_{P\sim\mu}\qty[\tv(P,Q)\leq2\epsilon+\tau]}{\max_\phi\Pr_{P\sim\mu}\qty[\abs{P[\phi]-Q[\phi]}>\tau]}\,.
\end{equation}

Applying this bound to the class of brickwork quantum circuit born distributions with respect to the measure of brickwork random quantum circuits they then show the following.

\begin{mytheorem}[Excerpt from Main Result of \cite{nietnerAveragecaseComplexityLearning2023}]\label{thm:from-avg-case}
    Let $n,d\in\mb N, \epsilon,\beta\in(0,1)$ and $\tau>0$ and let $\A$ be an algorithm that $\epsilon$-learns $\Cbrqc[,n,d]$ from $q$ many queries to $\stat_\tau(P_U)$ with probability at least $\beta$ with respect to $U\sim\mubrqc[,n,d]$. 
    Then:
    \begin{itemize}
        \item \textbf{Linear Depth:} 
        For $d\geq c n$ and $\epsilon\leq\frac{1}{300}-\tau$ it holds
        \begin{equation}
            q+1 \geq \qty(\beta - 3200\cdot 2^{-n})\cdot 2^{n-2}\tau^2\,,
        \end{equation}
        where $c$ is a constant that can be taken to be $c=1.2\cdot10^{20}$.
        \item \textbf{Infinite Depth:} 
        As $d\to\infty$ and for $\epsilon\leq e^{-1}-2^{-\frac{n}{2}-1}-\tau$ setting $\xi = e^{-1}-2^{-\frac{n}{2}-1}-\tau$ it holds

        \begin{equation}
            q+1 \geq \frac{\beta-2\exp\qty(-\frac{2^{n+2} \xi^2}{9 \pi^3})}{2\exp\qty(-\frac{2^n \tau^2}{9\pi^3})}\,.
        \end{equation}
    \end{itemize}  
\end{mytheorem}

We generalize this to the QSQ setting as follows.

\begin{theorem}[Hardness of Learning RQC's from $\qstat$]\label{thm:single-coopy-rqc-hard}
    Let $n,d\in\mb N, \epsilon,\beta\in(0,1]$ and $0<\tau\leq\epsilon$ and let $\A$ be an algorithm that, 
    when given access to $\rho\in\Qbrqc[,n,d]$ learns the corresponding Born distribution $P_\rho\in\Dbrqc[,n,d]$ to precision $\epsilon$ in total variation distance from $q$ many queries to $\qstat_\tau(\rho(U))$
    with probability at least $\beta$ over $U\sim\mubrqc[,n,d]$.
    Then,  
    \begin{itemize}
        \item \textbf{Linear Depth:} 
        For $\epsilon\leq\frac{1}{300}-\tau$ and $d\geq c n$ with $c=2^{67}$ 
        it holds
        \begin{equation}
            q+1 \geq \qty(\beta - 3200\cdot 2^{-n})\cdot \tau^2(2^{n}+1)\,.
        \end{equation}
        \item \textbf{Infinite Depth:} 
        As $d\to\infty$ and for $\epsilon\leq e^{-1}-2^{-\frac{n}{2}-1}-\tau$ setting $\xi = e^{-1}-2^{-\frac{n}{2}-1}-\tau$ it holds
        \begin{equation}
            q+1 \geq \frac{\beta-2\exp\qty(-\frac{2^{n+2} \xi^2}{9 \pi^3})}{2\exp\qty(-\frac{2^n \tau^2}{9\pi^3})}\,.
        \end{equation}
    \end{itemize}  
\end{theorem}

The proof follows from the following technical ingredient.

\begin{lemma}[Non-Trivial Learning of Approximate $2$-Designs]\label{lem:hardness-2-designs}
    For $N\in\mb N$, $\tau\in(0,1)$ and $\epsilon>0$, 
    let $\mu$ be a measure over $\U(N)$ that forms an additive $\epsilon$-approximate $2$-design.
    Then, for every $O\in\mb H^{N\times N}$ with $\norm{O}_{\mrm{op}}\leq1$ it holds 
    \begin{align}\label{eq:lem:hardness-2-design}
        \Var_{U\sim\mu}\qty[\tr[O\rho(U)]] \leq \frac{1}{N+1} +\epsilon+\epsilon^2
    \end{align}
    In particular, for the query complexity of non-trivial  learning $N$-dimensional quantum states it holds 
    \begin{align}
        \qnt[\mu, \tau] \geq
        \begin{cases}
            \tau^2 (N+1)\cdot \qty(1-o(1)) &,\quad \epsilon=o(N^{-1})\,, \\[8pt]
            \tau^2 \epsilon^{-1} \cdot\qty(1-o(1)) &,\quad \epsilon = \omega(N^{-1})\,,\\[8pt]
            \tau^2 \tfrac{N+1}{1+c} \cdot \qty(1+\tfrac{\epsilon}{N+1})^{-1} &, \quad \epsilon=\tfrac{1}{c}N^{-1}\,.
        \end{cases}
    \end{align}
\end{lemma}

\begin{proof}
    We start by bounding
    the Haar variance.
    By \Cref{lem:moments-lemma34} it holds
    \begin{align}
        \Ex_{U\sim\U(N)}\qty[\tr[O\rho(U)]] 
        = \tr[O\smash{\Ex_{U\sim\U(N)}\qty[\rho(U)]}]=\frac{\tr[O]}{N}\,,
    \end{align}
    and
    \begin{align}
        \tr[\qty(O\otimes O)\cdot\Ex_{U\sim\mu}\qty[\rho(U)\otimes\rho(U)]]
        =\frac{\tr[(O\otimes O)\cdot (\id\otimes\id + \mb F)]}{N(N+1)}
        =\frac{\tr[O^2] + \tr[O]^2}{N(N+1)}\,,
    \end{align}
    where we have written the projector onto the symmetric subspace as $P_{\vee}=\frac{1}{2}(\id\otimes\id+\mb F)$ with $\mb F$ the flip operator between the two copies of the Hilbert space.
    Thus, 
    \begin{align}
        \Var_{U\sim\U(N)} = \frac{\tr[O^2]N-\tr[O]^2}{N^2(N+1)} \leq \frac{1}{N+1}\,,
    \end{align}
    where the inequality uses $\tr[O^2]\leq N$ and $-\tr[O]^2\leq0$.

    Combining this with \Cref{lem:additive-design-variance-implication} then yields 
    \begin{align}
        \Var_{U\sim\mu}\qty[\tr[U\rho(U)]] \leq \frac{1}{N+1} + \epsilon+\epsilon^2\,.
    \end{align}

    The claim on $\qnt[\mu,\tau]$ follows from $(1+x)^{-1}\geq 1-x$ for $x\in[0,1]$.
\end{proof}

We are now set to
prove \Cref{thm:single-coopy-rqc-hard}.

\begin{proofof}[\Cref{thm:single-coopy-rqc-hard}]
    Let $N=2^n$.
    The stated problem can be cast as a learning problem
    \begin{equation}\label{eq:learning-born-from-qsq}
        \mc Z:\Qbrqc\to\set{B_\epsilon(P_\rho)\midvert \rho\in\Qbrqc}\doublecolon \rho\mapsto B_\epsilon(P_\rho)\,,
    \end{equation}
    which is equivalent to
    \begin{align}
        \mc Z':\Qbrqc\to\mc P(\mc S(N))\doublecolon\rho\mapsto \set{\sigma\in\mc S(N)\midvert \met_\mrm{diag}(\rho,\sigma)<\epsilon}\,,
    \end{align}
    where we have introduced the pseudo-distance
    \begin{align}
        \met_\mrm{diag}(\rho,\sigma) = \max_{\substack{\norm{O}_\mrm{op}\leq1\\ O\text{ diagonal}}}\frac{1}{2}\abs{\tr[\rho O] - \tr[\sigma O]}\,,
    \end{align}
    with $O\in\mb H^{N\times N}$.

    By \Cref{cor:complexity-restricted-linear-eps-learning}, fixing the reference state $\sigma=N^{-1}\id$, we lower bound the query complexity of learning $\mc Z'$ from $\qstat_\tau$ by 
    \begin{align}
        q+1\geq
        \frac{\beta-\Pr_{U\sim\mubrqc}\qty[\met_\mrm{diag}(\rho(U), \sigma)<2\epsilon+\tau]}{\max_{\norm{O}_\mrm{op}\leq1}\Pr_{U\sim\mubrqc}\qty[\abs{\tr[\rho(U) O]-\tr[\sigma O]}>\tau]} \\[8pt]
        = \qty(\beta-\Pr_{U\sim\mubrqc}\qty[\tv(P_U, \mc U)<2\epsilon+\tau]) \cdot \kfrac\qty(\mubrqc,\sigma,\tau)^{-1} \,.
    \end{align}
    Since $\met_\mrm{diag}(\rho,\sigma)=\tv(P_\rho, P_\sigma)$ we find that the numerator is equal to the corresponding numerator from the proof of \Cref{thm:from-avg-case}, which is bounded by a far from uniform bound.
    In the linear depth regime the far from uniform bound is proven via the $\delta$-approximate 8-design property with $\delta=O(N^{-3})$, implying that $\mubrqc$ forms a $\delta$-apprioximate 2-design.
    Thus, noting that $\mubrqc$ is an exact 1-design, the claim follows by bounding the denominator by \Cref{lem:hardness-2-designs}.

    At infinite depth, it suffices to note that the proof in \cite{nietnerAveragecaseComplexityLearning2023}, which goes via Lemma~22 and Lemma~23 together with Levy's Lemma, is performed for general bounded operators and not just diagonal operators.
    Hence, we can directly carry over the corresponding bounds.
\end{proofof}

\begin{remark}\label{rem:numerator-does-not-matter}
    As used in the previous proof, statistical query problems over output distributions of quantum states with respect to the measure induced by a measure over quantum states on the one hand, and quantum statistical query problems over quantum states with respect to this measure over quantum states and the same identification of solutions on the other hand, can be related. 
    In particular, the triviality as measured by $\mtt{triv}$ of both problems is equivalent.
    More generally, since $q_\mtt{nt}$ lower bounds the query complexity for all problems over the same source set for which $\mtt{triv}$ is sufficiently bounded away from one, any bound on $q_\mtt{nt}$ translates from one non-trivial problem to any other non-trivial problem over the same source set.
    Therefore, in the following, we restrict to learning quantum states, which is more natural in the QSQ setting, without mentioning the respective implications on learning the output distributions. 
\end{remark}

\subsubsection{Clifford Unitaries: Part I}\label{sec:single-copy-clifford}
\addsectionheader{sec:single-copy-clifford}

In the same spirit, we can show hardness of learning stabilizer states. 
Moreover, those results imply hardness of learning the output distribution as defined in \Cref{eq:learning-born-from-qsq} via \cite[Lemma 27]{nietnerAveragecaseComplexityLearning2023}. 

For any $n\in\mb N$ let $\Cl(2,n)$ denote the $n$-qubit Clifford group,
denote by $\mucl[,n]$ the uniform measure over $\Cl(2,n)$
and denote by $\mrm{Stab}(n)=\set{\rho=\ketbra{\psi}\midvert \ket\psi=C\ket0\,,\; C\in\Cl(2,n)}\subset\mc S(2^n)$
the set of $n$-qubit stabilizer states.
Then, we find.

\begin{theorem}[Hardness of Learning Stabilizer States from $\qstat$]\label{thm:single-stabilizer-hard}
    Let $n\in\mb N$ and $\beta\in(0,1]$. 
    Let $\A$ be an algorithm that learns $\mrm{Stab}(n)$ from $q$ queries to $\qstat_\tau$ with probability at least $\beta$ over $U\sim\mucl[,n]$.
    Then, it holds
    \begin{equation}
        q+1 \geq \beta\tau^2 2^{n-2}\,.
    \end{equation}
\end{theorem}

\begin{proof}
    We write the learning problem as 
    \begin{equation}\label{eq:stabilizer-learning-problem}
        \mc{Z}:\mrm{Stab}(n)\to\set{\set{\rho}\midvert \rho\in\mrm{Stab}(n)}\,.
    \end{equation}
    By definition, it holds $\mc Z_\rho=\set{\rho}=\mc Z(\rho)$.
    Thus, since for every pure state $\rho$ one can distinguish $\sigma=2^{-n}\id$ from $\rho$ by simply querying $\rho$, $\tr[\sigma\rho]=2^{-n}\neq1=\tr[\rho^2]$, 
    the problem as in \Cref{eq:stabilizer-learning-problem} is deterministically $1$-verifyable with respect to $\sigma$. 

    We apply \Cref{thm:random-lower-verifyable} with respect to $\sigma$ and obtain
    \begin{equation}\label{eq:bound-stabilizer-learning}
        q+1 \geq  \beta \cdot\kfrac\qty(\mucl,\sigma,\tau)^{-1} = \beta\cdot\kfrac\qty(\mb S^{N-1},\sigma,\tau)^{-1}\,,
    \end{equation}
    where the second step is due to $\mucl$ being an exact $2$-design. 
    \Cref{lem:hardness-2-designs} then yields the claim\footnotemark.

    \footnotetext{Alternatively one can reduce the expression to the diagonal setting.
    The Clifford group forms a unitary $2$-design and hence, the moment operators absorb the diagonalizing unitary for $O$.
    In particular, we can reduce the expression to that with $O$ being restricted to diagonal operators.
    The claim then follows from observing that this reduces the denominator to that of e.g. Eq.~(22) in \cite[Lemma 7]{nietnerAveragecaseComplexityLearning2023} with $\delta=0$.
    There, the corresponding concentration for this expression in the diagonal setting is proven via second moments and assuming a unitary $2$-design.}
\end{proof}

\begin{remark}
    The query complexity of the problem as in \Cref{eq:stabilizer-learning-problem} would remain unchanged when introducing an approximation parameter $\epsilon<\sfrac{1}{\sqrt{2}}$ in trace distance.  
    To see this, note that any pair of stabilizer states $\rho,\sigma$ have overlap $\abs{\braket{\psi}{\phi}}=2^{-\frac{k}{2}}$ with integer $k$.
    This implies by Fuchs-van de Graaf for pure stabilizer states $\rho\neq\sigma$ (see \Cref{eq:trace-dist-fidelity-pure}) that $\norm{\rho-\sigma}_{\tr}\geq\sqrt{2}$. 
\end{remark}

As a consequence of \Cref{thm:single-stabilizer-hard} we find the following corollary.

\begin{corollary}[Hardness of learning $\Qbrqc(\Cl(2,2))$ from $\qstat$]\label{cor:single-copy-clifford-circuits}
   Let $n,d\in\mb N$. Learning $n$-qubit depth $d$ stabilizer states $\Qbrqc[,n,d](\Cl(2,2))$ requires at least $q$ queries to $\qstat_\tau$, with 
   \begin{equation}
        q+1\geq 2^{\sfrac{d}{10}}\tau^2\,.
   \end{equation}
\end{corollary}

\begin{proof}
    Note that the statement is of a worst-case nature.
    In particular, it suffices to prove the query complexity for any measure over $\Qbrqc[,n,d]$ setting $\beta=1$.

    Any $n$-qubit Clifford unitary can be implemented by a depth $9n$ brickwork Clifford circuit \cite{bravyiHadamardfree2021a}.
    The claim thus follows from \Cref{thm:single-stabilizer-hard} using the measure induced by $\mucl[,\floor*{\frac{d}{9}},d]\otimes\mu_{\id}$, where $\mu_\id$ is the point measure on the identity on the remaining $n-\floor*{\frac{d}{9}}$ qubits.
\end{proof}

\subsection{Two-Copy Quantum Statistical Queries}\label{sec:two-copy-qsq}
\addsectionheader{sec:two-copy-qsq}

In the previous section we have seen that the results from \cite{nietnerAveragecaseComplexityLearning2023} do generalize from the setting of statistical query access to $P_\rho$ to the setting of quantum statistical query access to $\rho$.
In this section we will make a further extension to what we call the two-copy QSQ oracle.

\begin{definition}[Two Copy Quantum Statistical Queries $\tqstat$]\label{def:2qsq}
    Let $N>0$, $\tau>0$ and let $\rho\in\mc S(N)\subset\mb H^{N\times N}$ with $0<\rho<\id$ be an $N$ dimensional quantum state. 
    We denote by $\tqstat_\tau(\rho)$ the \emph{Two-Copy Quantum Statistical Query} (2QSQ) oracle of $\rho$ with tolerance $\tau$. 
    When queried with an observable $O\in\mb H^{N^2\times N^2}$ with $\norm{O}_\mrm{op}\leq1$ it returns a value $v$, which is promised to lie in the interval
    \begin{equation}
        v\in\qty[\tr[\rho\otimes\rho \cdot O]-\tau, \tr[\rho\otimes\rho \cdot O]+\tau]\,.
    \end{equation}
\end{definition}

The two-copy QSQ oracle is thus nothing but a QSQ oracle to two copies of the state
\begin{equation}
    \tqstat_\tau(\rho) = \qstat_\tau(\rho\otimes\rho)\,.
\end{equation}
The transition from $\stat(P_\rho)$ to $\qstat(\rho)$ corresponds to the enhanced power of measuring expectation values in any basis versus only in the computational basis.
Likewise, the transition from $\qstat(\rho)$ to $\tqstat(\rho)$ corresponds to the enhanced power of measuring expectation values in (simple) entangled bases, such as the Bell basis, versus non-entangled bases.

We now show how our formalism extends to the two-copy setting.
Let $\mc Q\subset\mc S(N)$ be a set of quantum states and let $\mc Z_1:\mc Q \to \mc P(\mc T)$ be a learning problem over a set of $N$ dimensional quantum states.
Naturally $\mc Z_1$ is equivalent to the learning problem
\begin{align}
    \mc Z_2: \mc Q^{\odot 2}\coloneq\set{\rho\otimes\rho\midvert \rho\in\mc Q}\subset\mc S(N^2) \to \mc Z_1(\mc Q)\subseteq\mc P(\mc T) \,,   
\end{align}
where we have introduced the operation $\odot$
\begin{align}
    \mc Q^{\odot k} = \set{\smash{\underbrace{\rho\otimes\cdots\otimes\rho}_{\times k}}\midvert\rho\in\mc Q}\,.\\[-5pt]
\end{align}
The $\qstat$ oracle maps learning problems over quantum states to linear learning problems.
The same can be said about the $\tqstat$ oracle: it maps quantum learning problems to linear learning problems. 
The linear functions corresponding to the $\qstat$ oracle are functions of the type 
\begin{align}
    f_\rho:\mb H^{N\times N}\to\mb R\doublecolon O\mapsto\tr[\rho O]\,.    
\end{align}
Similarly, the linear functions corresponding to $\tqstat$ are functions of the type
\begin{align}
    f_\rho^{\odot2}=f_{\rho\otimes\rho}:\mb H^{N^2\times N^2}\to\mb R\doublecolon O\mapsto\tr[(\rho\otimes\rho) O]\,.    
\end{align}
It is worth mentioning that problems remain in general non-trivial under $\odot$.
This is due to the tensor product structure.
In particular, 
\begin{align*}
    \norm{\rho\otimes\rho - \Ex_\sigma[\sigma\otimes\sigma]}_{\tr}
    \geq \norm{\tr_2[\rho\otimes\rho - \Ex_\sigma[\sigma\otimes\sigma]]}_{\tr} 
    = \norm{\rho - \Ex_\sigma[\sigma]}_{\tr}\,,
    \intertext{or similarly, for any $O\in\mb H^{N\times N}$}
    \tr[O\otimes\id(\rho\otimes\rho - \Ex_\sigma[\sigma\otimes\sigma])] 
    =\tr[O(\rho - \Ex_\sigma[\sigma])]\,.
\end{align*}

Before we generalize the hardness results from the previous section let us introduce a simple positive result for learning from $\tqstat$.

\begin{theorem}[Purity Testing]\label{thm:purity-test}
    Let $N\in\mb N$, $\delta>0$, let $\mc S_\mrm{pure}\subset\mc S(N)$ be the set of pure states
    and denote by $\mc S_\delta(N)=\set{\rho\in\mc S(N)\midvert \tr[\rho^2]\leq 1-\delta}$ the set of states that are at least $\delta$ mixed.
    Let $\dec(\mc S_\mrm{pure}(N), \mc S_\delta(N))$ denote the problem of deciding whether an unknown $\sigma \in \mc S_\mrm{pure}(N)\cup\mc S_\delta(N)$ is pure, or at least $\delta$ mixed.

    If $\tau<\sfrac\delta2$, then $\dec(\mc S_\mrm{pure}(N), \mc S_\delta(N))$ can be solved from a single query to $\tqstat_\tau$.
\end{theorem}

\begin{proof}
    Recall that for $\mb F$ the flip operator between two copies of Hilbert space it holds
    \begin{equation}
        \tr[(\rho\otimes\rho)\mb F] = \tr[\rho^2]\,.
    \end{equation}
    Thus, measuring the expectation of $\mb F$ to accuracy $\tau<\sfrac\delta2$ suffices to decide between pure and at least $\delta$ mixed.
\end{proof}

\subsubsection{Random Quantum Circuits: Part II}\label{sec:two-copy-rsq}
\addsectionheader{sec:two-copy-rsq}

We start by lifting \Cref{lem:hardness-2-designs} to the 2QSQ setting by means of requiring $4$-designs instead of $2$-designs, which then will imply the corresponding statements on $\Qbrqc$ and $\mubrqc$ from \Cref{def:brickwork-classes,def:rqc}.

\begin{theorem}[Hardness of Learning Unitary $4$-Designs from $\tqstat$]\label{thm:4-design-two-copy-hardness}
    Let $N\in\mb N$ and $\tau,\epsilon\in(0,1]$.
    Let
    $\mu$ be any measure over $\mc S(N)$ that forms an additive $\epsilon$-approximate projective $4$-design.
    Then it holds 
    \begin{align}
        \Var_{\rho\sim\mu}\qty[\tr[(\rho\otimes\rho)O]] = O\qty(N^{-1}) + \epsilon + \epsilon^2\,.
    \end{align}
    In particular, for the query complexity of non-trivial learning from $\tqstat$ it holds 
    \begin{align}
        \qnt[\mu^{\odot2},\tau] \geq 
        \begin{cases}
            \Omega\qty(\tau^2 N) &,\quad \epsilon=O(N^{-1})\\[8pt]
            \Omega\qty(\tau^2\epsilon^{-1}) &,\quad \epsilon=\omega(N^{-1})\,.
        \end{cases}
    \end{align}
\end{theorem}

\begin{proof}
    We first show the claim in case $\epsilon=0$ corresponding to a projective $4$-design.
    Then, for any $O\in\mb H^{N^2\times N^2}$ it holds 
    \begin{align}
        \Var_{\rho\stacksim{4}\mb S^{N-1}} \qty[\tr[(\rho\otimes\rho) O]] 
        =\Var_{\rho\sim\mb S^{N-1}} \qty[\tr[(\rho\otimes\rho) O]] 
        = \qty(M_4(O\otimes O)-M_2(O)^2)\,,
    \end{align}
    where we have introduced the moments 
    \begin{align}
        &M_r(A) = \Ex_{\rho\sim\mb S^{N-1}}\qty[\tr[\rho^{\otimes r}A]]
        =\tr[\smash{\underbrace{\Ex_{\rho}\qty[\rho^{\otimes r}]}_{K^{(r)}_{\mb S^{N-1}}}} A]\,,
        &&\text{where}
        &&&A\in\mb C^{N^r\times N^r}\,.\\
    \end{align}
    By \Cref{lem:moments-lemma34} it holds 
    \begin{align}\label{eq:proof:4design-2qsq-M_2}
        M_2(O) = \binom{N+1}{2}^{-1}\tr[P_\vee O] 
        = \binom{N+1}{2}^{-1}\frac{1}{2}\tr[\qty(\id\otimes\id + \mb F) O]\,, 
    \end{align}
    with $P_\vee$ the projector on the symmetric subspace and $\mb F$ the flip operator between the two copies of the Hilbert space.
    Similarly, 
    \begin{align}\label{eq:proof:4design-2qsq}
        M_4(O\otimes O) = \binom{N+3}{4}^{-1}\tr[P_{\vee^4} O\otimes O]
        = \binom{N+3}{4}^{-1}\frac{1}{4!}\sum_{\varsigma\in S_4}\tr[\Pi(\varsigma) O\otimes O]\,, 
    \end{align}
    with $P_{\vee^4}$ the symmetric projector on the four copies of the Hilbert space and $\Pi(\varsigma)$ the unitary representation on the copies of the Hilbert space corresponding to the permutation $\varsigma$.


    Now define $(n)_k=\frac{n!}{(n-k)!}$ and note that
    $\binom{X}{r}r! = (X)_r$.
    We bound
    \begin{align}
        &\binom{N+3}{4}^{-1}\frac{1}{4!} - \qty(\binom{N+1}{2}^{-1}\frac{1}{2!})^2 
        = (N+3)_4^{-1} - (N+1)_2^{-2}\\[10pt]
        &= \frac{1}{N^4} \qty(
            \qty(\prod_{j=1}^4 \qty(1+2\frac{k}{N}-\frac{jk}{kN}))^{-1} 
            - 
            \qty(\prod_{j=1}^2 \qty(1+2\frac{k}{N}-\frac{jk}{kN}))^{-2}
        )\\[10pt]
        &= -4 N^{-5} + O\qty(N^{-6})\,.\label{eq:4th-moment-estimate}
    \end{align}
    
    Defining $M_4^\otimes$ and $M_4^\times$ via
    \begin{align}
        M_4(O\otimes O)
        = \underbrace{\binom{N+3}{4}^{-1}\frac{1}{4!}\sum_{\varsigma_1,\varsigma_2\in S_2}\tr[\Pi(\varsigma_1) \otimes\Pi(\varsigma_2)O\otimes O]}_{\eqcolon M_4^\otimes(O\otimes O)} 
        \,+\, M_4^\times(O\otimes O)\,,
    \end{align}
    from \Cref{eq:4th-moment-estimate} then follows  
    \begin{equation}
        \abs{M_4^\otimes(O\otimes O)-M_2(O)^2} \leq 4 N^{-1} + O(N^{-2})\,.
    \end{equation}
    
    In order to estimate $M_4^\times(O\otimes O)$ note that for any $\norm{O}_\mrm{op}\leq1$
    it holds
    \begin{equation}
        \abs{M_4^\times(O\otimes O)}\leq M_4^\times(\id^{\otimes 4})\,.
    \end{equation}
    Since $\Var_\rho[\tr[(\rho\otimes\rho) (\id\otimes\id)]]=0$ it must hold 
    \begin{equation}
        \abs{M_4^\times(O\otimes O)}
        \leq M_2(\id\otimes\id)^2 - M_4(\id^{\otimes 4}) \leq 4N^{-1} + O(N^{-2})\,.
    \end{equation}
    
    Thus, we conclude that 
    \begin{equation}
        \Var_{\rho\sim\mb S^{N-1}}\qty[\tr[(\rho\otimes\rho)O]] \leq 8N^{-1} + O(N^{-2}) = O(N^{-1})\,,
    \end{equation}
    proving the claim for $\epsilon=0$.

    The case $\epsilon>0$ then follows
    by \Cref{lem:additive-design-variance-implication}, since
    \begin{align}
        \Var_{\rho\sim\mu} \qty[\tr[(\rho\otimes\rho)O]] 
        \leq \Var_{\rho\sim\mb S^{N-1}} \qty[\tr[(\rho\otimes\rho)O]] + \epsilon+\epsilon^2\,.
    \end{align}

    The claim on $\qnt[\mu^{\odot2}, \tau]$ then follows from expanding $(O(N^{-1}+\epsilon+\epsilon^2))^{-1}$ where the cases differentiate which of the contributions due to $\epsilon$ and $N^{-1}$ is the dominating one.
\end{proof}

This immediately implies the extension of the results from \Cref{sec:single-copy-rqc} to the 2QSQ-setting. 

\begin{corollary}\label{cor:rqc-two-copy-hard}
    Let $n,d\in\mb N, \epsilon,\in(\sfrac12,1],\beta\in(0,1]$ and $0<\tau\leq\epsilon$.
    Let $\A$ be an algorithm that, 
    when given access to $\rho(U)\in\Qbrqc[,n,d]$ learns the corresponding Born distribution $P_U\in\Dbrqc[,n,d]$ to precision $\epsilon$ in total variation distance from $q$ many queries to $\tqstat_\tau[\rho]$
    with probability at least $\beta$ over $U\sim\mubrqc[,n,d]$. 
    Then, for $d\geq c n$ and $\epsilon\leq\frac{1}{300}-\tau$ it holds
    \begin{equation}
        q+1 \geq  \Omega\qty((\beta-b_{c,n})\tau^2 2^n)\,,
    \end{equation}
    with $c=2^{67}$ as in \Cref{thm:single-coopy-rqc-hard}. 
\end{corollary}

\begin{proof}
    The claim follows immediately from combining the far from uniform bound as used in the proof of \Cref{thm:single-coopy-rqc-hard} via \cite[Corollary 11]{nietnerAveragecaseComplexityLearning2023} with \Cref{thm:4-design-two-copy-hardness}.
\end{proof}

\subsubsection{Clifford Unitaries: Part II}\label{sec:two-copy-clifford}
\addsectionheader{sec:two-copy-clifford}

The proofs for random unitary and random Clifford circuits in \Cref{sec:single-copy-qsq} were almost identical.
This is due to the fact that both classes form a unitary $2$-design.
As we have seen in the previous section \Cref{sec:two-copy-rsq}, however, we were using the $4$-design property to show hardness of learning random quantum circuits from $\tqstat$.
Although the Clifford group ``gracefully fails to be a $4$-design'' \cite{zhuCliffordFails2016} we are still able to extend those results to Clifford circuits.

\begin{theorem}[Two-Copy Hardness of Learning Stabilizer States]\label{thm:two-copy-clifford}
    Let $n\in\mb N$ and $\beta\in(0,1]$. 
    Let $\A$ be an algorithm that learns any $\rho(U)\in\mrm{Stab}(n)$ from $q$ queries to $\tqstat_\tau$ with probability at least $\beta$ over $U\sim\Cl(2,n)$.
    Then, it holds
    \begin{equation}
        q+1 \geq \Omega\qty(\beta\tau^2 2^n)\,.
    \end{equation}
\end{theorem}

Note that \Cref{thm:two-copy-clifford} straight forwardly lifts \Cref{cor:single-copy-clifford-circuits} to the 2QSQ setting with $2^{\sfrac{d}{10}}$ replaced by $\Omega(2^{\sfrac{d}{10}})$.

\begin{proof}
    Let $N=2^n$.
    We start by recalling Theorem 3 from \cite{kuengDistinguishing2016a}, which states in Equation~(32) 
    \begin{align}
        K_{\Cl(2,n)}^{(4)}(\ketbra{z})
        =\Ex_{U\sim\Cl(2,n)}\qty[\rho(U)^{\otimes 4}] 
        = N\binom{N+2}{3}^{-1}\qty((\alpha(z)-\beta(z))P_1 + \beta(z)P_{\vee^4})\,,
    \end{align}
    with $z=\ket{0}^{\otimes n}$ the input state, $-2(N(N+1))^{-1}\leq\alpha(z)\leq N^{-1}$, $\alpha(\ket{0}^{\otimes n})=N^{-1}$, 
    \begin{align}
        &\beta(z) = \frac{4(1-\alpha(z))}{(N+4)(N-1)}\,,
        &&\abs{\alpha(z)-\beta(z)}=O(N^{-1})\,,
    \end{align} 
    and $P_1=P_{\vee^4}Q$ with $Q$ a rank $N^2$ projector.
    Rewriting this yields
    \begin{align}
        K_{\Cl(2,n)}^{(4)}\qty(\ketbra{0}^{\otimes n})
        =   \Theta\qty(N^{-3}) P_1 + N\binom{N+2}{3}^{-1}\frac{4(1-N^{-1})}{(N+4)(N-1)} P_{\vee^4}\,.
    \end{align}
    For any bounded $O\in\mb H^{N^2\times N^2}$, $\norm{O}_\mrm{op}\leq1$ we now bound the variance 
    \begin{align}\label{eq:proof:variance-clifford-4th-moment}
        \Var_{U\sim\Cl(2,n)}\qty[\tr[O(\rho(U)\otimes\rho(U))]] 
        = \Ex_{U\sim\Cl(2,n)}\qty[\tr[O(\rho(U)\otimes\rho(U))]^2] - \Ex_{U\sim\Cl(2,n)}\qty[\tr[O(\rho(U)\otimes\rho(U))]]^2\,. 
    \end{align}
    By the $2$-design property, the last term coincides with the corresponding expectation value over the full unitary group $M_2(O)^2$ as in \Cref{eq:proof:4design-2qsq-M_2}. 
    Focussing on the first term then yields
    \begin{align}\label{eq:proof-clifford-4th-moment}
        \Ex_{U\sim\Cl(2,n)}\qty[\tr[O(\rho(U)\otimes\rho(U))]^2]
        = \tr[(O\otimes O) \Ex_{U\sim\Cl(2,n)}\qty[\rho(U)^{\otimes 4}]]\\[10pt]
        =O(N^{-1}) + \underbrace{N\binom{N+2}{3}^{-1}\frac{4(1-N^{-1})}{(N+4)(N-1)}}_{\eqcolon A(N)} \tr[(O\otimes O) P_{\vee^4}]\,,
    \end{align}
    where we have used that $Q$ and hence $P_1=P_{\vee^4} Q$ is a rank $N^2$ projector, which implies $\Theta(N^{-3})\mrm{tr}[(O\otimes O)P_1]= O(N^{-1})$.
    In order to compare \Cref{eq:proof-clifford-4th-moment,eq:proof:4design-2qsq} we rewrite $A(N)$ as
    \begin{align}
        N\binom{N+2}{3}^{-1}\frac{4(1-N^{-1})}{(N+4)(N-1)}
        =  \binom{N+3}{4}^{-1} \cdot\frac{1+3N^{-1}}{1+4N^{-1}}\,.
    \end{align}
    We find
    \begin{align}
        \abs{M_4(O\otimes O) - \Ex_{\Cl(2,n)}[\tr[O(\rho(U)\otimes\rho(U))]^2]} 
        \leq O(N^{-1}) 
        + \abs{M_4(O\otimes O) - A(N)\tr[(O\otimes O)P_{\vee^4}]} \\[10pt]
        = O(N^{-1}) + \abs{1-\frac{1+3N^{-1}}{1+4N^{-1}}} M_4(O\otimes O) = O(N^{-1}) \qty(1 +  M_4(O\otimes O)) = O(N^{-1})\,,
    \end{align}
    where we have used that $\abs{M_4(O\otimes O)}\leq\norm{O}_\mrm{op}^2\leq1$.
    Thus, from \Cref{eq:proof:variance-clifford-4th-moment,thm:4-design-two-copy-hardness} we find
    \begin{align}
        \Var_{U\sim\Cl(2,n)}\qty[\tr[O(\rho(U)\otimes\rho(U))]] 
        = M_4(O\otimes O) - M_2(O)^2 + O(N^{-1})
        = O(N^{-1})\,.
    \end{align}
    The hardness of learning stabilizer states from $\tqstat_\tau$ access is then bounded similar to 
    \Cref{eq:bound-stabilizer-learning} by $q+1\geq\beta \qnt*[\Cl(2,n)^{\odot 2},\tau]$ with
    \begin{align}
        \qnt[\Cl(2,n)^{\odot2},\tau]^{-1}\leq
        \max_{O}\Pr_{U\sim\Cl(2,n)}\qty[\abs{\tr[O(\rho(U)\otimes\rho(U))] - \tr[O\Ex_{U\sim\Cl(2,n)}[\rho(U)\otimes\rho(U)]]}>\tau] \\[10pt]
        \leq \tau^{-2} \Var_{U\sim\Cl(2,n)}\qty[\tr[O(\rho(U)\otimes\rho(U))]] 
        = \tau^{-2} O(N^{-1})\,,
    \end{align}
    with $O\in\mb H^{N^2\times N^2}$ and $\norm{O}_\mrm{op}\leq1$, which proves the claim.
\end{proof}

\subsection{Multi-Copy Quantum Statistical Queries}\label{sec:multicopy-qsq}
\addsectionheader{sec:multicopy-qsq}

Let us now extend the idea from the previous \Cref{sec:two-copy-qsq}.
Similar to the two-copy QSQ oracle $\tqstat$, which allows for entangling always two copies of the state, we define the multi-copy QSQ oracle as allowing for entangling multiple copies of the state per query.
Thus, the multi-copy QSQ oracle will, depending on the allowed amount of entangled copies, naturally interpolate between the statistical setting of QSQ's and the more general quantum learning, tomography or metrology settings.

\begin{definition}[Multi Copy QSQ's]
    Let $N,k\in\mb N$, $K=N^k$, $\tau>0$ and let $\rho\in\mc S(N)\subset\mb H^{N\times N}$ with $0<\rho<\id$ be an $N$ dimensional quantum state. 
    We denote by $\kqstat_\tau(\rho)$ the $k$\emph{-Copy Quantum Statistical Query} (QSQ) oracle of $\rho$ with tolerance $\tau$. 
    When queried with an observable $O\in\mb H^{K\times K}$ with $\norm{O}_\mrm{op}\leq1$ it returns a value $v$, which is promised to lie in the interval
    \begin{equation}
        v\in\qty[\tr[\rho^{\otimes k} \cdot O]-\tau, \tr[\rho^{\otimes k} \cdot O]+\tau]\,.
    \end{equation}
\end{definition}

As in case of the $\tqstat$ oracle, the $\kqstat$ oracle can be rewritten in terms of the $\qstat$ oracle as
\begin{equation}
    \kqstat_\tau(\rho) = \qstat_\tau\qty(\rho^{\otimes k})\,.
\end{equation}
Similarly, for any $k\in\mb N$ we can identify a learning problem over quantum states $\mc Z_1:\mc Q\subseteq\mc S(N)\to\mc P(\mc T)$ with a corresponding learning problem 
\begin{align}
    \mc Z_k:\mc Q^{\odot k}\subset\mc S\qty(K)\to\mc P(\mc T)\,,
\end{align}
with $K=N^k$.
The $\kqstat$ oracle then maps quantum learning problems to linear learning problems and the linear functions corresponding to the $\kqstat$ oracle are of the type 
\begin{align}
    f_\rho:\mb H^{K\times K}\to\mb R\doublecolon O\mapsto \tr[\rho^{\otimes k}O]\,.
\end{align}

It is easy to see from H\"offding bounds that any problem, which can be solved from $q$ queries to $\kqstat_\tau(\rho)$ can, with probability $1-\delta$ be solved from $O(\sfrac{qk}{\tau^2}\log(\sfrac{q}{\delta}))$ copies of the unknown state $\rho$ (this is similar to \Cref{thm:QSQ-to-PAC}). 
The number of copies in fact can be further reduced by the tools for shadow tomography to 
\begin{align*}
    \widetilde O\qty(\frac{k\log(\sfrac{1}{\delta})\cdot\log(q)^4\cdot\log(N)}{\tau^4})\,,
\end{align*}
where $\widetilde O$ hides polynomial factors in $\log\log(q), \log(k\log(N)), \log(\sfrac1\tau)$. See for example \cite{aaronsonShadow2018} and Equation~(B9) in \cite{jerbiQMLShadows2023} to map the observables to two output POVM's.

In reverse, the multi-copy QSQ framework also allows for implications about the sample complexity from entangled measurements by means of multi-copy QSQ results. 
The following lemma makes the previously mentioned notion of ``interpolation'' between statistical and entangled measurements rigorous.

\begin{lemma}[Entangled Measurements to Multi-Copy QSQ reduction]\label{lem:multi-copy-to-qsamp}
    Let $N,k\in\mb N$, $\delta\in[0,\sfrac12)$ and denote by $\mc Q\subseteq\mc S(N)$ a set of $N$ dimensional quantum states.
    Let $\mc Z:\mc Q\to\mc P(\mc T)$ be a learning problem over $\mc Q$ with $\mc T$ the set of solutions.
    Assume that for all $\mc T_i\neq\mc T_j\in\mc Z(\mc S)$ it holds $\mc T_i\cap\mc T_j=\varnothing$.
    If there is a quantum algorithm that, with probability $1-\delta$ learns $\mc Z$ from $k$ copies,
    then, for every $\tau<\sfrac12-\delta$ there exists a statistical algorithm that learns $\mc Z$ from $q$ queries to $\kqstat_\tau$ with 
    \begin{align}
        q = \ceil*{\log(\abs{\mc Z(\mc S)})}\,.
    \end{align}
    In particular, setting $\tau=\sfrac{1}{10}$, if learning $\mc Z$ requires at least $q>\ceil*{\log(\abs{\mc Z(\mc S)})}$ queries to $\kqstat_\tau$ to be solved, then $\mc Z$ cannot be solved with probability $1-\delta\geq\sfrac23$ given $k$ copies from entangled measurements.
\end{lemma}

We note that the disjointedness assumption $\mc T_i\cap\mc T_j=\varnothing$ can in general be enforced without changing the sample complexity of the quantum algorithm, for example by means of an $\epsilon$-net on $\mc T$. 

\begin{proofof}[\Cref{lem:multi-copy-to-qsamp}]
    Set $K=N^k$ and $L=\ceil*{\log(\abs{\mc Z(\mc S)})}$.
    Let $\set{\Pi_t\in\mb H^{K\times K}: t\in\mc T}$ denote the POVM corresponding to the algorithm $\A$ that learns $\mc Z$. 
    Let $\mtt{tree}(\mc Z(\mc S))$ be a tree decomposition of $\mc Z(\mc S)$, that is 
    \begin{align*}
        &\mtt{tree}(\mc Z(\mc S))=\set{\mc P_{l,b}\mc T\subseteq\mc Z(\mc S)\midvert 0\leq l\leq L\,, \;b\in\mb F_2^{l}}\,,\quad\text{with}\\[5pt]
        &\begin{aligned}
            &\mc P_{0,0}\mc T = \mc Z(\mc S)\,,
            &&\mc P_{l,b0}\mc T\cap\mc P_{l,b1}\mc T=\varnothing\,,
            &&&\mc P_{l,b0}\mc T\cup\mc P_{l,b1}\mc T = \mc P_{(l-1),b}\mc T\,,
            &&&&\text{for all $l\in[L]$ and $b\in\mb F_2^{l-1}$}\,,  
        \end{aligned}
    \end{align*}
    where we identify $\mb F_2^0=\set{0}$ and for $b=(b_1,\dots,b_{l-1})$ denoted by $b0=(b_1,\dots,b_{l-1},0)$ and similarly for $b1$.
    Note that each leaf $\mc P_{L,b}$ corresponds to a set in $\mc S(\mc S)$.

    For any pair $(l,bi)$ with $b\in\mb F_2^{l-1}$ and $i=0,1$ define the POVM elements
    \begin{align}
        &\Pi_{l-1,b} = \Pi_{l,b0} + \Pi_{l,b1}\,,
        &&\text{and}
        &&&\Pi_{L,b(s)} =\sum_{t\in \mc Z(s)}\Pi_t\,,
    \end{align}
    where the bit string $b(s)\in\mb F_2^L$ denotes the path in $\mtt{tree}(\mc Z(\mc S))$ that singles out the subset $\mc Z(s)\in\mc Z(\mc S)$.
    The statistical learning algorithm $\A_{\mrm{stat}}$ with access to $\kqstat_\tau$ then proceeds as follows.

    Given access to $\kqstat_\tau(\rho)$ for some unknown $\rho\in\mc Q$, in the first step $l=1$ the algorithm $\A_\mrm{stat}$ queries one of $\Pi_{1,0}$ and $\Pi_{1,1}$ and receives a value $v_{1,i}$ for the corresponding $i$. 
    At any given step $l\leq L$, the algorithm has a value $v_{(l-1),bi}$ with $b\in\mb F_2^{l-2}$ and $i=0$ or $1$.
    Let $v_{(l-1),bi}$ be greater than $\sfrac12$, then $\A_\mrm{stat}$ queries in the $l$'th step one of the observables $\Pi_{l,bi0}$ and $\Pi_{l,bi1}$. Else, if $v_{(l-1),bi}$ is smaller $\sfrac12$ it queries in the $l$'th step one of the observables $\Pi_{l,b\neg i0}$ and $\Pi_{l,b\neg i1}$.
    In the $L+1$'st step the algorithm has obtained one value $v_{L,bi}$.
    Assume without loss of generality  that $v_{L,bi}$ is greater $\sfrac12$ (else $v_{L,b\neg i}>\sfrac12$ will hold) and denote by $\mc Z(s)$ the corresponding solution set. Then $\A_\mrm{stat}$ returns any $t\in\mc Z(s)$.

    Let us now analyze the algorithm.
    By construction, $\A_\mrm{stat}$ makes at most $L$ queries. Thus, it remains to prove the correctness.
    Since for all $t\in\mc T$ it holds $0\leq\Pi_t\leq\id$ and $\sum_{t\in\mc T}\Pi_t=\id$, each query made is a valid query.
    Moreover, at each step 
    \begin{align}
        v_{l,b}
        \geq \tr[\Pi_{l,b}\rho^{\otimes k}] -\tau
        = \qty(\sum_{t\in\mc Z(s)\in\mc P_{l,b}\mc T} \tr[\Pi_t \rho^{\otimes k}]) -\tau\,.
    \end{align}
    Now, assume that in the $l$'th step, the correct solution set $\mc Z(s)\in\mc P_{l,bi}\mc T$.
    Then, by assumption and since all $\tr[\Pi_t\rho^{\otimes k}]\geq0$, we find $v_{l,bi}\geq \sum_{t\in\mc Z(s)}\tr[\Pi_t\rho]-\tau= 1-\delta-\tau>/sfrac12$ and similarly $v_{l,b\neg i}\leq \delta+\tau<\sfrac12$,
    which completes the proof.
\end{proofof}

An example application of \Cref{lem:multi-copy-to-qsamp} is \Cref{thm:purity-test}. 
In the same spirit, we make the following statement.

\begin{corollary}[Stabilizer Testing]\label{cor:stabilizer-testing}
    Let $\mc S_\mrm{pure}(2^n)$ be the set of pure $n$-qubit states and let $\mrm{Stab}(n)\subset\mc S_\mrm{pure}(2^n)$ be the set of $n$-qubit stabilizer states. 
    Further, let $\mc S_\epsilon^\mrm{Stab}(2^n)\subset\mc S_\mrm{pure}(2^n)$ be the set of pure states that are at least $\epsilon$-far from any stabilizer state in the sense that
    \begin{align}
        &\abs{\braket{S}{\psi}}^2\leq1-\epsilon^2 
        &&\text{for all} 
        &&&\ketbra{S}\in\mrm{Stab}(n)\,,\;\ketbra{\psi}\in\mc S_\epsilon^\mrm{Stab}(2^n)\,.
    \end{align}
    Then, for $\tau<\sfrac{\epsilon^2}{8}$ the problem $\dec(\mrm{Stab}(n), \mc S_\epsilon^\mrm{Stab}(2^n))$ of deciding whether an unknown pure state is a stabilizer state, or at least $\epsilon$-far from all stabilizer states, can be solved from a single query to $\mqstat{6}_\tau$.
\end{corollary}

\begin{proof}
    By Theorem~3.3 from \cite{grossSchurWeylDuality2021} their Algorithm~1 (see also their Figure~2 for the circuit description) performs a measurement on $6$ copies of the unknown state $\rho$ and, if $\rho\in\mrm{Stab}(n)$ returns $1$.
    However, if $\rho\in\mc S_\epsilon^\mrm{Stab}(2^n)$ then the algorithm returns $1$ with probability at most $\leq1-\sfrac{\epsilon^2}{4}$. 
    Denote by $\Pi_{0/1}$ the POVM corresponding to their algorithm. Then in other words
    \begin{align}
        \tr[\rho^{\otimes 6}\Pi_1] = 
        \begin{cases}
            1\,, & \rho\in\mrm{Stab}(n)\,,\\
            c_\rho\leq 1-\sfrac{\epsilon^2}{4}\,, & \rho\in\mc S_\epsilon^\mrm{Stab}(2^n)\,.
        \end{cases}
    \end{align}
    Thus, a single query to $\mqstat{6}_\tau(\rho)$ with $\tau<\sfrac{\epsilon^2}{8}$ suffices to decide whether $\rho$ is a stabilizer or at least $\epsilon$-far.
\end{proof}

\subsubsection{Haar Random States}

We start by bounding the complexity of non-trivial learning Haar random states given access to multi-copy quantum statistical queries $\kqstat_\tau(\rho)$ in terms of the copies $k$ and dimension $N$.

\begin{theorem}[Hardness of Learning Random States From $\kqstat$]\label{thm:hardness-haar-random}
    Let $N,k\in\mb N$ and $\tau>0$. 
    Let $\mc S(N)$ be the set of $N$ dimensional quantum states and denote by $U\sim\U(N)$ the Haar measure. 
    Then, the query complexity $q_\mrm{nt}$ of non-trivial learning $\rho(U)\in\mc S(N)$ with respect to $U\sim\U(N)$ is bounded by 
    \begin{align}
        \qnt[{\mb S^{N-1}}^{\odot k}, \tau] 
        \geq \exp\qty(\frac{N\tau^2}{9\pi^3k^2})\,.
    \end{align}
\end{theorem}

\begin{remark}
    \Cref{thm:hardness-haar-random} states that any non-trivial learning problem over Haar random states requires at least doubly exponentially many queries to $\kqstat$ for any $k<\sqrt{N}=2^{\sfrac{n}{2}}$.
    By \Cref{lem:multi-copy-to-qsamp} this directly translates to the setting of quantum samples where one is given access to copies of the unknown state $\rho$.
    In reverse, this implies that any efficient learning algorithm which works on all states must be trivial in the sense that $\triv[\mc Z,\mu]$ is exponentially close to one with respect to Haar random states.
    One example for this is Aaronson's shadow tomography \cite{aaronsonShadow2018}.
\end{remark}

The proof relies on Levy's lemma which can be stated as follows (see for example \cite{brandao_complexity_2021}).

\begin{lemma}[Levy's Lemma]\label{lem:levys-lemma}
    Let $f:\mb S^{N-1}\to\mb R$ be a real function on the complex unit sphere with Lipschitz constant $L$ with respect to the euclidean distance and denote by $z\sim\mb S^{N-1}$ the Haar measure on the sphere. Then it holds
    \begin{align}
        &\Pr_{z\sim\mb S^{N-1}}\qty[\abs{f(z) - \Ex_{w\sim\mb S^{N-1}}\qty[f(w)]}>\tau]
        \leq \exp\qty(-\frac{4N\tau^2}{9\pi^3L^2})
        &&\text{for any } \tau>0\,.
    \end{align}
\end{lemma}

\begin{proofof}[\Cref{thm:hardness-haar-random}]
    For any $\ket{\psi_0}\in\mb S^{N-1}$ the unitary Haar measure $U\sim\U(N)$ induces the spherical Haar measure $U\ket{\psi_0}\sim\mb S^{N-1}$.
    Now, let $K=N^k$ and for an arbitrary $O\in\mb H^{K\times K}$ define the function
    \begin{align}
        f:\mb S^{N-1}\to\mb R\doublecolon \ket{\psi}\mapsto \mel*{\psi^{\otimes k}}{O}{\psi^{\otimes k}}\,.
    \end{align}
    To bound the Lipschitz constant we proceed as in \cite[Lemma 22]{nietnerAveragecaseComplexityLearning2023}.
    First note that
    \begin{align}
        \abs{f(\psi)-f(\varphi)} 
        = \abs{\tr[O\qty(\ketbra{\psi}^{\otimes k} - \ketbra{\varphi}^{\otimes k})]}
        \leq \norm{O}_\mrm{op} \norm{\ketbra{\psi}^{\otimes k} - \ketbra{\varphi}^{\otimes k}}_{\tr}\\ 
        \leq 2 \norm{O}_\mrm{op} \norm{\ket{\psi}^{\otimes k} - \ket{\varphi}^{\otimes k}}_{2}\,,
    \end{align}
    where the last inequality is due to Fuchs-van de Graaf
    \begin{align}
        \frac{1}{2}\norm{\ketbra{\psi}^{\otimes k} - \ketbra{\varphi}^{\otimes k}}_{\tr} 
        = \sqrt{1-\abs{\braket{\psi}{\varphi}}^{2k}} 
        = \sqrt{1+\abs{\braket{\psi}{\varphi}}^{k}}\sqrt{1-\abs{\braket{\psi}{\varphi}}^{k}}\\[10pt]
        \leq \sqrt{2}\sqrt{1-\mrm{Re}(\braket{\psi}{\varphi}^k)} 
        = \norm{\ket\psi^{\otimes k}-\ket\varphi^{\otimes k}}_2\,.
    \end{align}
    Next, observe that 
    \begin{align}
        \ket\psi^{\otimes k}-\ket\varphi^{\otimes k} 
        = \sum_{i=0}^{k-1} \ket\psi^{\otimes i}\otimes\qty(\ket\psi-\ket\varphi)\otimes \ket\varphi^{\otimes (k-i-1)}
    \end{align}
    such that by the triangle inequality it holds
    \begin{align}
        \abs{f(\psi)-f(\varphi)} 
        \leq 2k\norm{O}_\mrm{op}\norm{\ket\psi-\ket\varphi}_2\,.
    \end{align}
    Thus, the claim follows from setting the reference state $\sigma_k=\Ex_U[\rho(U)^{\otimes k}]$ together with Levy's \Cref{lem:levys-lemma}
    \begin{align}
        \qnt[{\mb S^{N-1}}^{\odot k}, \tau] 
        \geq \qty(\max_O\Pr_{U\sim\U(N)}\qty[\abs{\tr[O\qty(\rho(U)^{\otimes k} - \sigma_k)]}>\tau])^{-1}
        \geq \exp\qty(\frac{N\tau^2}{9\pi^3k^2})\,,
    \end{align}
    where the $\max$ is over all $O\in\mb H^{K\times K}$ with $\norm{O}_\mrm{op}\leq1$.
\end{proofof}

\subsubsection{Random Quantum Circuits: Part III}\label{sec:multi-copy-rqc}
\addsectionheader{sec:multi-copy-rqc}

Let us now de-randomize the hardness result for learning Haar random states.

\begin{theorem}[Hardness of Learning $t$-Designs From $\kqstat$]\label{thm:t-design-multi-copy-hardness}
    Let $k,N\in\mb N$, denote by $K=N^k$, let $\tau,\epsilon\in(0,1]$ and let $\mu$ be an additive $\epsilon$-approximate projective $2k$-design.
    Then for every $O\in\mb H^{K\times K}$ with $\norm{O}_\mrm{op}\leq1$ it holds 
    \begin{align}
        \Var_{\rho\stacksim{\epsilon,2k}\mb S^{N-1}}\qty[\tr[\rho^{\otimes k}O]] \leq 8\frac{k^2}{N}+\epsilon+\epsilon^2 + O\qty(\frac{k^4}{N^2})\,.
    \end{align}
    In particular, for the query complexity of non-trivial learning from $\kqstat_\tau$ it holds
    \begin{align}
        \qnt[\mu^{\odot k},\tau] =
        \begin{cases}
            \Omega\qty(\frac{\tau^2 N}{k^2})&,\quad \epsilon=O\qty(\frac{k^2}{N})\\[8pt]
            \frac{\tau^2}{\epsilon}\qty(1 - o\qty(\frac{k^2}{N\epsilon})) &,\quad \epsilon=\omega\qty(\frac{k^2}{N})\,.
        \end{cases}
    \end{align}
\end{theorem}

An immediate consequence of \Cref{thm:t-design-multi-copy-hardness} is the following corollary.

\begin{corollary}\label{cor:mcopy-qsamp-implication-rqc}
    Let $k,n\in\mb N$ with $\log k=o(n)$.
    Then, $\epsilon$-learning depth 
    $d= O(n\ln(k)^5k^6)$ 
    brickwork quantum circuit states from entangled measurements and with success probability at least $1-\delta\geq\sfrac23$ requires, for every $\epsilon\leq\sfrac23$ and for every $n\geq3$ at least $k$ copies.
\end{corollary}

\begin{proof}
    We apply \Cref{thm:t-design-multi-copy-hardness} and show a simplified version of \Cref{lem:multi-copy-to-qsamp}. We refer to \Cref{rem:mcopy-qsamp-remark} for the direct application of \Cref{lem:multi-copy-to-qsamp}.
    
    Let $\mc Q$ be the set of depth $d$ brickwork quantum states on $n$ qubits.
    Note that brickwork random quantum circuits at sufficiently large depth $d$ form a $2k$-design.
    Let $N=2^n$, $K=N^k$ and denote
    \begin{align}
        \sigma_{\vee^k} = \Ex_{\rho\sim\mb S^{N-1}}\qty[\rho^{\otimes k}]=\binom{N+k-1}{k}^{-1} P_{\vee^k}\,.
    \end{align}
    Since every $\rho^{\otimes k}\in\mc S(N)^{\odot k}$ is pure and because
    \begin{align}
        \tr[\qty(\rho^{\otimes k})\sigma_{\vee^k}] = \binom{N+k-1}{k}^{-1}=O(N^{-k})\,,
    \end{align} 
    it holds that $\dec(\mc S(N)^{\odot k}, \sigma_{\vee^k})$ and hence $\dec(\mc Q^{\odot k}, \sigma_{\vee^k})$ is non-trivial.
    
    By \Cref{thm:t-design-multi-copy-hardness}, solving $\dec(\mc Q^{\odot k}, \sigma_{\vee^k})$ requires strictly more than one query to $\kqstat_\tau$ for $\tau\geq\sqrt{2\epsilon}=\frac{1}{\sqrt{2}}$ and $n$ sufficiently large.

    Now, for the sake of contradiction, let $\mc A$ be an algorithm for $\epsilon$-learning from $k-1$ copies and denote by $\set{\Gamma_\rho\midvert \rho\in\mc S(N)}$ the corresponding POVM on $k-1$ copies.
    Define by $M$ the following observable on $k$-copies: First measure $\Gamma$ on $k-1$ copies. 
    Given output $\rho$ measure the projector $\Pi^\mrm{loc}=\ketbra{\psi}$ onto $\rho$'s largest eigenvector on the $k$'th copy. 
    If the measurement returns $1$ return ``$\mc Q$'' and else ``$\set{\sigma_{\vee^k}}$''. 

    When acting on $k-1$ copies of some $\varsigma\in\mc Q$ the measurement $\Gamma$ will measure, with probability at least $1-\delta$ some $\rho$ which is at most $\epsilon$ far in trace distance from the true state $\varsigma$.
    In this case it holds by the Fuchs-van de Graaf \Cref{eq:fuchs-van-de-graaf}
    \begin{align}
        \mel{\psi}{\varsigma}{\psi} =1-\frac14\norm{\varsigma-\ketbra\psi}_{\tr}^2
        \geq 1-\frac{\epsilon^2}{4}\,.
    \end{align}
    Thus, when $M$ acts on $k$ copies of $\varsigma$ we measure $1$ with probability at least $p_\mc Q\geq(1-\delta)(1-(\frac{\epsilon}{2})^2)$. 

    In contrast, since for the partial trace on the first $k-1$ qubits  $\tr_{[k-1]}[\sigma_{\vee^k}]=N^{-1}\id$ is the maximally mixed state, the final measurement will  be $1$ with probability at most $p_\sigma\leq\delta + \frac{1}{N}(1-\delta)$.
    This implies that a single query with $M$ to $\mqstat{k}_\tau$ suffices to distinguish between $\mc Q^{\odot k}$ and $\sigma_{\vee^k}$ provided $p_\mc Q-p_\sigma>2\tau$.
    Setting $\tau=\sfrac{1}{10}$ this contradicts the hardness result from \Cref{thm:t-design-multi-copy-hardness}, such that by \Cref{lem:multi-copy-to-qsamp} no such learning algorithm $\mc A$ can exist.
\end{proof}

\begin{remark}\label{rem:mcopy-qsamp-remark}
    Learning the best approximation with respect to an $\epsilon$-net implies an $\epsilon$-learner.
    Similarly, an $\epsilon$-learner implies a learner with respect to an $\epsilon$-net with no additional queries (which may be computationally unfeasible).
    Thus, we could also directly prove \Cref{cor:mcopy-qsamp-implication-rqc} by applying \Cref{lem:multi-copy-to-qsamp} to a learner with respect to an $\epsilon$-net:
    The $\ceil{\log{\abs{\mc Z(\mc S)}}}$ would then correspond to the size of the net which, by the Solovay-Kitaev theorem would induce a linear overhead for any polynomially sized circuit, which again contradicts the hardness result due to \Cref{thm:t-design-multi-copy-hardness}. 
\end{remark}

\begin{proofof}[\Cref{thm:t-design-multi-copy-hardness}]
    We first consider the case $\epsilon=0$ and proceed analogous to the proof of \Cref{thm:4-design-two-copy-hardness}.
    Let $K=N^k$, assume $\mu$ is a unitary $2k$-design and let $O\in\mb H^{K\times K}$ with $\norm{O}_\mrm{op}\leq1$.
    We write the variance as
    \begin{align}
        \Var_{\rho\sim\mu}\qty[\tr[O\rho(U)^{\otimes k}]]
        = M_{2k}(O\otimes O) - M_k(O)^2
    \end{align}
    where we defined
    \begin{align}
        M_r(A) 
        = \tr[A\cdot\Ex_{\rho\sim\mb S^{N-1}}\qty[\rho(U)^{\otimes r}]] 
        = \tr[A\cdot K_{\mb S^{N-1}}^{(r)}]
        =\binom{N+r-1}{r}^{-1}\tr[A P_{\vee^r}]
        \,,
    \end{align}
    and used \Cref{lem:moments-lemma34} with $P_{\vee^r}$ the projector onto the symmetric subspace with respect to the cyclic group on the copies of Hilbert space.

    We then partition the cyclic group as $S_{2k}=S_k\times S_k\cup S_{2k}\setminus S_k\times S_k=S_\otimes\cup S_\times$ where $S_\otimes$ contains all permutations $\varsigma=(\varsigma_1,\varsigma_2)$ that can be represented as $\Pi_{2k}(\varsigma)=\Pi_k(\varsigma_1)\otimes\Pi_k(\varsigma_2)$, whereas all permutations in $S_\times$ contain crossings between the first and second $k$ copies.
    Then, we rewrite
    \begin{align}
        M_{2k}(A) = \binom{N+2k-1}{2k}^{-1}(2k!)^{-1}\qty(\tr[A\sum_{\varsigma\in S_\otimes}\Pi_{2k}(\varsigma)] + 
        \tr[A\sum_{\varsigma\in S_\times}\Pi_{2k}(\varsigma)])
        \eqcolon M_{2k}^\otimes(A) + M_{2k}^\times(A)\,.
    \end{align}
    We first focus on $M_{2k}^\otimes (O\otimes O)-M_k(O)^2$.
    Denote by $(n)_k=\frac{n!}{(n-k)!}$ 
    and rewrite
    \begin{align}
        \mrm{diff} 
        = \binom{D+2k-1}{2k}\frac{1}{(2k)!} - \qty(\binom{D+k-1}{k}\frac{1}{k!})^2 = (D+2k-1)_{2k}^{-1} - (D+k-1)_k^{-2}\\[10pt]
        = N^{-2k}\qty(\qty(\prod_{i=1}^{2k}\qty(1+2\tfrac{k}{N}-\tfrac{i}{k}\tfrac{k}{N}))^{-1} - \qty(\prod_{i=1}^{ k}\qty(1+\tfrac{k}{N}-\tfrac{i}{k}\tfrac{k}{N}))^{-2})\\[10pt]
        =\frac{1}{N^{2k}}\cdot
        \frac{
            \prod_{i=1}^{ k}\qty(1+\tfrac{k}{N}-\tfrac{i}{k}\tfrac{k}{N})^2 - \prod_{i=1}^{2k}\qty(1+2\tfrac{k}{N}-\tfrac{i}{k}\tfrac{k}{N})
        }{
            \prod_{i=1}^{ k}\qty(1+\tfrac{k}{N}-\tfrac{i}{k}\tfrac{k}{N})^2\prod_{i=1}^{2k}\qty(1+2\tfrac{k}{N}-\tfrac{i}{k}\tfrac{k}{N})
        }\,.
    \end{align}
    For every $i\in[k]$ it holds $1\leq (1+\frac{k}{N}-\frac{i}{k}\frac{k}{N})<1+\frac{k}{N}$. Similarly, for every $i\in[2k]$ it holds $1\leq (1+2\frac{k}{N}-\frac{i}{k}\frac{k}{N})<1+2\frac{k}{N}$.
    Thus, it holds for the denominator
    \begin{align}
        1 \leq 
        \prod_{i=1}^{ k}\qty(1+\tfrac{k}{N}-\tfrac{i}{k}\tfrac{k}{N})^2\prod_{i=1}^{2k}\qty(1+2\tfrac{k}{N}-\tfrac{i}{k}\tfrac{k}{N})\,.
    \end{align}
    In order to bound the numerator first note that $1+\frac{k}{N}-\frac{i}{k}\frac{k}{N} \leq 1+2\frac{k}{N}-\frac{i}{k}\frac{k}{N}$ and thus $\mrm{diff}\leq 0$.
    Then we rewrite 
    \begin{align}
        \prod_{i=1}^k\qty(1+\tfrac{k}{N}-\tfrac{i}{k}\tfrac{k}{N})^2 \geq 1\,,
    \end{align}
    and 
    \begin{align}\label{eq:2k-prefactor-bound}
        \prod_{i=1}^{2k}\qty(1+2\tfrac{k}{N}-\tfrac{i}{k}\tfrac{k}{N})
         \leq \qty(1+2\tfrac{k}{N})^{2k}
         \leq 1 + 4\tfrac{k^2}{N} + O(\tfrac{k^4}{N^2})\,,
    \end{align}
    where we have used $\ln(1+x)\leq x$ and thus
    \begin{align}
        (1+x)^l
        =e^{l\ln(1+x)}
        \leq e^{lx}
        =1+lx+O\qty(l^2x^2)\,.
    \end{align}
    Combining this yields $0 \leq -\mrm{diff} \leq 4\tfrac{k^2}{N^{2k+1}} + O\qty(\tfrac{k^4}{N^{2k+2}})$,
    which implies 
    \begin{align}
        0
        \leq M_k(O)^2 - M_{2k}^\otimes(O\otimes O) 
        = -\mrm{diff}\cdot\tr[(O\otimes O)\cdot(P_{\vee^k}\otimes P_{\vee^k})] 
        \leq 4\tfrac{k^2}{N} + O\qty(\tfrac{k^4}{N^2})\,.
    \end{align}
    
    It remains to bound $M_{2k}^\times(O\otimes O)$.
    Note that for any $\norm{O}_\mrm{op}\leq1$ it holds 
    \begin{align}
        \abs{M_{2k}^\times(O\otimes O)}
        \leq \abs{M_{2k}^\times (\id_{2k})}\,.
    \end{align}
    Since $\Var_\rho\qty[\tr[\rho^{\otimes k}\id_k]]=0$ it thus holds
    \begin{align}
        \abs{M_{2k}^\times(O\otimes O)}
        \leq \abs{M_{2k}^\times (\id_{2k})}
        \leq \abs{M_{2k}^\otimes(\id_{2k}) - M_k(\id_k)^2}
        \leq 4\tfrac{k^2}{N} + O\qty(\tfrac{k^4}{N^2})\,.
    \end{align}
    We conclude the $\epsilon=0$ case with 
    \begin{align}
        \Var_{\rho\sim\mb S^{N-1}}\qty[\tr[\rho^{\otimes k}O]]\leq 8\frac{k^2}{N}+O\qty(\frac{k^4}{N^2})\,.
    \end{align}

    We now consider the case $\epsilon>0$. 
    By \Cref{lem:additive-design-variance-implication}
    we find
    \begin{align}
        \Var_{\rho\stacksim{\epsilon,2k}\mb S^{N-1}}\qty[\tr[O\rho(U)^{\otimes k}]] 
        \leq \Var_{\rho\sim\mb S^{N-1}}\qty[\tr[O\rho(U)^{\otimes k}]] +\epsilon+\epsilon^2\,.
    \end{align}
    
    The claim on $\qnt[\mu^{\odot k},\tau]$ then follows from expanding $(1+x)^{-1}$ in the different regimes.
\end{proofof}

\stopcontents[quantum]

\section{Parametrized Learning Algorithms}\label{sec:variational}
\addsectionfooter{sec:variational}
\addsectionheader{sec:variational}
\startcontents[variational]
\vspace{1cm}
\printcontents[variational]{}{1}{}
\vspace{1cm}

\subsection{Set-Up}\label{sec:parametrized-setup}
\addsectionheader{sec:parametrized-setup}

We now shift our attention to what we refer to as parametrized algorithms.
Let $\Theta$ be a parameter space. 
In general, we assume $\Theta\subset\mb R^m$ compact for some sufficiently large $m$.
Let $l:\Theta\to\mb R$ be a function that is minimized by some (potentially non-unique) $\vartheta_{\min}$\footnotemark{}.
\footnotetext{
    The existence of such a minimizer $\vartheta_{\min}$ is in general already due to the compactness of $\Theta$. 
    For any $l\in\mc L$ there always exists a $\vartheta_{\min}$ that minimizes $l$ because $\Theta$ is compact. 
    Practically however the minimum is often because $l$ is the composition of a convex function and a parametrization map.
    }
We refer to $l$ as a \emph{Loss Function}.
An oracle $\mc O(l)$ that, when queried with some $\vartheta\in\Theta$ returns the value $l(\vartheta)$, potentially up to some tolerance, is referred to as a \emph{Loss Function Oracle}.
The task of the algorithm is then to find an estimate $\hat\vartheta$ that satisfies an optimality criterion.
This optimality criterion is either defined with respect to the \emph{Access Loss Function} $l$,
or with respect to another loss function $l_\mrm{op}:\Theta\to\mb R$ referred to as the \emph{Operational Loss Function}.
The access loss function encodes the access $\A$ is given to the problem instance and the operational loss function encodes the actual task of the learning problem.
The optimality criterion can then be formalized in terms of a function $\mrm{opt}$ as 
\begin{align}\label{eq:optimality-condition}
    \vartheta\quad\text{such that}\quad l_\mrm{op}(\vartheta) \leq \mrm{opt}(\Theta, l_\mrm{op}, \epsilon) 
    \quad\text{with}\quad
    \mrm{opt}(\Theta, l_\mrm{op},\epsilon) = \min_{\phi\in\Theta} l_\mrm{op}(\phi) + \epsilon\,.
\end{align}
A parametrized algorithm $\A$ then is one which can be written as an optimization routine $\A_\mrm{opt}$ that interacts with $\mc O(l)$ and finds a $\hat\vartheta$ that satisfies the optimality condition \Cref{eq:optimality-condition}
\begin{align}
    \qty(\A_\mrm{opt}\underset{\vartheta}{\overset{l(\vartheta)}{\circlearrowleft}} \mc O(l) ) \longrightarrow  \hat\vartheta\quad \text{with}\quad l_\mrm{op}(\hat\vartheta)\leq \mrm{opt}(\Theta, l_\mrm{op}, \epsilon)\,.
\end{align}
If $\Theta$ is continuous and if $\A_\mrm{opt}$ uses a variational optimization routine we refer to $\A$ as a variational algorithm.

In practice, the loss function access often comes with two more properties:
\begin{lossproperties}
    \item \label{item:boundedloss} The loss function is bounded $l:\Theta\to[a,b]\subset\mb R$.
    \item \label{item:impreciselossfunctionoracle}  The loss function oracle $\mc O(l)$, when queried with some $\vartheta\in\Theta$ returns $l(\vartheta)$ up to a tolerance $\widetilde\tau$. 
\end{lossproperties}
\Cref{item:boundedloss} is often simply due to the fact that the numerical accuracy of any practical algorithm is bounded and, as such, the range is bounded to some compact interval to avoid numerical instabilities.
While the numerical inaccuracy also implies some inaccuracy as in \Cref{item:impreciselossfunctionoracle}, the dominant contribution to $\widetilde\tau$ usually comes from statistical errors in the evaluation of the loss function.
Notably, if \Cref{item:boundedloss,item:impreciselossfunctionoracle} are satisfied we can rescale the loss function to the interval $[0,1]$, such that the corresponding oracle $\mc O(l)$ becomes an evaluation oracle $\eval_\tau(l)$ for $l$ with tolerance $\tau=\frac{\widetilde\tau}{b-a}$. 

As such, we can employ the formalism from \Cref{sec:formalism} in order to obtain bounds on the complexity of parametrized algorithms.
We define the corresponding learning problem as follows.

\begin{definition}[Parametrized Learning]\label{def:lossfunctionlearning}
    Let $\epsilon>0$, $m\in\mb N$, $\Theta\subset\mb R^m$ be a compact set referred to as the parameter space and let $\mc L$ be a set of functions $l:\Theta\to[0,1]\subset\mb R$ referred to as the \emph{Set of Loss Functions}.
    We refer to the elements $l\in\mc L$ as \emph{Access Loss Functions}. 
    For each $l\in\mc L$ let $l_\mrm{op}:\Theta\to\mb R$ be the corresponding \emph{Operational Loss Function} and let 
    $\mrm{opt}(\Theta, l_\mrm{op},\epsilon)$ define an \emph{Optimality Criterion} as in \Cref{eq:optimality-condition}.
    Then, \emph{Parametrized Learning} is defined as the learning problem
    \begin{align}
        \mc Z^\epsilon: \mc L\to \mc P(\Theta)\doublecolon l\mapsto\set{\hat\vartheta\in\Theta\midvert l_\mrm{op}\qty(\hat\vartheta)\leq \mrm{opt}(\Theta, l_\mrm{op}, \epsilon)}\,.
    \end{align}
\end{definition}

Note that by the definition of the learning problem we implicitly defined a class of loss functions $\mc L$ with respect to which the success of the algorithm is defined.
This is, here we apply the standard perspective from problem complexity where an algorithm is required to be successful on every problem contained in a class of instances.  

As already argued in the introduction \Cref{sec:intro-parametrized} and further explained in \Cref{sec:derandomize-random-init}, we only consider algorithms for which the optimization is deterministic and where any randomness is due to the evaluation of the loss function and a possible random initialization of the parameters.
By modelling stochastic evaluations by the evaluation oracle with a finite tolerance we assume that the model and its parameters are not fine-tuned to exploit the randomness due to the evaluation as in \cite{abbe_poly-time_2020}.
Moreover, in \Cref{sec:derandomize-random-init} we show how to analyze a random initialization in terms of deterministic bounds.

The deterministic average-case query complexity of parametrized learning $\mc Z^\epsilon$ with respect to a measure $\mu$ over $\mc L$ from $\tau$ accurate queries to $l$ can then be lower bounded by \Cref{thm:deterministicavglowerbound} as
\begin{equation}\label{eq:complexity-lossfunction-learning}
    q\geq
    \frac{\beta-\sup_{\vartheta\in\Theta}\Pr_{l\sim\mu}\qty[l_\mrm{op}(\vartheta)\leq\mrm{opt}(\Theta, l_\mrm{op}, \epsilon)]}{\sup_{\vartheta\in\Theta}\Pr_{l\sim\mu}\qty[\abs{l(\vartheta)-g(\vartheta)}>\tau]}\,,
\end{equation}
where $g$ is any function $g:\Theta\to[0,1]$.

In case $l=l_\mrm{op}$ and $\min_\vartheta l(\vartheta)=0$ for all $l$, there is an interesting connection between the complexity of non-trivial learning and the fraction of trivial learning.
The complexity of non-trivial learning is usually bounded by Chebyshev's inequality, setting $g(\vartheta)=\Ex_l[l(\vartheta)]$ as
\begin{equation}
    \Pr_{l\sim\mu} \qty[\abs{l(\vartheta)-g(\vartheta)}>\tau]
    \leq
    \frac{
        \Var_{l\sim\mu} \qty[l(\vartheta)] 
    }{
        \tau^2
    }
    \,.
\end{equation}
Thus, if this quantity is small for all $\vartheta$, leading to a large complexity of non-trivial learning,
then $l(\vartheta)$ concentrates around its mean $g(\vartheta)=\Ex_l[l(\vartheta)]$.
The fraction of non-trivial learning is therefore small if $g(\vartheta)$ is sufficiently greater $\epsilon$ for all $\vartheta$.
And conversely, if $g(\vartheta)$ is in expectation over $\vartheta$ sufficiently below $\epsilon$, the learning problem becomes average-case trivial.

\begin{corollary}
    Let $\mc Z^\epsilon$ be a parametrized learning problem
    with $l=l_\mrm{op}$ and such that for all $l\in\mc L$ it holds $\min_\vartheta l(\vartheta)=0$. 
    Denote by $l_{\min}=\min_\vartheta\Ex_l[l(\vartheta)]$ and by $\sigma^2=\max_\vartheta\Var_l[l(\vartheta)]$.
    Then, if $l_{\min}>\epsilon$ it holds
    \begin{equation}
        \sup_{\vartheta\in\Theta}\Pr_{l\sim\mu} \qty[l(\vartheta) \leq \epsilon] 
        \leq \sup_{\vartheta\in\Theta}\Pr_{l\sim\mu} \qty[\abs{l(\vartheta)-\Ex_l[l(\vartheta)]} \geq \Ex_l[l(\vartheta)]-\epsilon ]
        \leq
        \frac{\sigma^2}{(l_{\min}-\epsilon)^2}\,.
    \end{equation}
    In particular, the query complexity for learning $\mc Z^\epsilon$ can be lower bounded as 
    \begin{equation}
        q\geq
        \sigma^{-2}\tau\cdot\qty(\beta - \frac{\sigma^2}{(l_{\min}-\epsilon)^2})
        \,.
    \end{equation} 
    Similarly, for $l_{\min}\leq\Ex_l[l(\vartheta)]=o(\epsilon)$ sufficiently small the average-case problem is trivial, and there exists a trivial algorithm with success probability at least
    \begin{equation}
        \sup_{\vartheta\in\Theta}\Pr_{l\sim\mu}\qty[l(\vartheta)<\epsilon] 
        \geq 1 - \frac{\Ex_{l\sim\mu}\qty[l(\vartheta)]}{\epsilon}\,,
    \end{equation}
    with respect to $l\sim\mu$.
\end{corollary}

\subsection{Narrow Gorge and Barren Plateau}\label{sec:narrow-barren}
\addsectionheader{sec:narrow-barren}

In this subsection we will connect the bounds we obtained from applying the formalism from \Cref{sec:formalism} to parametrized learning as defined in \Cref{def:lossfunctionlearning} 
with other complexity witnesses in learning.

Hardness of learning from loss functions is in practice often witnessed in terms of narrow gorges and barren plateaus. 
While each of these phenomena focuses on different aspects of the loss function landscape,
they often come hand in had \cite{Arrasmith_2022}.

Before going into details of narrow gorges and barren plateaus let us explain some strengths and weaknesses of the different approaches.
Our formalism as introduced in \Cref{sec:formalism} has the nice feature of yielding unconditional bounds. This is, the lower-bounds given hold for \emph{any} learning algorithm $\A$, independent of the assumptions on $\A$.
However, in practice it is often either not clear, which set of loss functions $\mc L$ one is actually interested in,
or those sets $\mc L$ for which one can actually derive bounds on the relevant probabilities are not immediately connected to practically relevant sets at best, or way too large to realistically hope for an efficient learning algorithm, at worst.
The phenomena of barren plateau and narrow gorge thus come from a different, more practically motivated intuition.
Those phenomena consider only a single loss function $l$ and analyze how generic certain properties of $l$ are with respect to a probability measure $\vartheta\sim\nu$ over $\Theta$.
As such, the presence of these phenomena does not per se rule out any learning algorithm, but emphasizes the need for careful initialization, updates or novel methods to stabilize the learning process with respect to a given loss function.

Simply put, a loss function admits a narrow gorge if the set of minimizers is concentrated in a subset of (exponentially) vanishing measure. 
As such, the minimizers quite literally live in a ``narrow gorge''.
More formally we define it as follows.

\begin{definition}[Narrow Gorge]
    Let $n$ be a scaling parameter, $l^{(n)}:\Theta^{(n)}\to\mb R$ be a loss function, $\nu^{(n)}$ a measure over $\Theta^{(n)}$ and refer to $\Ex_{\vartheta\sim\nu^{(n)}}[l^{(n)}(\vartheta)]$ as the mean loss.
    We say that $l^{(n)}$ admits a \emph{Narrow Gorge} with respect to $\nu$ if:
    \begin{narrowgorge}
        \item \label{item:narrowgorge-nontrivialloss} The minimal loss $l^{(n)}(\vartheta_{\min})$ is bounded away from the mean loss 
        \begin{equation}
            \Delta(n)=\Ex_{\vartheta\sim\nu^{(n)}}\qty[l^{(n)}(\vartheta)] - l^{(n)}(\vartheta_{\min}) = \Omega\qty(\overpoly)\,.
        \end{equation}
        \item \label{item:narrowgore-lossconcentration} The loss is sharply concentrated around its mean. In particular, for any $\delta>0$ it holds 
        \begin{equation}
            \Pr_{\vartheta\sim\nu^{(n)}}\qty[\abs{l^{(n)}(\vartheta) - \Ex_{\vartheta\sim\nu^{(n)}}\qty[l^{(n)}(\vartheta)]}>\delta]\leq\frac{2^{-\Omega(n)}}{\delta^2}\,.
        \end{equation}
    \end{narrowgorge}
\end{definition}

In particular, the valid solutions $\hat\vartheta$ to $l$ for a sufficiently small $\epsilon=\Omega(\overpoly)$ with $\Delta(n)-\epsilon=\Omega(\overpoly)$ are by \Cref{item:narrowgorge-nontrivialloss} inverse polynomially bounded away from the expected loss. 
Thus, the mass of this solution set with respect to $\nu$ is by \Cref{item:narrowgore-lossconcentration} exponentially small.

The similarity to the bounds as in \Cref{eq:complexity-lossfunction-learning} is immediate. 
\Cref{item:narrowgore-lossconcentration} is a statement about the concentration of the loss function with respect to $\vartheta\sim\nu$ just as the denominator in \Cref{eq:complexity-lossfunction-learning},
which corresponds to the complexity of non-trivial learning,
is a statement about the concentration of the loss function with respect to $l\sim\mu$.
Similarly, \Cref{item:narrowgorge-nontrivialloss} is a statement about the non-triviality of the loss function with respect to $\vartheta$,
just as the numerator in \Cref{eq:complexity-lossfunction-learning} is a statement about the non-triviality of the learning problem with respect to $l$.

\subsubsection{Narrow Gorge and Hardness of Learning}\label{sec:narrow-and-hardness}
\addsectionheader{sec:narrow-and-hardness}

Let us work out this relationship in more detail. 
To begin with, let $n$ be a scaling parameter and $\epsilon=\Omega(\overpoly)$. 
For $l\in\mc L$ let $l_\mrm{reg}(\vartheta)=l(\vartheta)-\min_\phi l(\phi)$ be the regularized loss function.
Consider the case $l_\mrm{op}=l_\mrm{reg}$.
Set $g(\vartheta)=\Ex_{l\sim\mu}[l(\vartheta)]$ and assume that for $\beta=1-\Omega(\overpoly)$ and any $\tau>0$ the obtained lower bound from \Cref{eq:complexity-lossfunction-learning} yields $q=\sfrac{2^{\Omega(n)}}{\tau^2}$.
Since the numerator in \Cref{eq:complexity-lossfunction-learning} is bounded by one, the complexity is due to the complexity of non-trivial learning as characterized by the denominator
\begin{align}\label{eq:numerator-complexity-lfl}
    &\sup_{\vartheta\in\Theta}\Pr_{l\sim\mu}\qty[\abs{l(\vartheta)-\Ex_{l\sim\mu}[l(\vartheta)]}>\tau]\leq \frac{2^{-\Omega(n)}}{\tau^2}\,.
\end{align}
Due to the supremum this holds for any $\vartheta\in\Theta$. 
As such, the expression in \Cref{eq:numerator-complexity-lfl} can be understood as dual to \Cref{item:narrowgore-lossconcentration}.
While \Cref{item:narrowgore-lossconcentration} states that a given loss function $l$ concentrates around its mean, 
\Cref{eq:numerator-complexity-lfl} states that the loss of a given parameter $\vartheta$ concentrates around its mean with respect to the ensemble of loss functions $l\sim\mu$.

Similarly, from $q=\sfrac{2^{\Omega(n)}}{\tau^2}$ and $\beta=1-\Omega(\overpoly)$ we deduce
\begin{align}
    \sup_{\vartheta\in\Theta}\Pr_{l\sim\mu}\qty[l_\mrm{reg}(\vartheta)\leq \epsilon]=1-\Omega\qty(\overpoly)\,.
\end{align}
This implies that  for any $\vartheta\in\Theta$ there is a subset of loss functions $\mc L^\vartheta$ of size at least $\Omega(\overpoly)$ with respect to $\mu$ such that for any $l\in\mc L^\vartheta$ the loss is not within optimality by $\vartheta$. 
The previous equation implies
\begin{align}
    &\Ex_{l\sim\mu}\qty[l_\mrm{reg}(\vartheta)] \geq  \epsilon\Omega\qty(\overpoly)
    \intertext{and hence}
    &\Ex_{l\sim\mu}\qty[l_\mrm{reg}(\vartheta)] - \min_l l_\mrm{reg}(\vartheta) = \epsilon\Omega\qty(\overpoly) = \Omega\qty(\overpoly)\,.\label{eq:dual-NG1}
\end{align}
Thus, \Cref{eq:dual-NG1} is dual to \Cref{item:narrowgorge-nontrivialloss}. 
In particular, where \Cref{item:narrowgorge-nontrivialloss} states that the mean loss for fixed $l$ is bounded away from the minimal loss with respect to $l$, \Cref{eq:dual-NG1} states that the mean regularized loss of $\vartheta$ under $l\sim\mu$ is bounded away from the minimal regularized loss of $\vartheta$.
We summarize this as follows.

\begin{corollary}[Hardness of Learning Implies Dual Narrow Gorge]\label{cor:hardness-and-narrowgorge}
    Let $n$ be a scaling parameter,
    $\epsilon=\Omega(\overpoly)$, $\tau>0$ and let $\mc Z^\epsilon$ be as in \Cref{def:lossfunctionlearning} with $l_\mrm{op}=l_\mrm{reg}=l-\min_\phi l(\phi)$ for all $l\in\mc L$ and assume that $\mc Z^\epsilon$ is a surjective learning problem.
    Denote by $\widehat{\mc Z}^\epsilon$ the dual learning problem and identify any $\vartheta\in\Theta$ with the \emph{Dual Loss Function}
    \begin{equation}
        \vartheta^{\star\star}:\mc L\to[0,1]\doublecolon l\mapsto \vartheta^{\star\star}(l)=l(\vartheta)\,.
    \end{equation}
    We can interpret $\widehat{\mc Z}^\epsilon$ as a parametrized learning problem with loss function set $\mc L^\star=\Theta^{\star\star}=\{\vartheta^{\star\star}\mid \vartheta\in\Theta\}\simeq \Theta$.
    Let $\mu$ be a measure over $\mc L$. 
    If the query complexity for learning $\mc Z^\epsilon$ with probability $\beta=1-\Omega(\overpoly)$ over $\mu$ from $\tau$ accurate evaluation queries is, by \Cref{eq:complexity-lossfunction-learning} lower bounded by $q=\sfrac{2^{\Omega(n)}}{\tau^2}$,
    then every dual loss function $\vartheta^{\star\star}$ in the dual parametrized learning problem $\widehat{\mc Z}^\epsilon$ satisfies \Cref{item:narrowgore-lossconcentration} and an $\Omega(\overpoly)$ fraction of dual loss functions admits a narrow gorge.
\end{corollary}

\subsubsection{Barren Plateau and Hardness of Learning}\label{sec:barren-and-hardness}
\addsectionheader{sec:barren-and-hardness}

Let us now turn towards barren plateaus.
The phenomenon of a barren plateau is phrased in a similarly scenic language as the narrow gorge.
A barren plateau refers to the phenomenon when the gradient of the loss function is concentrated around zero.
This is much in spirit with the vanishing gradient problem and the shattered gradient problem in deep learning \cite{ShatteredGradientsProblem2018}. 
If the loss landscape is dominated by a flat plateau, or more generally multiple flat plateaus optimization is hard for any gradient based algorithm.
Due to the flatness any such algorithm, once inside such a plateau will in general not be able to exit the plateau without a more violent intervention.
We use the following definition to formalize barren plateaus.

\begin{definition}[Barren Plateau]
    Let $n$ be a scaling parameter, $\vartheta\in\Theta^{(n)}$ be a vector of $m=O(\poly(n))$ continuous variables, $l^{(n)}:\Theta^{(n)}\to\mb R$ be a loss function, let $\nu^{(n)}$ be a measure over $\Theta^{(n)}$ and for $i\in[m]$ denote by $\partial_i l$ the partial derivative of $l$ with respect to the $i$'th component of $\vartheta$.
    We say that $l^{(n)}$ admits a \emph{Barren Plateau} with respect to $\nu$ if:
    \begin{barrenplateau}
        \item \label{item:barrenplateau} Any partial derivative of the loss function is concentrated around zero. In particular, for any $i\in[m]$ and any $\tau>0$ it holds
        \begin{equation}
            \Pr_{\vartheta\sim\nu}\qty[\abs{\partial_i l(\vartheta)}>\tau] = \frac{2^{-\Omega(n)}}{\tau^2}\,.
        \end{equation}
    \end{barrenplateau}
\end{definition}

Interestingly, it was shown in \cite{Arrasmith_2022} that for most variational quantum-classical algorithms the two phenomena, the barren plateau and the narrow gorge, do coincide.
In particular, they have shown that \Cref{item:narrowgore-lossconcentration} implies \Cref{item:barrenplateau} and vice versa.
While their proof used slightly different definitions of the two phenomena the logic, applied to our definitions, is identical.
Moreover, we would like to point out that their result can be easily extended to an even broader class of loss functions, such as QAOA
by applying the results from \cite{wierichsGeneralParametershiftRules2022} in the proof of \cite[Lemma 2]{Arrasmith_2022}.
As such, we can directly re-use the discussion from \Cref{sec:narrow-and-hardness} and arrive at the following observation.

\begin{observation}[Hardness of Learning Implies a Barren Plateau]\label{obs:hardness-implies-barren}
    Let $n, \epsilon, \mc Z^{\epsilon}$ and $\widehat{\mc Z}^{\epsilon}$ be as in \Cref{cor:hardness-and-narrowgorge} and make the same identification between $\vartheta$ and $\vartheta^{\star\star}$.
    Let $\mu$ be a measure over $\mc L$ and assume that, for any $\tau>0$, the query complexity of non-trivial learning $\mc Z^\epsilon$ with respect to $\mu$ is lower bounded by the inverse of \Cref{eq:numerator-complexity-lfl}.
    If all $l\in\mc L$ can be written as a (sum of) quantum loss function, for which the gradient can be evaluated by the generalized parameter shift rule from polynomially many circuit evaluations as in \cite{wierichsGeneralParametershiftRules2022},
    then every dual loss function $\vartheta^{\star\star}$ admits a barren plateau.
\end{observation}

\subsubsection{Self Learning}\label{sec:selflearningdef}
\addsectionheader{sec:selflearningdef}

The previous sections have shed light on the connection between the hardness bounds as introduced in this work, 
and practically relevant phenomena such as the barren plateau and the narrow gorge.
In particular, we found that hardness due to our bounds often directly implies a narrow gorge in many instances and, for most quantum loss functions, a barren plateau
in any instance of the dual problem.
This motivates the study of learning problems that are self-dual as defined in \Cref{def:duallearningproblem}. 
A natural example of this is what we define as \emph{Self-Learning}.

\begin{definition}[Self-Learning]\label{def:self-learning}
    Let $\mc M$ be a class of objects endowed with a metric $\met_\mc M$.
    The radius of $\mc M$ with respect to $\met_\mc M$ is given by
    \begin{align}
        &R= \frac{1}{2}\max_{o,w\in\mc M}\met_\mc M(o,w)\,,
        &&\text{and we assume that}
        &&&L:[0,2R] \to[0,1]
    \end{align}
    is a monotonous function.
    For $m\in\mb N$ let $\Theta\subset\mb R^m$ and let 
    \begin{align}
        \Pi:\Theta\to\mc M\doublecolon\vartheta\mapsto\Pi(\vartheta)=o_\vartheta
    \end{align}
    be a surjective map referred to as the parametrization of $\mc M$.
    To every $\phi\in\Theta$ we assign the loss function 
    \begin{align}
        l^\phi:\Theta\to[0,1]\doublecolon\vartheta\mapsto l^\phi(\vartheta)=L\circ\met_\mc M\qty(\Pi(\vartheta),\Pi(\phi))\,.
    \end{align}
    By
    \begin{align}
        \mc L=\mc L(\Theta,\Pi,L,\met_\mc M) = \set{l^\phi\midvert \phi\in\Theta}
    \end{align} 
    we reffer to the corresponding set of loss functions. 
    The parametrized learning problem $\mc Z^\epsilon$ with parameter space $\Theta$ corresponding to $\mc L$ with $l_\mrm{op}=l$ as defined in \Cref{def:lossfunctionlearning} is then referred to as \emph{Self-Learning}.        
\end{definition}

It is immediate to see that self-learning is self-dual:
To see this, note that, by the symmetry of the metric $\met_\mc M$, for any $\vartheta\in\Theta$ it holds $\mc Z^\epsilon(\vartheta)=\mc Z^\epsilon_\vartheta$.
In particular,
\begin{align}
    \mc Z^\epsilon(\vartheta) 
    =\set{\phi\in\Theta\midvert L\qty(\met_\mc M(o_\phi, o_\vartheta))\leq\epsilon}  
    =\set{\phi\in\Theta\midvert L\qty(\met_\mc M(o_\vartheta, o_\phi))\leq\epsilon}
    =\mc Z^\epsilon_\vartheta\,.
\end{align}
Thus, the dual learning problem $\widehat{\mc Z}^\epsilon$ is, in fact, equal to $\mc Z^\epsilon$.
This implies, by \Cref{sec:narrow-and-hardness,sec:barren-and-hardness},
that the bound as in \Cref{eq:complexity-lossfunction-learning} has direct implications on the existence of a narrow gorge and often also a barren plateau in every instance of $\mc L$.

\subsection{Examples}\label{sec:parametrized-examples}
\addsectionheader{sec:parametrized-examples}

In this section we will exemplify our formalism when applied to parametrized learning problems.

\subsubsection{Working with Random Initialization}\label{sec:derandomize-random-init}
\addsectionheader{sec:derandomize-random-init}

As already stated, the results considered where all in terms of deterministic algorithms. 
While most practically used routines for the parameter optimization are deterministic, randomness often is still present in terms of random initialization and in terms of stochastic evaluation of the loss function.
We assume that common algorithms are not fine-tuned to be able to exploit the stochastic evaluation in the sense of \cite{abbe_poly-time_2020}.
To be precise, we assume that the tolerance in the evaluation oracle sufficiently models stochastic evaluations.
On the other hand, as we will show in this section, random initialization only has a mild effect on our bounds.

We model a random initialization as follows. 
First, we assume that the optimizer is independent of the random initialization. That is, the optimizer does depend on the random initial value only through the results to the queries made.
For a loss function class $\mc L$ with elements $l:\Phi\times\Theta\to[0,1]$ we assume that $\phi\sim\nu$ is a random assignment for the first variable $\phi\in\Phi$.
A measure $\mu$ on $\mc L$ hence induces a measure 
$\mu^\phi$ on $\mc L^\phi = \set{l(\phi,\cdotspace)\midvert l\in\mc L}$.
If an algorithm, that initializes $\phi\sim\nu$ can solve $\mc L^\phi$ query efficiently with probability $\Omega(\overpoly)$ over $\nu$, then the algorithm can be boosted to solve it with overwhelming probability by running $r=O(n)$ independent runs with polynomially many queries and returning the best $\vartheta$ obtained from these runs.
In order to show hardness one therefore needs to show that the probability of successfully solving $\mc L^\phi$ with respect to $\nu$ is negligible.

The same holds for the average-case setting with respect to a measure $\mu$ over $\mc L$.
In order to show hardness of learning a $\beta$ fraction of $\mc L$ it suffices to show that the probability of learning a $\beta$ fraction of $\mc L^\phi$ with respect to $\mu^\phi$ is negligible with respect to $\nu$.
In particular, for every polynomial $q$ it needs to hold
\begin{align}\label{eq:random-init-first-bound}
    \Pr_{\phi\sim\nu}\qty[\qty(\beta-\triv[\mc Z_\phi^\epsilon,\tau])\qnt[\mu^\phi,\tau]\leq q]=\mtt{neg}(n)
\end{align}
where $n$ is the scaling parameter, $\mtt{neg}(n)$ is a negligible function in $n$ and $\mc Z_\phi^\epsilon$ denotes the parametrized learning problem induced by $\mc L^\phi$ for some $\epsilon>0$.

We first make the following assumption.
\begin{assumption}\label{assump:random-init-phi-triv-independent}
    For every $l\in\mc L$ it holds that for every $\phi,\phi'\in\Phi$ and every $\vartheta\in\Theta$ there exists a $\vartheta'\in\Theta$ such that $l(\phi,\vartheta)=l(\phi',\vartheta')$.
\end{assumption}
This is a natural assumption in many settings.
It holds for example if $\Theta=\Phi=[0,2\pi)^m$ and $l:\Phi\times\Theta\to[0,1]$ is defined by $l(\phi,\vartheta)=l'(\phi+\vartheta\mod2\pi)$ with $l':\Theta\to[0,1]$.
\Cref{assump:random-init-phi-triv-independent} implies that
\begin{align}
    \triv[\mc Z_\phi^\epsilon]=\triv[\mc Z_{\phi'}^\epsilon]=\triv[\mc Z^\epsilon]    
\end{align} 
is independent of $\phi$.
\Cref{eq:random-init-first-bound} can then be rewritten as
\begin{align}
    \Pr_{\phi\sim\nu}\qty[\qnt[\mu^\phi,\tau]\leq q \qty(\beta-\triv[\mc Z^\epsilon,\tau])^{-1}]=\mtt{neg}(n)\,.
\end{align}
One way to bound this probability is as follows. 
Using the variance bound on $\qnt[\mu^\phi,\tau]$ from \Cref{cor:variance-bound-qnt}
together with Markov's inequality we then find that it suffices to show
\begin{align}
    q\frac{\Ex_{\phi\sim\nu}\qty[\max_\vartheta\Var_{l\sim\mu}\qty[l(\phi,\vartheta)]]}{\tau^{2} \qty(\beta-\triv[\mc Z^\epsilon,\tau])}
    \leq\mtt{neg}(n)\,,
\end{align}
in order to prove \Cref{eq:random-init-first-bound}.

We now extend this analysis to the general case, in which \Cref{assump:random-init-phi-triv-independent} does not necessarily hold.
This is the case if, for example, the random initialization fixes a subset of parameters which effectively only allows the algorithm to discover a strict subset of parameters which depends on $\phi\sim\nu$.
We make the weaker assumption:
\begin{assumption}\label{assump:random-init-triv-unlikely}
    Let $t$ and $\delta\in(0,1)$, then it holds that 
    \begin{align}
        \Pr_{\phi\sim\nu}\qty[\triv[\mc Z_\phi^\epsilon]>t]\leq \delta\,.
    \end{align}
\end{assumption}
This intuitively states that the trivial fraction is, with probability at least $1-\delta$ over $\phi\sim\nu$ is upper bounded by $t$.

Under \Cref{assump:random-init-triv-unlikely} we can almost repeat the previous reasoning.
The left-hand side (LHS) of \Cref{eq:random-init-first-bound} can be upper-bounded by
\begin{align}
    \text{LHS}
    \leq(1-\delta)\Pr_{\phi\sim\nu}\qty[\qty(\beta-t)\qnt[\mu^\phi,\tau]\leq q] + \delta
    \leq \frac{q(1-\delta)}{\beta-t}\Ex_{\phi\sim\nu}\qty[\max_\vartheta\Var_{l\sim\nu}\qty[l(\phi,\vartheta)]] + \delta\,.
\end{align}
Thus, hardness is again implied for a sufficiently small expected variance and $\beta>t$, $\delta\leq\mtt{neg}(n)$.

\begin{envbox}[opacityfill=.11]{Red}\begin{remark}\label{rem:random-init}
    To summarize this we remark, that the query complexity of deterministic algorithms with random initializations crucially depends on the expected variance
    \begin{align}
        \Ex_{\phi\sim\nu}\qty[\max_\vartheta\Var_{l\sim\mu}\qty[l(\phi,\vartheta)]]
        = \max_\vartheta
        \Ex_{l\sim\mu}\qty[\Ex_{\phi\sim\nu}\qty[l(\phi,\vartheta)^2]] - \Ex_{l,l'\sim\mu}\qty[\Ex_{\phi\sim\nu}[l(\phi,\vartheta)l'(\phi,\vartheta)]]\,,
    \end{align}
    and the ``randomized fraction of trivial learning'' which, for $t\in(0,1)$ is given by
    \begin{align}
        \mtt{rtriv}\qty(\mc Z^\epsilon,\mu,\nu,t)
        =\Pr_{\phi\sim\nu}\qty[\triv[\mc Z^\epsilon_\phi,\mu^\phi]>t]\,.
    \end{align}
    In particular, a simple bound on the query complexity for average-case learning by deterministic algorithms with random initialization and $\mtt{rtriv}$ sufficiently bounded away from $\beta$ is given by  
    \begin{align}
        q=\Omega\qty(\tau^2 \min_{\vartheta, l, l'}\qty(\Ex_{\phi\sim\nu}\qty[l(\phi,\vartheta)(l(\phi-\vartheta) - l'(\phi-\vartheta))]^{-1})) \,.
    \end{align}
\end{remark}\end{envbox}

\paragraph[Example]{Example:} (Random Initialized VQEs) Let us consider the following popular example related to the typical incarnation of barren plateaus \cite{mccleanBarren2018}. 
In variational quantum eigensolvers (VQE) the aim is to find a state $\rho(\vartheta)$ parametrized by $\vartheta\in\Theta=[0,2\pi)^m$ that minimizes the energy of a Hamiltonian.
We write $\rho(\phi,\vartheta)=\rho(\phi+\vartheta\mod2\pi)$ and assume that $\rho(\phi,\vartheta)$ forms a $2$-deisng for $\phi\sim\nu$ and for all $\vartheta$.
Note, that \Cref{assump:random-init-phi-triv-independent} holds.
Let $n\in\mb N$, $N=2^n$ and $\mc H\subset\mb C^{N\times N}$ be a class of traceless bounded Hamiltonians $\norm{H}_\mrm{op}\leq c$.
The corresponding set of loss functions consists of functions 
\begin{align}
    l_H:\Theta\times\Theta\to[0,1]\doublecolon(\phi,\vartheta)\mapsto \frac1c\tr[\rho(\phi,\vartheta)H]+\frac12\,.
\end{align}
In order to lower bound the query complexity with respect to random initialization $\phi\sim\nu$ and some measure $\mu$ over $\mc H$ we upper-bound for every $\vartheta\in\Theta$
\begin{align}
    \Ex_{\phi\sim\nu}\qty[\Var_{H\sim\mu}\qty[l(\phi,\vartheta)-\frac12]]
    = \frac{1}{c^2}\Ex_{\phi\sim\nu}\qty[
        \Ex_{H\sim\mu}\qty[
            \tr[\rho(\phi,\vartheta)H]^2
        ]
        -
        \Ex_{H\sim\mu}\qty[
            \tr[\rho(\phi,\vartheta)H]
        ]^2
    ]\\[8pt]
    \leq \frac{1}{c^2}\Ex_{H\sim\mu}\qty[
        \Ex_{\phi\sim\nu}\qty[
            \tr[\rho(\phi,\vartheta)H]^2
        ]
    ]
    = \frac{1}{c^2}\Ex_{H\sim\mu}\qty[
        \frac{\tr[H]^2+\tr[H^2]}{N(N+1)}
    ]
    \leq \frac{1}{N+1}\,,
\end{align}
where the last inequality is due to the tracelessness and the boundedness of $H$.

\begin{remark}\label{rem:random-init2}
    Since the above derivation did not impose any assumption on $\mc H$ but tracelessness, it follows immediately that learning must be hard in this model.
    To see this, note that learning $\mc H=\set{H_1,H_2}$ for traceless Hamiltonians $H_{1/2}$ is exponentially hard, even if $H_{1/2}$ are very simple Hamiltonians with a non-zero trivial fraction.
    The latter is implied for example if $H_1$ and $H_2$ have well separated ground states and which can be approximated sufficiently well by the parametrization.
    This type of hardness result is the typical implication due to barren plateaus and narrow gorges: Optimization with a random initialization due to the measure measuring the barren plateau must be hard.    
\end{remark}

\subsubsection{Learning Basis States and Pauli Operators}\label{sec:basis-states}
\addsectionheader{sec:basis-states}

As a warm-up we now show that self-learning computational basis states is hard. 
To this end, let
\begin{equation}
    \mc B= \mc B^{(n)} = \set{\rho \midvert \rho=\ketbra z\,,\; z\in\mb F_2^n}\,,
\end{equation}
be the set of computational basis states and let $\mc B$ be endowed with the trace distance $\met_{\tr}(\rho,\sigma)=\frac12\norm{\rho-\sigma}_{\tr}$.
Let $\mc Z$ be the corresponding self-learning problem for $\mc B$ with $L$ and $\Pi$ defined as 
\begin{align}\label{eq:L-learning-basis-states}
    &L:[0,1]\to[0,1]\doublecolon x\mapsto 1-\sqrt{1-x^2}\,,
    &&\text{and}
    &&&\Pi:\mb F_2^n\to\mc B\doublecolon z\mapsto\ketbra z\,.
\end{align}

\begin{theorem}[Self-Learning Basis States]\label{thm:self-learning-basis}
    Let $n\in\mb N$, $\epsilon\in[0,1),\beta>0, \tau\in[0,1]$ and let $\mu$ be the uniform measure over $\mb F_2^n$ denoted as $z\sim\mb F_2^n$. 
    Then, self-learning a $\beta$-fraction of $\mc B$ with respect to $\met_{\tr}$ and $L$ as in \Cref{eq:L-learning-basis-states} 
    requires at least $(\beta-2^{-n})2^n$ many queries to $\eval_\tau(l^z)$ with $l^z=L\circ\met_{\tr}(\cdotspace, \ketbra z)\in\mc L$ the unknown loss function. 
    In particular, every loss function $l^z$, with $z\in\mb F_2^n$ admits a narrow gorge.
\end{theorem}

\begin{remark}\label{rem:basis-state-hardness}
    The problem of learning basis states from loss function queries is equivalent to the problem of finding the ground state of the Hamiltonian $H=\id-\ketbra{z}$ from ``energy queries'' $E(y)=\mel{y}{H}{y}=1-\delta_{xy}$. 
    This problem is about as trivial as it can be when given a proper description of the problem in terms of the bit string $z$.
    However, when accessing the problem instance only through a poorly chosen loss function, the problem renders information theoretically as complex as it can be.
    As such, this basic theorem highlights the care that has to be taken in the choice of loss function.
\end{remark}

\begin{proofof}[\Cref{thm:self-learning-basis}]
    We begin with the following observation.
    By the Fuchs-van de Graaf equality  it holds for any $\rho=\ketbra{\psi}{\psi}, \sigma=\ketbra{\phi}{\phi}$ in terms of the fidelity $F(\rho,\sigma)=\tr[\sqrt{\sqrt{\rho}\sigma\sqrt{\rho}}]^2$
    \begin{equation}
        \met_{\tr}(\rho,\sigma) = \sqrt{1-F(\rho,\sigma)} = \sqrt{1-\abs{\braket{\psi}{\phi}}^2}\,.
    \end{equation}
    Thus, since $\mc B$ only contains pure state we find 
    \begin{equation}
        L\circ\met_{\tr}(\rho,\sigma) = 1 - \abs{\braket{\psi}{\phi}}\,.
    \end{equation}

    We now analyze the probability of trivial learning $\triv[\mc Z^\epsilon,\mu]$.
    For every $z, y\in\mb F_2^n$ it holds $\braket{z}{y}=\delta_{zy}$. 
    Hence, for every $0\leq\epsilon<1$ and every $y\in\mb F_2^n$ we find 
    \begin{equation}
        \triv[\mc Z^\epsilon,\mu]=\sup_y\Pr_{z\sim\mb F_2^n}\qty[l^z(y)<\epsilon]=
        \Pr_{z\sim\mb F_2^n}\qty[\smash{\underbrace{\abs{\braket{z}{y}}}_{=\delta_{zy}}}>1-\epsilon] = \frac{1}{2^n}\,.
    \end{equation}

    Next, we analyze the complexity of non-trivial learning $\qnt[\mu,\tau]$.
    For every $y\in\mb F_2^n$ and $\tau\in[0,1]$ it holds similar as above that
    \begin{align}
        \qnt[\mu,\tau]^{-1}=\max_x\Pr_{z\sim\mb F_2^n}\qty[\abs{\braket{z}{x}}>\tau]=\frac{1}{2^n}\,,
    \end{align}
    where we have set $g(z)=0$ for all $z$.
    
    Thus, learning a $\beta$ fraction of $\mc B$ with respect to the uniform measure requires by \Cref{eq:complexity-lossfunction-learning} at least 
    \begin{align}
        q \geq (\beta - \triv[\mc Z^\epsilon,\mu])\qnt[\mu,\tau] = (\beta - 2^{-n})2^{n}
    \end{align}
    queries.

    The remainder follows due to \Cref{cor:hardness-and-narrowgorge}.
\end{proofof}

Leaving the realm of self-learning we can also extend the prior result to more general access to the loss function.

\begin{corollary}[Learning Basis States]
    Let $\epsilon\in(0,1)$ with $1-\epsilon=\Omega(1)$, $\tau>0$, $\beta=2^{-o(n)}$, $\Theta\subset\mb R^m$ for some $m>0$, let $\mc Q\subset\mc S(2^n)$ be a set of $n$-qubit quantum states and let $\Pi:\Theta\to\mc Q$ be a parametrization of $\mc Q$.
    Let $L$ as before and denote by $\mc Z^\epsilon$ the parametrized learning problem over 
    \begin{equation}
        \mc L=\mc L^{(n)} = 
        \set{l:\Theta\to[0,1]\doublecolon \vartheta\mapsto l(\vartheta)\midvert l(\vartheta)=l^z(\vartheta)=L\circ\met_{\tr}(\Pi(\vartheta), \ketbra z)}\,,
    \end{equation}
    with $l_\mrm{op}=l$.
    Then, learning $\mc Z^\epsilon$ requires at least $\tau^2 2^{\Omega(n)}$ many $\tau$-accurate evaluation queries of the unknown loss function $l^z$.
\end{corollary}

\begin{proof}
    The proof is similar to that of \Cref{thm:self-learning-basis}.
    Note that for any $\vartheta$
    \begin{equation}
        \Pr_{z\sim\mb F_2^n}\qty[\abs{\braket{\psi(\vartheta)}{z}}>\tau]
        \leq
        \frac{\Var_{z\sim\mb F_2^n}[\braket{\psi(\vartheta)}{z}]}{\tau^2}
        \leq
        \frac{\Ex_{z\sim\mb F_2^n}[\braket{\psi(\vartheta)}{z}\braket{z}{\psi(\vartheta)}]}{\tau^2}\,.
    \end{equation}
    The second moment can then be bounded as
    \begin{align}
        \Ex_{z\sim\mb F_2^n}\qty[\braket{\psi(\vartheta)}{z}\braket{z}{\psi(\vartheta)}] =  
        \sum_{x,y,z}\frac{\overline{c}_xc_y}{2^n} \delta_{xz}\delta_{zy} =
        \sum_{z}\frac{\abs{c_z}^2}{2^n} = \frac{1}{2^n}\,,
    \end{align}
    which yields $\Pr_z[\abs{\braket{\psi(\vartheta)}{x}}>\tau]\leq 2^{-n}\tau^{-2}$.

    In fact, we can use the same bound for the fraction of non-trivial learning to obtain 
    \begin{equation}
        \Pr_{z\sim \mb F_2^n}[\abs{\braket{\psi(\vartheta)}{x}}>1-\epsilon]\leq 
        \frac{2^{-n}}{(1-\epsilon)^{2}}\,.
    \end{equation}

    Thus, at least
    \begin{equation}
        q\geq 2^n\tau^2\qty(\beta - \frac{2^{-n}}{1-\epsilon})
        = \tau^2 2^{\Omega(n)}
    \end{equation}
    queries to $\eval_\tau(l^z)$ are necessary to learn the unknown $z$.
\end{proof}

Note that we can identify the basis state projector in terms of the Pauli $Z$ operator as 
\begin{align}
    \ketbra{z}=\bigotimes_i\ketbra{z_i}=\frac{1}{2^n}\bigotimes_{i=1}^n(\id+(-1)^{z_i}Z)\,.
\end{align} 
This motivates another reformulation of the above results in terms of Pauli operators which is tightly related to Cerezo et al.'s work on barren plateaus and global cost functions \cite{cerezoCostFunction2021a}.

\begin{theorem}[Hardness of Learning Global Cost Functions]\label{thm:globalcostfunctionandhardness}
    Let $n\in\mb N$, denote by $\mc G$ the set of length $n$ Pauli strings.
    For any parametrization of $n$-qubit quantum states $\ket{\psi}$ and every $P\in\mc G$ let $l^P(\psi)=\frac12(\ev{P}{\psi}+1)$ be the associated loss function, $\mc L=\set{l^P\midvert P\in\mc G}$ and denote by $\mc Z^\epsilon$ the corresponding parametrized learning problem with $l=l_\mrm{op}$.
    We assume that for every $P\in\mc G$ there exists a state in the parametrization such that $\ev{P}{\psi}=-\Omega(\overpoly)$
    Then, learning any $\beta$ fraction of $\mc Z^\epsilon$ with respect to the uniform distribution over $\mc G$ requires at least $q\geq\beta\tau^22^{n+1}$ many queries.
\end{theorem}

Note that the assumption on the parametrization is automatically fulfilled if the parametrization has an $\Omega(\overpoly)$ fidelity with the states $\ket{z}$ with $z\in\{0,1,+,-, i,-i\}^{\times n}$.

\begin{proof}
    For any $x=(a,b)\in\mb F_2^{2n}$ represent the corresponding Pauli string in its Weyl representation
    \begin{align}
        W_x=i^{a\cdot b}X^{a_1}Z^{b_1}\otimes\cdots\otimes X^{a_n}Z^{b_n}\,.
    \end{align}
    Note, that 
    \begin{align}
        \ketbra{\psi} = 2^{-n}\sum_{x\in\mb F_2^{2n}} \ev{W_x}{\psi} W_x\,,
    \end{align}
    and hence 
    \begin{align}
        1 = \tr[\ket{\psi}\!\!\braket{\psi}\!\!\bra{\psi}] = 2^{-2n} \sum_{x,y\in\mb F_2^{2n}} \ev{W_x}{\psi}\!\!\ev{W_y}{\psi} \tr[W_x W_y] = 2^{-n} \sum_{x\in\mb F_2^{2n}} \ev{W_x}{\psi}^2\,.
    \end{align}
    This implies
    \begin{align}
        \Var_{P\sim\mc G}\qty[\ev{P}{\psi}] \leq \Ex_{P\sim\mc G}\qty[\ev{P}{\psi}^2] 
        = \Ex_{x\sim\mb F_2^{2n}}\qty[\ev{W_x}{\psi}^2] 
        = 2^{-2n}\sum_{x\in\mb F_2^{2n}} \ev{W_x}{\psi}^2 
        = 2^{-n}\,.
    \end{align}
    In particular, for every $\ket\psi$ the loss $l^P(\psi)=\frac12(\ev{P}{\psi}+1)$ concentrates exponentially close to $\frac12$ for $P\sim\mc G$ with $\Var_P[l^P(\psi)]\leq 2^{n+1}$.
    This directly implies that $\qnt[\mc G,\tau]\geq\tau^{2}2^{n}$. 
    Moreover, it follows that $\triv[\mc Z^\epsilon,\mc G]=O(2^{-n})$ for every $\epsilon=\Omega(\overpoly)$ by the assumption implies that the parametrization contains for every $P$ a state with loss inverse polynomially smaller than $\frac12$.
    Thus, the claim follows.
\end{proof}

\begin{remark}\label{rem:regardingPauliLearning}
    Thinking about $\ev{P}{\psi}=\tr[P\rho]=\lrangle{P,\rho}_{HS}$ in terms of the Hilbert Schmidt inner product, noting that the Pauli strings form a basis for Hermitian operators, makes the connection to learning basis states even more apparent.
    The hardness of learning basis states holds for computational basis states and even any exponentially sized subset thereof.
    In contrast, for Pauli strings, it is crucial to consider $\id, Z$ as well as the non-diagonal $X$ and $Y$ operators to show hardness of learning. 
    This can be easily seen by observing that any $\id, Z$ string can be directly read out from computational basis states.
    Similarly, below, we show the less obvious insight that Pauli strings composed of $Z$ and $X$ operators only can be learned efficiently,
    which implies that non-commutativity of the local Pauli variables alone is no sufficient condition for hardness of learning. 
\end{remark}

\begin{theorem}[Learning $ZX$-Pauli Strings]\label{thm:learningZXStrings}
   Let $\mc G_{ZX}=\set{Z,X}^{\otimes n}$ be the set containing all length $n$ strings composed of $Z$ and $X$ operators only and let $\ket{\psi}$ be a parametrization that contains the set of stabilizer states.
   Then, there is an algorithm for learning an unknown $P\in\mc G_{ZX}$ from $n+1$ evaluation queries to $l^P(\psi)=\frac12(\ev{P}{\psi}+1)$ with tolerance $\tau\leq\frac15$.
\end{theorem}

\begin{proof}
    For any $x\in\mb F_2^n$ denote 
    \begin{align}
        P_x = A_1\otimes A_2\otimes\cdots\otimes A_n\in\mc G_{ZX}\quad\text{with}\quad
        A_i=
        \begin{cases}
            Z,& x_i=0\,,\\
            X,& x_i=1\,.
        \end{cases}
    \end{align}
    To begin with, let $\ket{S}$ be the stabilizer state corresponding to the generating set (omitting identity tensor factors)
    \begin{align}
        S=\set{Z_1\otimes Z_2\otimes\cdots\otimes Z_n, Y_1\otimes Y_2,\dots, Y_{n-1}\otimes Y_n}\,.
    \end{align}
    We then find that 
    \begin{align}
        \abs{\ev{P_x}{S}} = 
        \begin{cases}
            1, & \abs{x}=0\,,\\
            0, & \abs{x}=1\,,
        \end{cases}
    \end{align}
    where $\abs{x}$ denotes the parity of $x$.
    We now define the generating sets 
    \begin{align*}
        S_i^e &= \set{Z_1\otimes Z_2\otimes\cdots\otimes Z_n, Y_1\otimes Y_2,\dots, Y_{i-1}\otimes X_i, X_i\otimes Y_{i+1}, Y_{i+1}\otimes Y_{i+2},\dots ,Y_{n-1}\otimes Y_n}\,,\\[5pt]
        S_i^o &= \set{Z_1\otimes\cdots\otimes Z_i\otimes X_{i+1}\otimes Z_{i+2}\otimes\cdots\otimes Z_n, Y_1\otimes Y_2,\dots, Y_{i-1}\otimes X_i, X_i\otimes Y_{i+1}, Y_{i+1}\otimes Y_{i+2},\dots ,Y_{n-1}\otimes Y_n}\,.
    \end{align*}
    Using $[X\otimes Y,X\otimes A]\neq[X\otimes Y,Z\otimes A]=0$ for $A\in Z,X$, we find for $\abs{x}=0$ that it holds
    \begin{align}
        \abs{\ev{P_x}{S_i^e}} = 
        \begin{cases}
            0 , & x_i=1\,,\\
            1 , & x_i=0\,.
        \end{cases}
    \end{align}
    Similarly, for $\abs{x}=1$ we find 
    \begin{align}
        \abs{\ev{P_x}{S_i^o}} = 
        \begin{cases}
            0 , & x_i=1\,,\\
            1 , & x_i=0\,.
        \end{cases}
    \end{align}
    Thus, the algorithm first queries $l^P(S)$ and then $l^P(S_i^a)$ for $i\in[n]$ and with $a=e,o$ depending on the output of the first query.
    The tolerance is sufficiently fine because each exact evaluation is either $0, \frac12$ or $1$, such that at worst we need to distinguish the case $0.3\leq v\leq0.7$ from $v\leq0.2\lor 0.8\leq v$. 
\end{proof}

\subsubsection{Parametrized Quantum States}\label{sec:PQCs}
\addsectionheader{sec:PQCs}

Let us now use the formalism from this work to derive a simple proof of hardness of self-learning for parametrized quantum states.
While worst-case hardness is implied by hardness of learning basis states, this section shows average-case hardness. 
More, since parametrized quantum circuits are a continuous class that usually admits a parameter shift rule, hardness of self-learning is directly connected to the existence of a barren plateau.

\begin{definition}[Quantum Circuit Architecture]\label{def:quantum-circuit-architecture}
    Let $k\in\mb N$. 
    A $k$-local \emph{Quantum Circuit Architecture} (QCA)
    is a family of maps that, for every system size $n$ and depth $d$ maps from a parameter space $\Theta$ to the space of $n$-qubit unitaries that can be decomposed as a depth $d$ quantum circuits composed of $k$-local gates.
    In particular, the QCA is defined in terms of, for every $n$ and $d$, a parametrization map  
    \begin{equation}
        U:\Theta\to\mrm{U}\qty(2^n)\,.
    \end{equation}
\end{definition}

Using the \Cref{sec:narrow-and-hardness,sec:barren-and-hardness} we then reproduce as a corollary the main result of \cite{mccleanBarren2018}. 
For any $n,k$ and $d\in\mb N$ and $k$-local quantum circuit architecture with parametrization $U:\Theta\to \U(n)$ as defined in \Cref{def:quantum-circuit-architecture} let
\begin{equation}
    \mc Q = \mc Q_\QCA^{(n)} = \set{\rho\midvert \rho=\ketbra{ \psi(\vartheta)}{ \psi(\vartheta)}\,,\; \ket \psi(\vartheta)=U(\vartheta)\ket0\,,\; \vartheta\in\Theta}
\end{equation}
be the set of $n$-qubit states with respect to the architecture $\QCA$.
We endow $\mc Q$ with the trace distance $\met_{\tr}(\rho,\sigma)=\frac12\norm{\rho-\sigma}_{\tr}$ and
denote by $\ket{\psi(\cdotspace)}:\Theta\to\mc Q$ the parametrization on $\mc Q$ induced by $U$.   
Then, let $\mc Z^\epsilon$ be the corresponding self-learning problem with $L$ defined as 
\begin{equation}
    L:[0,1]\to[0,1]\doublecolon x\mapsto 1-\sqrt{1-x^2}\,.
\end{equation}
Then it holds.

\begin{theorem}
    Let $\epsilon\in(0,1)$ with $1-\epsilon=\Omega(\overpoly)$, $\tau>0$, $\beta=2^{-o(n)}$, $\mu$ be a measure over $\Theta$ and assume that $\mu$ and $ U$ are such that $\ket{\psi(\vartheta)}$ with $\vartheta\sim\mu$ forms a complex spherical $(1,1)$-design. 
    Then, self-learning a $\beta$ fraction of $\mc Q$ with respect to $\met_{\tr}$ and $L$ requires at least $\Omega(\tau^2 2^n)$ many $\tau$-accurate evaluation queries.
    In particular, for an $\Omega(\overpoly)$ fraction of $\phi\in\Theta$ with respect to $\mu$ it holds that the loss function $l^\phi=L\circ\met_{\tr}( U'(\cdotspace),  U'(\phi))$ does admit a narrow gorge.
    Further, if $\QCA$ admits an efficient parameter shift rule, then $l^\phi$ admits a barren plateau for every $\phi\in\Theta$.
\end{theorem}

As the proof will use $(1,0)$- and $(1,1)$-moments only, we remark that the above claim immediately carries over to additive $\delta$-approxiamte spherical $(1,1)$-designs when $\delta=O(2^{-n})$.
Moreover, we remark that most common quantum circuit architectures form a $(1,1)$-design already at depth $d=1$.

\begin{proof}
    Recall the first observation in the proof of \Cref{thm:self-learning-basis}
    \begin{equation}
        L\circ\met_{\tr}(\ketbra\psi,\ketbra\phi) = 1-\abs{\braket{\psi}{\phi}} 
    \end{equation}

    As such, by \Cref{eq:complexity-lossfunction-learning,cor:hardness-and-narrowgorge,obs:hardness-implies-barren} it suffices to find good upper bounds for these two quantities:
    \begin{itemize}
        \item The fraction of trivial learning
        \begin{equation}
            \max_{\vartheta}\Pr_{\phi\sim\mu}\qty[\abs{\braket{\psi(\phi)}{\psi(\vartheta)}}>1-\epsilon]\,.
        \end{equation}
        \item The complexity of non-trivial learning
        \begin{equation}
            \max_{\vartheta}\Pr_{\phi\sim\mu}\qty[\abs{\braket{\psi(\phi)}{\psi(\vartheta)}}>\tau]\,,
        \end{equation}
        where we have set $g=1$.
    \end{itemize}

    Note that, for the first claim, it suffices to prove 
    \begin{equation}\label{eq:proof-PQC-FOM}
        \max_{\vartheta}\Pr_{\phi\sim\mu}\qty[\abs{\braket{\psi(\phi)}{\psi(\vartheta)}}>\tau]\leq\frac{2^{-n}}{\tau^2}\,,
    \end{equation}
    since \Cref{eq:proof-PQC-FOM} directly implies
    \begin{equation}
        \max_{\vartheta}\Pr_{\phi\sim\mu}\qty[\abs{\braket{\psi(\phi)}{\psi(\vartheta)}}>1-\epsilon]\leq \frac{2^{-n}}{(1-\epsilon)^2}\,,
    \end{equation}
    which then, by \Cref{eq:complexity-lossfunction-learning}, yields the first claim
    \begin{equation}
        q\geq
        2^{n}\tau^2\qty(2^{-o(n)} - \frac{2^{-n}}{(1-\epsilon)^2}) = \Omega\qty(\tau^2 2^n)\,.
    \end{equation}

    We now prove \Cref{eq:proof-PQC-FOM}. Let $\vartheta\in\Theta$ arbitrary. 
    Then, by Chebyshev's inequality
    \begin{equation}\label{eq:proof-PQC-var}
        \Pr_{\phi\sim\mu}\qty[\abs{\braket{\psi(\phi)}{\psi(\vartheta)}}>\tau]\leq\frac{\Var_{\phi\sim\mu}[\braket{\psi(\phi)}{\psi(\vartheta)}]}{\tau^2}=\frac{1}{\tau^2}\qty(M_2-\abs{M_1}^2)\,,
    \end{equation}
    where we introduce the first and second moments
    \begin{align}
        &M_1= \Ex_{\phi\sim\mu}[\braket{\psi(\phi)}{\psi(\vartheta)}]  
        &&\text{and} 
        &&&M_2 = \Ex_{\phi\sim\mu}[\braket{\psi(\phi)}{\psi(\vartheta)}\braket{\psi(\vartheta)}{\psi(\phi)}]\,.
    \end{align}

    By the $(1,1)$-design assumption we can easily bound these using the corresponding Haar moments 
    (see \Cref{def:spherical-design} and text below)
    and obtain 
    \begin{align}
        &M_1 = \Ex_{\psi\sim\mb S^{n-1}} \qty[\bra{\psi}]\ket{\chi} = 0\,,
        &&M_2 = \bra{\chi}\Ex_{\psi\sim\mb S^{n-1}}\qty[\ketbra{\psi}]\ket{\chi} = \frac{1}{2^n} \braket{\chi} =\frac{1}{2^n}\,, 
    \end{align}
    where we have set $\ket\chi=\ket{\psi(\vartheta)}$.
    We plug this into \Cref{eq:proof-PQC-var} and obtain
    \begin{equation}
        \Pr_{\phi\sim\mu}\qty[\abs{\braket{\psi(\phi)}{\psi(\vartheta)}}>\tau]\leq \frac{1}{\tau^2 2^n}\,.
    \end{equation}

    The claim on the narrow gorge and the barren plateau then follows immediately from \Cref{cor:hardness-and-narrowgorge,obs:hardness-implies-barren}.
\end{proof}

\subsubsection{Linear and Data Re-Uploading QML Models}\label{sec:data-re-up}
\addsectionheader{sec:data-re-up}

Let us now turn towards self-learning of variational quantum machine learning (vQML) models.
We will consider a very generic architecture that contains most current vQML models, in particular data re-uploading models.
For a detailed discussion of current vQML models, we refer to e.g. \cite{jerbiQML2022}.

Let $n,D,k,m\in\mb N$, $k>1$, let $\ket{\psi_0}=\bigotimes_{i=1}^n\ket{\psi_{0i}}$ be an $n$-qubit tensor product state and $O$ an $n$-qubit observable with operator norm $\norm{O}_\mrm{op}\leq1$, $\Theta\subset\mb R^{m\cdot D\cdot \floor{\sfrac{n}{k}}}$ a compact set and let $\mc X$ be some set. 
$\mc X$ deals as the data set, so for example $\mc X=\mb F_2^b$ for some $b$ if we consider functions over length $b$ bit strings, 
or, $\mc X\subseteq \mb R$ in case of real functions.
Consider the vQML architecture represented by the following function class
\begin{equation}\label{eq:def-vqml-architecture}
    \mc F = \mc F_{O,U,W,\psi_0}^{(n,D)}= \set{f:\mc X\to[-1,1]\doublecolon x\mapsto f(x) \midvert f=f_\vartheta\,,\; \vartheta\in\Theta}\,,
\end{equation}
where for any $\vartheta\in\Theta$ the function $f_\vartheta$ is defined as 
\begin{equation}
    f_\vartheta(x) = \mel**{\psi_0\;}{
        \qty(\prod_{l=1}^D U_l(\vartheta_l)^\dagger W_l(x)^\dagger) O \qty(\prod_{l=D}^1  W_l(x)U_l(\vartheta_l))
    }{\;\psi_0}\,.
\end{equation}
Here, each $U_l(\vartheta_l)$ is one layer of $k$-local unitaries 
\begin{equation}
    U_l(\vartheta_l) = \bigotimes_{i=1}^{\floor{\sfrac{n}{k}}} U_{li}(\vartheta_{li})\,,
\end{equation}
where each $\vartheta_{li}\in\mb R^m$.
Moreover, each $W_l(x)$ is an encoding layer, which we assume to consist of on-site unitaries
\begin{equation}
    W_l(x) = \bigotimes_{i=1}^n W_{li}(x)\,.
\end{equation}
Note that, depending on the choice $U_{li}$ and the $W_{li}$,
most linear, as well as data re-uploading models can be phrased in terms of this architecture.

For a data generating distribution $x\sim\mc D$ over $\mc X$ 
we then define the $\mc D$-inner product on $\mc F$ as 
\begin{equation}
    \lrangle{\cdotspace, \cdotspace}_{\mc D}:\mc F\times\mc F\to[-1,1]\doublecolon (f,g)\mapsto \lrangle{f,g}_{\mc D}=\Ex_{x\sim\mc D}\qty[f(x)g(x)]\,,
\end{equation}
and the corresponding $L^2$-distance as 
\begin{equation}
    \met_{L^2(\mc D)}:\mc F\times\mc F\to[0,1]\doublecolon (f_\phi, f_\vartheta)\mapsto \frac{1}{2}\norm{f_\phi-f_\vartheta}^2_{L^2(\mc D)} = \frac{1}{2}\Ex_{x\sim\mc D}\qty[\qty(f_\phi(x)-f_\vartheta(x))^2]\,.
\end{equation}

In the following, we will consider the task of $\epsilon$-self-learning $\mc F$ with respect to $L=\mrm{id}$ and $\met_{L^2(\mc D)}$.
In particular, the corresponding set of loss functions is given by
\begin{equation}
    \mc L = \set{l^\phi:\Theta\to[0,1]\doublecolon \vartheta\mapsto l^\phi(\vartheta)=L\circ\met_{L^2(\mc D)}(f_\vartheta,f_\phi) \midvert \phi\in\Theta}\,.
\end{equation}

To start with let us derive the following useful lemma.

\begin{lemma}\label{lem:function-selflearning-variance}
    Let $\gamma>0$, $\mc F$ be a class of functions $f_\vartheta:\mc X\to[-1,1]$ with parameter space $\vartheta\in\Theta$, $\mc D$ a probability distribution over $\mc X$ and $\mu$ a probability measure over $\Theta$. 
    Denote by $\mc L$ the class of loss functions corresponding to self-learning $\mc F$ with respect to $\met_{L^2(\mc D)}$.
    Then, for every $\vartheta\in\Theta$ it holds 
    \begin{equation}\label{eq:lem-function-selflearning-variance}
        \Var_{\phi\sim\mu}\qty[l^\phi(\vartheta)]  
        \leq 4 \cdot\sqrt{\Var_{\phi\sim\mu}\qty[f_\phi]}
        \leq 4 \cdot\sqrt{\Ex_{\substack{\phi\sim\mu\\x\sim\mc D}}\qty[f_\phi(x)^2]}\,,
    \end{equation}
    where we used the variance with respect to the $L^2(\mc D)$ inner product
    \begin{align}
        \Var_{\phi\sim\mu}\qty[f_\phi]=\Ex_{\phi\sim\mu}\qty[\lrangle{f_\phi,f_\phi}_{\mc D}]-\lrangle{\Ex_{\phi\sim\mu}\qty[f_\phi],\Ex_{\phi\sim\mu}\qty[f_\phi]}_{\mc D}\,.
    \end{align}
\end{lemma}

\begin{remark}\label{rem:function-learning-triviality}
    Informally, the lemma states that concentration of $f_\phi$ with respect to $\phi\sim\mu$ implies concentration of the loss function $l^\phi(\vartheta)$ with respect to $\phi\sim\mu$.
    As such, function classes, for which non-trivially learning is hard,
    are often average-case trivial.
\end{remark}

\begin{proofof}[\Cref{lem:function-selflearning-variance}]
    To begin with, note that 
    \begin{align}\label{eq:shift-is-fine-vqml}
        l^\phi(\vartheta) 
        = \frac{1}{2} \Ex_{x\sim\mc D}\qty[\qty(f_\phi(x) - f_\vartheta(x))^2] 
        = \frac{1}{2} \Ex_{x\sim\mc D}\qty[\qty(\tilde f_\phi(x) - \tilde f_\vartheta(x))^2]
        = \tilde{l}^\phi(\vartheta)\,,
    \end{align}
    with $\tilde f_\vartheta(x) = f_\vartheta(x) - \Ex_{\phi\sim\mu}[f_\phi(x)]$ the ``centered'' version of $f_\vartheta$. 
    Thus, $\Var_\phi[l^\phi(\vartheta)]=\Var_\phi[\tilde{l}^\phi(\vartheta)]$.
    
    Using
    \begin{align}
        l^\phi(\vartheta) 
        = \frac{1}{2}\norm{f_\vartheta}_{L^2(\mc D)}^2 + \frac{1}{2}\norm{f_\phi}_{L^2(\mc D)}^2 - \lrangle{f_\phi,f_\vartheta}_{\mc D} \,,
    \end{align}
    and that for constant $y$ it holds $\Var[X+y]=\Var[X]$, we bound
    \begin{align}
        \Var_{\phi\sim\mu} \qty[l^\phi(\vartheta)]
        &= \Var_{\phi\sim\mu} \qty[
            \frac{1}{2}\norm{f_\phi}^2_{L^2(\mc D)}
            -\lrangle{f_\phi,f_\vartheta}_{L^2(\mc D)}
        ] \\[10pt]
        &= \frac{1}{2}\Var_{\phi\sim\mu} \qty[\norm{f_\phi}^2_{L^2(\mc D)}]
        +\Var_{\phi\sim\mu} \qty[\lrangle{f_\phi,f_\vartheta}_{\mc D}]
        -\Cov_{\phi\sim\mu} \qty[\norm{f_\phi}_{L^2(\mc D)}^2,\lrangle{f_\phi,f_\vartheta}_{\mc D}]\,.
    \end{align}

    As for all $\phi$ and $x$ it holds $f_\phi(x)\leq 1$ we bound
    \begin{align}
        &\abs{\Ex_{\phi\sim\mu} \qty[\lrangle{f_\phi,f_\vartheta}_{\mc D}^2]} 
        \leq\abs{\Ex_{\phi\sim\mu} \qty[\lrangle{f_\phi,f_\vartheta}_{\mc D}]} 
        \leq \Ex_{\phi\sim\mu} \qty[\norm{f_\phi}_{L^1(\mc D)}] 
        = \Ex_{\substack{\phi\sim\mu\\ x\sim\mc D}} \qty[\abs{f_\phi(x)}]\,,\label{eq:vqml-lem-1}
        \intertext{as well as}
        &\abs{\Ex_{\phi\sim\mu} \qty[\norm{f_\phi}^4_{L^2(\mc D)}]} 
        \leq\abs{\Ex_{\phi\sim\mu} \qty[\norm{f_\phi}^2_{L^2(\mc D)}]} 
        \leq \Ex_{\phi\sim\mu} \qty[\norm{f_\phi}_{L^1(\mc D)}] 
        = \Ex_{\substack{\phi\sim\mu\\ x\sim\mc D}} \qty[\abs{f_\phi(x)}]\,, \label{eq:vqml-lem-2}
        \intertext{and similarly}
        &\abs{\Ex_{\phi\sim\mu}\qty[\norm{f_\phi}^2_{L^2(\mc D)}\lrangle{f_\phi,f_\vartheta}_{\mc D}]}
        \leq \Ex_{\phi\sim\mu}\qty[\norm{f_\phi}^2_{L^2(\mc D)}]\leq \Ex_{\substack{\phi\sim\mu\\ x\sim\mc D}} \qty[\abs{f_\phi(x)}]\,.\label{eq:vqml-lem-3}
    \end{align}
    Which can be bound by Jensen's inequality $\Ex[\abs{X}]^2\leq\Ex[\abs{X}^2]$ as
    \begin{equation}
        \Ex_{\substack{\phi\sim\mu\\ x\sim\mc D}} \qty[\abs{f_\phi(x)}]
        \leq \sqrt{\Ex_{\substack{\phi\sim\mu\\ x\sim\mc D}} \qty[f_\phi(x)^2]}\,.
    \end{equation}

    Now, using $\Var[X]\leq\Ex[X^2]$ and $\abs{\Cov[X,Y]}\leq \abs{\Ex[XY]} + \abs{\Ex[X]\Ex[Y]}$ together with the triangle inequality and \Cref{eq:shift-is-fine-vqml,eq:vqml-lem-1,eq:vqml-lem-2,eq:vqml-lem-3} we conclude 
    \begin{align}
        \Var_{\phi\sim\mu}\qty[l^\phi(\vartheta)]
        =\Var_{\phi\sim\mu}\qty[\tilde l^\phi(\vartheta)]
        \leq 4 \cdot\sqrt{\Ex_{\substack{\phi\sim\mu\\ x\sim\mc D}}\qty[\tilde f_\phi(x)^2]} 
        = 4\cdot \sqrt{\Var_{\phi\sim\mu}\qty[f_\phi]}\,,
    \end{align} 
    which completes the claim.
\end{proofof}

Let us now focus on two specific vQML models.  
To begin with, we define linear models.

\begin{definition}[Linear Models]\label{def:linear-models}
    Let $D_E,D_P\in\mathbb N$ and let $D=D_E+D_P$. 
    Let $\mc F$ be as in \Cref{eq:def-vqml-architecture} where $U_{l}(\vartheta_{l})=U_{l}$ are independent of $\vartheta$ for all $1\leq l\leq D_E$,
    and where all $W_l(x)=\id$ for all $D_E+1\leq l\leq D$.
    Then, we refer to $\mc F$ as a \emph{Linear Variational Quantum Machine Learning Model}, or, in short, a linear model.
    We refer to $D_E$ as the \emph{Encoding Depth} and $D_P$ as the \emph{Parametrization Depth}.
    Defining the \emph{Encoding Unitary} $W(x)$ and the \emph{Variational Unitary} $U(\vartheta)$ as
    \begin{align}
        &W(x) = \prod_{l=D_E}^1 W_l(x)U_l &&\text{and} &&&U(\vartheta)=\prod_{l=D}^{D_E+1} U_l(\vartheta)\,,
    \end{align}
    we can then write every $f\in\mc F$ as $f_\vartheta(x)=\tr[\rho_x O_\vartheta]$, where
    \begin{align}
        &\rho_x=W(x)^\dagger\ketbra{\psi_0} W(x) &&\text{and} &&&O_\vartheta=U(\vartheta)OU(\vartheta)^\dagger\,.
    \end{align}
    In particular, $f_\vartheta(x)=\lrangle{\rho_x,O_\vartheta}_{HS}$ can be written in terms of the Hilbert-Schmidt inner product.
\end{definition}

\begin{theorem}[Linear Models and Barren Plateaus: Part I]\label{thm:bp-linear-models}
    Let $\mc F$ be the function class corresponding to a linear model, $\delta=2^{-\Omega(n)}$, assume that $U(\vartheta)\sim U\circ\mu$ forms an additive $\delta$-approximate 2-desing and let $\mc D$ be a distribution over $\mc X$.
    Then, the query complexity for non-trivial self-learning $\mc F$ from $\tau$-accurate loss function queries with respect to $\met_{L^2(\mc D)}$ and wit respect to te measure on $\mc F$ induced by $\mu$ is lower bounded by 
    \begin{align}
        \qnt[\mu,\tau] = \frac{2^{\Omega(n)}}{\tau^2}\,.
    \end{align}
    In particular, if the $U_{li}(\vartheta_{li})$ have equidistant spectrum, then the loss function $l^\phi$ admits a barren plateau for every $\phi\in\Theta$.
\end{theorem}

\begin{proof}
    We can write the observable as $O=N^{-1}\tr[O]\id+O_\times$ with $\tr[O_\times]=0$.
    By \Cref{lem:function-selflearning-variance} it suffices to show 
    \begin{equation}
        \Var_{\phi\sim\mu} \qty[f_\phi] = 2^{-\Omega(n)}\,.
    \end{equation}
    By assumption $U(\phi)\sim U\circ\mu$ is an additive $\delta$-approximate unitary 2-design.
    For $\delta=0$ and denoting $2^n=N$ it holds
    \begin{align}
       \Ex_{\phi,\vartheta\sim\mu}\qty[\lrangle{f_\phi,f_\vartheta}_\mc D] 
       = \Ex_{x\sim\mc D}\qty[\Ex_{\phi}[f_\phi(x)],\Ex_{\phi}[f_\phi(x)]]
       = \frac{\tr[O]^2}{N^2}\,
    \end{align}
    and 
    \begin{align}
        \Ex_{\substack{\phi\sim\mu\\ x\sim\mc D}} \qty[f_\phi(x)^2]
        =  \Ex_{x\sim\mc D}\qty[\Ex_{\phi\sim\mu } \qty[\tr[U(\phi)^\dagger\rho(x)U(\phi)O]]^2]
        = \frac{\tr[O^2]+\tr[O]^2}{N(N+1)}\,.
        \label{eq:vqml-lin-mod-second-mom}
    \end{align}
    Using $\tr[O^2]\leq N$ and $\tr[O]\leq N$ this yields
    \begin{align}
        \Var_{\phi\sim\mu} \qty[f_\phi] = \frac{\tr[O^2]N-\tr[O]^2}{N^2(N+1)} \leq \frac{1}{N+1}\,.
    \end{align}
    The claim for $\delta>0$ then follows due to \Cref{lem:additive-design-variance-implication}
    while the claim about the barren plateau is because te equidistant spectrum implies by \cite{wierichsGeneralParametershiftRules2022} an efficient parameter-shift rule.
\end{proof}

Similarly, the same holds if the encoding forms an additive approximate $2-design$.

\begin{corollary}[Linear Models and Barren Plateaus: Part II]\label{cor:bp-linear-models-part-2}
    Let $\delta=2^{-\Omega(n)}$ and let $\mc F$ be the function class corresponding to a linear model, where $W(x)\sim W\circ\mc D$ forms an additive $\delta$-approximate 2-design where $\mc D$ is the distribution over $\mc X$.
    Then, the query complexity for non-trivial self-learning $\mc F$ from $\tau$-accurate loss function queries with respect to $\met_{L^2(\mc D)}$ is lower bounded by 
    \begin{align}
        \qnt[\mu,\tau] \geq \frac{2^{\Omega(n)}}{\tau^2}\,.
    \end{align}
    In particular, if the $U_{li}(\vartheta_{li})$ have equidistant spectrum, then the loss function $l^\phi$ admits a barren plateau for every $\phi\in\Theta$.
\end{corollary}

\begin{proof}
    To see this observe that due to the linearity and for $\delta=0$
    \begin{align}
        \Ex_{\phi\sim\mu}[f_\phi(x)] = \tr[\rho_x \Ex_{\phi\sim\mu}[O_\phi]] = \tr[\rho_x \tilde O]\,.
    \end{align}
    Without loss of generality, assume that $\tr[O]=0$ and hence $\tr[\tilde O]=\tr[O_\phi]=0$ for all $\phi$ (else, shift $O\mapsfrom O-N^{-1}\tr[O]\id$ which leaves the variance unchanged).
    Then, we find 
    \begin{align}
        &\Var_{\phi\sim\mu}[f_\phi] 
        = \Ex_{\stackrel{\phi\sim\mu}{x\sim\mc D}}\qty[\tr[\rho_x O_\phi]^2] - \Ex_{\stackrel{\phi,\vartheta\sim\mu}{x\sim\mc D}}\qty[\tr[\rho_x O_\phi]\tr[\rho_x O_\vartheta]]\\[8pt]
        &= \frac{1}{N(N+1)}\qty(\Ex_{\phi\sim\mu}\qty[\tr[O_\phi^2]+\tr[O_\phi]^2] - \Ex_{\phi,\vartheta\sim\mu}\qty[\tr[O_\phi O_\vartheta] + \tr[O_\phi]\tr[O_\vartheta]])\\[8pt]
        &= \frac{1}{N(N+1)}\qty(\Ex_{\phi\sim\mu}\qty[\tr[O^2]] - \tr[\tilde O^2])
        \leq \frac{2}{N(N+1)}\label{eq:linear-model-2-design-encoding}
    \end{align}
    The claim for $\delta>0$ follows from \Cref{lem:additive-design-variance-implication}
    and the claim on the barren plateau follows the same argument as in the previous proof.
\end{proof}

Next, we investigate data re-uploading models in general.

\begin{definition}[Data Re-Uploading Models]\label{def:data-re-up}
    Let $D\in\mathbb N$ and let $\mc F$ be as in \Cref{eq:def-vqml-architecture}.
    Then we refer to $\mc F$ as a \emph{Data Re-Uploading Model} over $\mc X$ of depth $D$ with observable $O$. 
\end{definition}

\begin{theorem}[Data Re-Uploading Models and Barren Plateaus]\label{thm:data-re-up-barren}
    Let $D\geq1$ and $\delta=2^{-\Omega(n)}$. 
    Let $\mc F$ be a data re-uploading model of depth $D$ with observable $O$. Let $\mu=\bigotimes_{l,i}\mu_{li}$ be independently distributed over the local variables and
    assume that locally $U_{li}(\vartheta_{li})$ forms a $2$-design and either  
    \begin{itemize}
        \item $O$ is a Pauli string with $\Omega(n)$ support, 
        or
        \item $O$ is traceless and the unitary $U(\vartheta)=\prod_{l=D}^1U_l(\vartheta_l)$ forms an additive $\delta$ approximate 2-design for $\vartheta\sim\mu$. 
    \end{itemize}
    Then, the query complexity of non-trivial self-learning $\mc F$ from $\tau$-accurate loss function queries  with respect to $\met_{L^2(\mc D)}$ is lower bounded by
    \begin{align}
        \qnt[\mu,\tau] = \frac{2^{\Omega(n)}}{\tau^2}\,.
    \end{align}
    In particular, if the $U_{li}(\vartheta_{li})$ have equidistant spectrum, then the loss function $l^\phi$ admits a barren plateau for every $\phi\in\Theta$.
\end{theorem}

\begin{proof}
    The proof is analogous to that of \Cref{thm:bp-linear-models}.
    Since each $k$-local unitary $U_{li}(\vartheta_{li})$ forms a $2$-design the average is invariant under the transformation $O\mapsto X^{\otimes n}O X^{\otimes n}$.
    Therefore, for every $x$ it holds $\Ex_{\phi\sim\mu}[f_\phi(x)]=0$, which implies 
    \begin{align}
        \Var_{\phi\sim\mu}[f_\phi]=\Ex_{\substack{\phi\sim\mu\\x\sim\mc D}}[f_\phi(x)^2]\,.
    \end{align}
    Next, we note
    \begin{align}\label{eq:proof-datareup-1}
        \Ex_{\substack{\phi\sim\mu\\ x\sim\mc D}} \qty[f_\phi(x)^2]  
        = \Ex_{\substack{\phi\sim\mu\\ x\sim\mc D}} \qty[\mel**{\psi_0}{\prod_{l=1}^DU_l(\phi_l)^\dagger O \prod_{l=D}^1 U_l(\phi_l)}{\psi_0}^2] 
        = \Ex_{\phi\sim\mu} \qty[\tr[\rho(\phi)\otimes\rho(\phi) O\otimes O ]]\,,
    \end{align}   
    where the local unitaries $W_{li}(x)$ have been canceled by the 2-design property of the local $U_{li}(\vartheta_{li})$ and where 
    \begin{equation}\label{eq:rho-phi}
        \rho(\phi) 
        = \prod_{l=D}^1 U_l(\phi_l) \ketbra{\psi_0} \prod_{l=1}^DU_l(\phi_l)^\dagger  
        = U(\phi)\ketbra{\psi_0}U(\phi)^\dagger\,.
    \end{equation}

    The case where the $U(\phi)$ form a $\delta$ approximate 2-design then follows immediately from \Cref{thm:bp-linear-models}.

    We now consider the case that $O$ is a Pauli string with support $\Omega(n)$.
    To upper bound the second moment of $f_\phi(x)$ consider the last layer of unitaries together with the observable $O\otimes O$ in \Cref{eq:proof-datareup-1}. 
    Let $U_{li}$ be a local unitary in the last layer that is connected to a non-trivial element in the Pauli string and let $P_{li}$ be the corresponding Pauli substring.
    By \Cref{lem:unitary-second-moments} we find
    \begin{align}
        \Ex_{U_{li}\sim\mu_{li}}\qty[U_{li}^{\otimes 2} P_{li}^{\otimes 2} U_{li}^{\otimes 2\,\dagger}]
        = \frac{2K}{(K-1)(K+1)}\mb F + \frac{2}{(K-1)(K+1)}\id
        = \Theta\qty(\tfrac{1}{K}) \mb F + \Theta\qty(\tfrac{1}{K^2}) \id = M
    \end{align} 
    where we have used and defined $\tr[P_{li}^{\otimes2}]=0$ and $\tr[P_{li}^{\otimes2}\mb F]=2^k=K$, defined $M=\Theta(\frac{1}{K})\mb F+\Theta(\frac{1}{K^2})\id$ and hence $\norm{M}\mrm{op}\leq \frac{2K}{(K-1)(K+1)}$ and where $\id$ and $\mb F$ refer to the identity, respectively flip on $\mb C^K\otimes\mb C^K$.

    By assumption, there exist $\Omega(n)$ many $U_{li}$'s connected to non-trivial Pauli elements.
    Hence, we can simplify \Cref{eq:proof-datareup-1} as 
    \begin{align}
        \Ex_{\phi\sim\mu}\qty[\tr[\rho(\phi)^{\otimes 2} O^{\otimes 2}]] 
        = \tr[\sigma M^{\otimes \Omega(n)}] 
        \leq \norm{M}_\mrm{op}^{\Omega(n)}
        = 2^{-\Omega(n)}\,,
    \end{align}
    with $\sigma$ the averaged state on the first $D-1$ layers and $M$ the observable averaged over the last layer as above.

    We conclude with $\qnt[\mu,\tau]=2^{\Omega(n)}$ in either case.

    The claim about the barren plateau follows from the same arguments as in the previous proofs.
\end{proof}

\subsubsection{A (Very) Simple Hard Class}\label{sec:simple-hard-class}
\addsectionheader{sec:simple-hard-class}

In this section we compare our method to Anschuetz and Kiani's statistical dimension due to their SQ framework \cite{anschuetz2022}.

We will significantly strengthen their Proposition 4 in \cite[Supplementary Note~3]{anschuetz2022}.
We restrict ourselves to Proposition 4 and remark that
similar statements can be made for all other concept classes considered in \cite[Supplementary Note~3]{anschuetz2022} by invoking our \Cref{cor:bp-linear-models-part-2}.

Consider the function class $\mc F_S$ consisting of functions $f_{U}:\mc X\to[-1,1]$,
with $\mc X$ a set of $n$-qubit quantum states,
defined as 
\begin{align}\label{eq:anschuetz}
    f_{U}(\rho)
    = \tr[\rho\qty(\bigotimes_{i=1}^n U_i^\dagger) S \qty(\bigotimes_{i=1}^n U_i)] = \tr[\rho U^\dagger S U]
\end{align}
where the vector of unitaries $U=(U_1,\dots,U_n)\in\mc U$ is from some set of single qubit unitaries and $S$ is a fixed signed non-identity Pauli string $S=\lambda P_1\otimes\cdots\otimes P_n$ with $\lambda\in\set{-1,1}$ and $P_i\in\set{\id, X, Y, Z}$, with $(P_1,\dots, P_n)\neq(\id,\dots, \id)$.
Let $\mc D$ be a distribution over $\mc X$ such that $\rho\sim\mc D$ forms a projective 2-design.

Note that this class is closely related to linear models as in \Cref{def:linear-models}.
$\mc F_S$ can be understood as a linear model where the encoding is such that $W(x)$ forms a $2$-design.
Then it holds.

\begin{theorem}\label{thm:anschuetzprop4}
    Let $S$ be a non-identity stabilizer and let $\mc F_S$ be as defined in \Cref{eq:anschuetz} for any $\mc U$ with $\abs{\mc U}\geq2$ and let $\mc D$ as above. 
    Denote by $\mc L_S$ the set of loss functions corresponding to self-learning $\mc F_S$ with respect to $\met_{L^2(\mc D)}$.
    Then, for every probability measure $\mu$ over $\mc U$ it holds that the query complexity of non-trivial self-learning $\mc F_S$ from $\tau$-accurate loss function queries with respect to $\mu$ and $\met_{L^2(\mc D)}$ is lower bounded by
    \begin{align}
        \qnt[\mu,\tau] = \frac{2^{\Omega(n)}}{\tau^2} \,.
    \end{align}
    In particular, if $\mc U$ is continuous and every $U_i$ has an equidistant spectrum, every loss function $l^U\in \mc L$ admits a barren plateau.
\end{theorem}

\begin{proof}
    The claim follows immediately from \Cref{cor:bp-linear-models-part-2} with $\delta=0$.
\end{proof}

\begin{remark}
    Note that by \Cref{eq:linear-model-2-design-encoding}
    any $f_U$ is exponentially concentrated around $\Ex_U[f_U]$ in $L^2(\mc D)$-norm.
    Thus, we immediately find $\Var_U[\lrangle*{f_U,f_Q}]\leq 2^{-\Omega(n)}$, 
    which
    in turn implies the same lower bounds on learning $\mc F_S$ from quantum correlational queries of the form $Q^\dagger S Q$ to $\qcsq_\tau(U^\dagger S U)$.
    This then corresponds to applying the parameter shift perspective in \cite{anschuetz2022} to parametrized learning (see their Eq.~(16) and below).
\end{remark}  

\begin{remark}
    By \Cref{thm:anschuetzprop4} learning an unknown $U\in\mc U$ requires at least $\sfrac{2^{2n}}{\tau^2}$ many queries to $\eval_\tau$ for any non-trivial $\mc U$.
    Thus, let $\mc U=\set{\id, X^{\otimes n}}$. 
    Then we immediately find deciding whether $U=\id$ or $X^{\otimes n}$ requires exponentially many queries.
    In contrast, a similar bound can, due to the set size of $\mc U$, never be obtained from the statistical dimension in \cite{anschuetz2022}.
\end{remark}

\subsubsection{Remark on Optimizing Local Hamiltonians}\label{sec:remark-local-hamiltonians}
\addsectionheader{sec:remark-local-hamiltonians}

In the example about random initializations in \Cref{sec:derandomize-random-init} we considered variational quantum eigensolvers (VQE).
VQEs aim at finding low energy states of a given Hamiltonian, where a class of Hamiltonians $\mc H$ induces a class of loss functions $\mc L$.
A practically relevant and often considered special case is that where $\mc H$ consists of local Hamiltonians $H=\sum_i h_i$ only, with terms $h_i=\id_{[n-k]}\otimes h_i^{(\mrm{loc})}$ that act non-trivially only on $k$ qubits.
It is easy to see that from any parametrization that contains the computational basis states in every Pauli basis 
\begin{align}
    \set{\ketbra{a_1,\dots,a_n}\midvert a\in\set{0,1,+,-,i,-i}^{\times n}}\subseteq\set{\rho(\vartheta)\midvert \vartheta\in\Theta}\subseteq\mc S(N) 
\end{align}
one efficiently learn $H$ by estimating the $4^k\binom{n}{k}$ many non-trivial coefficients in the Pauli decomposition of $H$.
In particular, for constant $k$ one can from polynomially many queries to $\eval_\tau(l_H)$ learn each $h_i^{(\mrm{loc})}$ to accuracy $O(\tau)$ in $\ell_\infty$-norm and, hence, for $\tau=O(\overpoly)$ sufficiently small, in any norm.
Therefore, our framework can only yield polynomial lower bounds for VQE optimization of local Hamiltonians, independent of the measure $H\sim\mu$.

\stopcontents[variational]

%
%

\newpage
\addcontentsline{toc}{part}{Appendix}
\appendix
\part*{Appendix}
\startcontents[appendix]
\vspace{1cm}
\printcontents[appendix]{}{1}{}
\vspace{1cm}

\section{Boostable Learning Problems}\label{app:boostable}
\addsectionheader{app:boostable}
\addsectionfooter{app:boostable}

Boostable learning problems are, in short, learning problems, for which a weak learner can query efficiently be boosted to a strong learner. 
Since weak learning is only defined for specific learning problems -- specifically PAC learning as in \Cref{prob:pac-learning} with $\epsilon=1-\gamma$ for a small $\gamma>0$ -- we formalize boosting via the following notion.

\begin{definition}[Boostable Learning Problems and Exact Learning]\label{def:boostablelearning}
    Let $0\leq\epsilon_0\leq\epsilon_1\leq\infty$ and let $\mc I=[\epsilon_0,\epsilon_1]$ (respectively $\mc I=[\epsilon_0,\epsilon_1)$ if $\epsilon_1=\infty$).
    A \emph{Parameter Family of learning Problems} with parameter $\epsilon$ and parameter set $I$ is a family of learning problems $\set{\mc Z^\epsilon:\mc S\to\mc \mc P(\mc T) \midvert\epsilon\in\mc I}$.
    We say $\mc Z^\epsilon$ is \emph{Boostable} if there exists a $c\geq1$, such that for any $\epsilon\in\mc I$ and any
    $s,r\in\mc S$ such that $\mc Z^\epsilon(s)\cap \mc Z^\epsilon(r)\neq\varnothing$
    it holds that 
    \begin{equation}\label{eq:boostablelearning}
        \mc Z^\epsilon(r)\cup\mc Z^\epsilon(s)\subseteq \mc Z^{c\bar\epsilon}(s)\,,
    \end{equation}
    where $\bar\epsilon=\min\set{\epsilon, c^{-1}\epsilon_1}$.
    The special cases $\epsilon_0=\epsilon_1$ and equivalently $c=1$, are  referred to as \emph{Exact Learning Problems}.
\end{definition}

The intuition of this condition is that the solutions of related source objects become increasingly similar at a sufficient scale.
In particular, let $s,r$ be any two source objects with related solutions in the sense of $\mc Z^\epsilon(s)\cap\mc Z^\epsilon(r)\neq\varnothing$.
Then, the solutions to $r$ at scale $\epsilon$ are also valid solutions to $s$ at scale $c\bar\epsilon$.

\begin{example}
    For any $\mc S\subset\widetilde{\mc S}$ and $s^*\in\widetilde{\mc S}\setminus\mc S$ the decision problem $\dec(\mc S, s^*)$ from \Cref{ex:decissionproblem} is an exact learning problem. 
    Similarly, any learning problem that can be defined by a parameter $\epsilon\in(0,1)$, a map $f:\mc S\to\mc T$ and a metric $m$ on $\mc T$ such that $\mc Z(s)=\{t\in\mc T\;\vert\; m(t,f(s))<\epsilon\}$ is, by the triangle inequality for $m$ boostable.
    This implies that PAC learning is boostable.
\end{example}

The following lemma will make the naming clear.



\begin{lemma}[Boosting]\label{lem:boosting}
    Let $\set{\mc Z^\epsilon \midvert \epsilon\in\mc I}$ be a boostable learning problem and denote by $\bar\epsilon=\min\set{\epsilon,c^{-1}\epsilon_1}$. 
    Assume that, for $\epsilon_0\leq\epsilon\leq\epsilon_1$ there exists an algorithm $\A$ with
    oracle access $\O$ that learns $\mc Z^{\epsilon}$ with probability $\alpha=1/2+\gamma$ from $q$ queries to $\O$. 
    Then there exists an algorithm that 
    learns $\mc Z^{c\bar\epsilon}$ with probability $\alpha'=1-\delta'$ from 
    \begin{align}\label{eq:boosting}
        q'=\frac{q}{2\gamma^2}\cdot\ln\qty(\frac{1}{\delta'})
    \end{align}
    many queries to $\O$. In particular, there exists an algorithm with success probability $1-\exp(-n)$ from $\frac{qn}{2\gamma^2}$ queries.
\end{lemma}

\begin{proof}
    Assume $\A$ learns $\mc Z^{\epsilon}$ from $q$ queries to $\O$ with probability at least $\alpha=1/2+\gamma$. For any boosting parameter $b>0$ we define the following boosting algorithm $\mc B$. 
    
    Run $\A$ $b$ times. This results in $b$ solutions $t_i\in\mc T$, $i=1,\dots,b$. 
    Search for any $s\in\mc S$ such that at least $\ceil*{\frac{b}{2}}$ solutions are contained in $\mc Z^{\epsilon}(s)$.
    While this step is computationally inefficient it does not involve any further queries to $\O$.
    If there exists such an $s$ return any $t\in\mc Z^{\epsilon}(s)$. 

    We now claim that With at most $b\cdot q$ queries to $\O$ algorithm $\mc B$ successfully learns $\mc Z^{\bar\epsilon}$
    with probability at least $1-\exp(-2b\gamma^2)$.
    
    To see this first observe that by construction $\mc B$ makes no more than $b\cdot q$ queries.
    To verify the second part of the claim assume we run $\mc B$ on an unknown $r\in\mc S$ via access to $\O(r)$. 
    Define the random variable $X_i$ as the event ``$t_i\not\in\mc Z^{\epsilon}(r)$''.
    In particular, $X_i=1$ if $t_i$ is not a valid solution to $r$. 
    Then $\Pr[X_i=1]=\Ex[X_i]=\sfrac12-\gamma=\overline{\alpha}$.
    Defining the random variable $S_b=\sum_{i=1}^b X_i$ we obtain from H\"offding's Theorem
    \begin{align}
        \Pr_{\A,\O}\qty[\frac{S_b}{b}\geq\frac12]=
        \Pr_{\A,\O}\qty[\frac{S_b}{b}-\overline\alpha\geq\gamma]=
        \Pr_{\A,\O}\qty[S_b-\overline\alpha\cdot b\geq\gamma\cdot b]\leq\exp\qty(-2b\gamma^2)\,.
    \end{align}
    Therefore, with probability $1-\exp(-2b\gamma^2)$ for at least half the solutions it holds $t_i\in\mc Z^{\epsilon}(r)$.
    This implies that for any $s\in\mc S$ such that $\mc Z^{\epsilon}(s)$ contains at least $\ceil*{\sfrac{b}{2}}$ solutions it must hold $\mc Z^{\epsilon}(s)\cap \mc Z^{\epsilon}(r)\neq\varnothing$. 
    By \Cref{eq:boostablelearning} we then find $\mc Z^{\epsilon}(r)\subseteq \mc Z^{\bar\epsilon}(s)$, which proves the claim.
    In particular,
    \begin{align}
        b=\ln(\frac{1}{\delta'})\cdot\frac{1}{2\gamma^2}
    \end{align} 
    suffices to obtain a failure probability of at most $\delta'$ proving \Cref{eq:boosting}.
\end{proof}

With the previous \Cref{lem:boosting}, we arrive at the following corollary.

\begin{corollary}[Random Average-Case Lower Bound for Boostable Problems]\label{cor:boostablerandomlower}
    Assume a given boostable learning problem $\set{\mc Z^\epsilon \midvert \epsilon_0\leq\epsilon\leq\epsilon_1}$ with $c\geq1$ and denote $\bar\epsilon=\min\set{\epsilon,c^{-1}\epsilon_1}$. 
    Let $\mu$ be a measure over $\mc S$ and suppose for some $\epsilon$ there is a random algorithm $\A_\epsilon$ that learns $\mc Z^\epsilon$ with probability $\alpha$ over $\A_\epsilon$'s randomnes and probability $\beta$ with respect to $s\sim\mu$ from $q$ many $\tau$ accurate evaluation queries. Then for any $f\in M^\mc X$ and any $\delta'<1-\alpha$ it must hold
    \begin{align}
        q\geq
        \frac{2(\alpha-\sfrac12)^2}{\ln\qty(\sfrac{1}{\delta'})}\cdot
        \frac{(1-\delta')\cdot\beta - \sup_{t\in\mc T}\Pr_{s\sim\mu}\qty[s\in\mc Z^{c\bar\epsilon}_t]}{\max_{x\in\mc X}\Pr_{s\sim\mu}\qty[\met\qty(s(x),f(x))>\tau]}\,.
    \end{align} 
\end{corollary}

\begin{proof}
    If $0<\alpha,\beta,\gamma$ with $\sfrac12+\gamma=\alpha$ and $\A_\epsilon$ is an algorithm such that 
    \begin{align}
        \Pr_{s\sim\mu} \qty[\Pr_\A\qty[\A_\epsilon^{\eval_\tau(s)}\to\mc Z^\epsilon(s) \;\text{using at most $q$ queries}]\geq\alpha] \geq \beta\,,
    \end{align}
    then by the previous \Cref{lem:boosting} for any $\delta'<1-\alpha$ and $\bar\epsilon=\min\set{\epsilon, c^{-1}\epsilon_1}$ there is an algorithm $\mc B$ such that
    \begin{align}
        \Pr_{s\sim\mu} \qty[\Pr_\mc B\qty[\mc B^{\eval_\tau(s)}\to\mc Z^{c\bar\epsilon}(s) \;\text{using at most $\frac{q}{2(\alpha-\sfrac12)^2}\cdot\ln\qty(\sfrac{1}{\delta'})$ queries}]\geq1-\delta'] \geq \beta\,.
    \end{align}
    This implies  
    \begin{align}
        \Pr_{s\sim\mu, \mc B} \qty[\mc B^{\eval_\tau(s)}\to\mc Z^{c\bar\epsilon}(s) \;\text{using at most $\frac{q}{2(\alpha-\sfrac12)^2}\cdot\ln\qty(\sfrac{1}{\delta'})$ queries}] \geq (1-\delta')\cdot\beta\,.
    \end{align}
    By \Cref{cor:jointavglowerbound} for any $f\in M^\mc X$ it must hold 
    \begin{align}
        \frac{q}{2(\alpha-\sfrac12)^2}\cdot\ln\qty(\sfrac{1}{\delta'}) \geq
        \frac{(1-\delta')\cdot\beta - \sup_{t\in\mc T}\Pr_{s\sim\mu}\qty[s\in\mc Z^{c\bar\epsilon}_t]}{\max_{x\in\mc X}\qty[\met\qty(s(x),f(x))>\tau]}\,,
    \end{align}
    or equivalently
    \begin{align}
        q\geq
        \frac{2(\alpha-\sfrac12)^2}{\ln\qty(\sfrac{1}{\delta'})}\cdot
        \frac{(1-\delta')\cdot\beta - \sup_{t\in\mc T}\Pr_{s\sim\mu}\qty[s\in\mc Z^{c\bar\epsilon}_t]}{\max_{x\in\mc X}\qty[\met\qty(s(x),f(x))>\tau]}\,.
    \end{align}
\end{proof}

\section{Omitted Definitions for Mirror Descent}\label{app:convex}
\addsectionheader{app:convex}
\addsectionfooter{app:convex}
As in \Cref{sec:linearlearningupper} we denote by $\mc V$ a linear space over $\mb K$ where $\mb K$ is either $\mb R$ or $\mb C$, 
and denote by $\lrangle{\cdotspace, \cdotspace}$ the inner product in $\mc V$ with the convention of anti-linearity in the first argument.
Let $\norm{\cdotspace}$ be a norm on $\mc V$, $\dualnorm{\cdotspace}$ its dual norm and
let $\mc W\subseteq\mc V$ be some open convex set. 
We refer to $\mc W$ as the primal space.

\begin{definition}[$\alpha$-Regularizer]\label{def:alpharegularizer}
    A function $R:\mc W\to\mb K$ is called an $\alpha$-\emph{regularizer} if it is strictly $\alpha$-convex with respect to $\norm{\cdotspace}$. In particular, for all $t\in[0,1]$ and $x,y\in\mc W$ it holds
    \begin{equation}
        R(tx+(1-t)y)\leq tR(x) +(1-t)R(y) -\frac{\alpha t(1-t)}{2}\norm{x-y}^2\,.
    \end{equation}
    We say that $R$ is a $0$-regularizer if $R$ is strictly convex, this is $R(tx+(1-t)y)< tR(x) +(1-t)R(y)$.
\end{definition}

\begin{definition}[Bregman Divergence]\label{def:bregmandivergence}
    Let $\overline{\mc W}$ be the closure of $\mc W$ and let $R:\overline{\mc W}\to\mb R$ be continuously differentiable. Then, the map 
    \begin{equation}
        D_R:\overline{\mc W} \times \overline{\mc W}\to \mb R_{\geq0} \doublecolon (x,y)\mapsto R(x)-R(y) - \lrangle{\nabla R(x), x-y}
    \end{equation}  
    is called the \emph{Bregman divergence} of $R$.
\end{definition}

Let us introduce the following regularizers and corresponding Bregman divergences on $\mb R^N, \mb R^N$ and $\mb H^{N\times N}$ respectively:
\begin{align}
    &R_{S}(x) = \sum_{i=1}^N x_i\ln(x_i)\,, && &&& D_{S}(x,y) = \sum_{i=1}^N\qty(x_i\ln\tfrac{x_i}{y_i} -x_i +y_i)\\[8pt]
    &R_2(x) = \tfrac{\norm{x}_2^2}{2}\,, && &&& D_2(x,y) = R_2(x-y) = \tfrac{\norm{x-y}_2^2}{2}\\[8pt]
    &R_{N}(X) = \tr[X\ln X]\,, && &&& D_N(X,Y) = \tr[X\ln X-X\ln Y] + \tr[X-Y]\,.
\end{align}
Here, $R_S$ is the negative Shannon entropy and $D_S$ is the Kullback-Leibler divergence, or relative entropy and $R_N$ denotes the negative von Neumann entropy and $D_N$ corresponds to the quantum relative entropy. 
A function $f$ is $\alpha$-strictly convex with respect to a norm $\norm{\cdotspace}$, if its Bregman divergence satisfies 
\begin{align}
    D_f(x,y)\geq\frac{\alpha}{2}\norm{x-y}^2\,.
\end{align}
Thus, by Pinsker's inequlity it holds that $\zeta_S=1$ with respect to $\norm{\cdotspace}_1$, trivially $\zeta_2=1$ with respect to $\norm{\cdotspace}_2$ and by the quantum Pinsker inequality $\zeta_N=\frac{1}{\ln2}$ with respect to $\norm{\cdotspace}_{\tr}$.

\begin{definition}[Mirror Map]\label{def:mirrormap}
    A map $R:\mc W\to\mb R$ is called a \emph{mirror map} if the following hold:
    \begin{itemize}
        \item $R$ is an $\alpha$-regularizer
        \item $R$ is differentiable on $\mc W$
        \item $\set{\nabla R(x)\midvert x\in\mc W}=\mc V$, we say the \emph{dual space} of $R$ is all of $\mc V$.
        \item The gradient diverges on the boundary $\lim_{x\to\partial\mc W}\norm{\nabla R(x)}=\infty$. 
    \end{itemize}
\end{definition}

\begin{definition}[Constraint Set]\label{def:constraintset}
    A set $\mc K$ is a \emph{constraint set} with respect to $\mc W$ if the following hold:
    \begin{itemize}
        \item $\mc K$ is closed convex 
        \item $\mc K\subseteq\overline{\mc W}$
        \item $\mc K\cap\mc W\neq\varnothing$.
    \end{itemize}
\end{definition}

\begin{definition}[Bregman Projection]\label{def:bregmanprojection}
    The \emph{Bregman projection} with respect to the mirror map $R$ over primal space $\mc W$ and with respect to constraint set $\mc K$ is defined as 
    \begin{equation}
        \Pi^R_{\mc K\cap\mc W}:\mc W\to\mc K\cap\mc W\doublecolon y\mapsto \argmin_{x\in\mc K\cap\mc W}D_R(x,y)\,.
    \end{equation}
\end{definition}

In general the $\argmin$ in \Cref{def:bregmanprojection} might not be well-defined as $\mc K\cap\mc W$ is not a closed set.
However, $D_R(\cdotspace, y)$ is $\alpha$-convex.
Moreover, due to the convexity of $R$ and as $\nabla R$ diverges on $\partial W$ the minimum cannot be attained at 
$\mc K \setminus \mc W \cap \mc K$ 
such that the $\argmin$ will indeed be attained in $\mc K \cap \mc W$. See also Lemma 3.11.3 in \cite{Portella2019}.

\section{Metric Variance}\label{app:metric-variance}
\addsectionheader{app:metric-variance}

The metric mean and variance generalize the mean and variance to general metric spaces. 
This is necessary in spaces, in which convex combination, or addition and scalar multiplication more generally, is not defined, or not defined everywhere. 
For all cases we have considered in this work the conventional definitions suffice and the metric mean and variance will reduce to the conventional notions $\Var[X]=\Ex[X^2-\Ex[X]^2]$.
However, for the sake of generality -- and simply because of the beauty of this perspective on the mean and variance -- we present here the more general framework for the metric mean and variance. 

\begin{definition}[Metric Variance and Metric Mean]
    Let $(M,\met)$ be a metric space and let $X$ be an $M$-valued random variable which is distributed according to $\lambda\in\Delta(M)$.
    \begin{itemize}
        \item We define the \emph{Metric Variance} of $X$ as
        \begin{align}
            \Var_{X\sim\lambda}[X] = \frac12 \Ex_{X_1, X_2\sim\lambda}\qty[\met(X_1,X_2)^2]\,.
        \end{align}
        \item Similarly, by
        \begin{align}
            \Ex_{X\sim\lambda}[X]
            =\argmin_{v\in M}\Ex_{X\sim\lambda}[\met(v, X)^2]\,,
        \end{align}
        we denote the \emph{Metric Mean} of $X$, if the $\argmin$ uniquely exists.
    \end{itemize}
    
\end{definition}

In the above definition we have borrowed the definition of the variance in metric spaces from \cite{leistnerVariance17} which is related by a factor of $2$ to the $2$-diameter as defined in \cite[Definition~5.2]{memoliGromovWasserstein2011}.
Crucially, for $M\subseteq\mb C^n$ with $\met(a,b)=\norm{a-b}_{2}$, the metric variance reduces to the conventional variance.
The metric mean is not necessarily well-defined: it may be non-unique or non-existent.
As such, it is astonishing that the metric variance does exist in any metric space.

Crucially, the connection between the variance and concentration represented by Chebyshev's inequality -- probably the most practical interpretation of the variance -- does carry over to the metric variance as the following lemma shows.

\begin{lemma}[Chebyshev's Inequality for the Metric Variance]\label{lem:metric-chebyshev}
    Let $X$ be a random variable over a metric space $(M,\met)$ distributed according to $\mu$. Then it holds
    \begin{align}
        \Pr_{X_1,X_2\sim\mu}\qty[\met(X_1,X_2)>2\tau] \leq \frac{\Var[X]}{\tau^2}\,.
    \end{align} 
\end{lemma}

\begin{proof}
    The claim follows immediately from Markov's inequality
    \begin{align}
        \Pr_{X_1,X_2\sim\mu}\qty[\met(X_1,X_2)>2\tau] 
        = \Pr_{d\sim\met\circ(\mu\otimes\mu)}\qty[d^2>(2\tau)^2]
        \leq \frac{\Ex_{d\sim\met\circ(\mu\otimes\mu)}[d^2]}{\tau^2}
        = \frac{\Var[X]}{\tau^2}\,.
    \end{align}
\end{proof}

Now, \Cref{cor:variance-bound-qnt} can be straightforwardly generalized by Chebyshev's inequality.

\begin{corollary}[Variance Bound for $q_\mtt{nt}$]\label{cor:variance-bound-qnt-metric}
    It holds
    \begin{align}
        \qnt[\mu,\tau] \geq \frac{\tau^2}{4}\cdot\Var_{s\sim\mu}[s(x)]^{-1}\,.
    \end{align}
    Further, if the metric mean $\overline{X}=\Ex_{s\sim\mu}[s(x)]$ exists for all $x\in\mc X$, and it holds $\met(v, \overline{X})^2\leq \frac12\Ex_X[\met(v, X)^2]$ for all $v\in M$, this can be improved to
    \begin{align}
        \qnt[\mu,\tau] \geq \tau^2\cdot\Var_{s\sim\mu}[s(x)]^{-1}\,.
    \end{align}
\end{corollary}

Note that $\met(v, \overline{X})^2\leq \frac12\Ex_X[\met(v, X)^2]$ holds in any convex metric space, and in words it merely states, that the deviation from the mean is, taking into account the weight due to $\mu$, symmetric around the mean.

\begin{proof}
    By Chebyshev's inequality \Cref{lem:metric-chebyshev}, for every $x\in\mc X$ most $s\sim\mu$ are $2\tau$ close to $v\in M$. Thus, there is a function $f:\mc X\to M$ such that for every $x$ the value $s(x)$ as $s\sim\mu$ will concentrate $2\tau$ close to $f(x)$ and hence $\kfrac(\mu, f, \tau)\leq \tau^{-2}\Var[s(x)]$.
    
    The second claim follows from 
    \begin{align}
        \Pr_X\qty[\met(X,\overline{X})>\tau] 
        = \Pr_X\qty[\met(X,\overline{X})^2>\tau^2]
        \leq \tfrac{1}{\tau^2}\Ex_X\qty[\met(X,\overline{X})^2]
        \leq \tfrac{1}{2\tau^2}\Ex_{X,Y}\qty[\met(X,Y)^2]
        = \tfrac{1}{\tau^2}\Var[X]\,.
    \end{align}
\end{proof}

\stopcontents[appendix]

%
%

\clearsectionfooter
\printbibliography

\end{document}